\documentclass[12pt]{article}

\usepackage{setspace,graphicx,epstopdf,amsmath,amsfonts,amssymb,amsthm}

\usepackage{verbatim}
\makeatletter
\def\includeversion#1{%
  \expandafter\def\csname #1\endcsname{}%
  \expandafter\def\csname end#1\endcsname{}%
  \expandafter\let\csname not-#1\endcsname\comment
  \expandafter\let\csname endnot-#1\endcsname\endcomment
  \expandafter\def\csname if#1\endcsname##1##2{##1}
}
\def\excludeversion#1{%
  \expandafter\let\csname #1\endcsname\comment
  \expandafter\let\csname end#1\endcsname\endcomment
  \expandafter\def\csname not-#1\endcsname{}%
  \expandafter\def\csname endnot-#1\endcsname{}%
  \expandafter\def\csname if#1\endcsname##1##2{##2}
}
\makeatother

\usepackage{marginnote,datetime,enumitem,rotating,fancyvrb}
\usepackage{hyperref,float}
\usepackage[longnamesfirst]{natbib}
\usepackage{booktabs}
\usepackage{placeins}
\usepackage{bbm}
\usdate

\usepackage[title,titletoc]{appendix}

\usepackage{tocloft}

\usepackage{adjustbox}

\usepackage[justification=centering]{caption}
\usepackage[margin=1in]{geometry}%

\usepackage{lscape}

\usepackage{indentfirst} 
\usepackage{endnotes}    
\makeatletter
\newtheoremstyle{jf}
{6pt}
{6pt}
{\itshape}
{}
{}
{:}
{.5em}
{}

\theoremstyle{jf}

\renewcommand{\@seccntformat}[1]{{\csname the#1\endcsname}.\hspace{1em}}

\def\thesection       {\Roman{section}}

\def\thetable         {\Roman{table}}

\renewcommand{\section}{\@startsection
  {section}{1}{0mm}{-3.5ex \@plus -1ex \@minus -.2ex}{2.3ex \@plus.2ex}{\centering\normalfont\Large\bfseries}}

\renewcommand{\subsection}{\@startsection
  {subsection}{2}{0mm}{-3.25ex \@plus -1ex \@minus -.2ex}{1.5ex \@plus.2ex}{\normalfont\large\itshape}}

\renewcommand{\p@subsection}{\thesection .}
\renewcommand{\p@subsubsection}{\thesection .}

\def\appendix{\par
  \setcounter{section}{0}
  \setcounter{subsection}{0}%
  \gdef\thesection{\@Alph\c@section}%
  \renewcommand{\theequation}{\thesection\arabic{equation}}%
  \renewcommand{\@seccntformat}[1]{{Appendix \csname the##1\endcsname}.\hspace{1em}}
  \renewcommand{\section}{\setcounter{equation}{0}\@startsection
    {section}{1}{0mm}{-3.5ex \@plus -1ex \@minus -.2ex}{2.3ex \@plus.2ex}{\centering\normalfont\Large\bfseries}}
}
\makeatother
\usepackage[labelfont=bf,labelsep=period]{caption}   
\usepackage{bm}
\usepackage{subcaption}

\usepackage{caption}
\usepackage{booktabs}
 \usepackage{threeparttable}
  \usepackage{longtable}
  \usepackage{array}
\newcolumntype{C}[1]{>{\centering\arraybackslash}p{#1}}
\usepackage[table]{xcolor}
\usepackage{makecell}
\usepackage{setspace}

\usepackage{fouriernc}
\usepackage[T1]{fontenc}

\usepackage{chapterbib} 
\usepackage[longnamesfirst]{natbib} 
\usepackage{tikz}
\usepackage{pgfplots}
\pgfplotsset{compat=newest}
\usepgfplotslibrary{groupplots}
\usetikzlibrary{intersections}
\usepgfplotslibrary{fillbetween}
\usetikzlibrary{patterns}

\usepackage[most]{tcolorbox}
\usepackage{enumitem}  

\usepackage{listings}

\lstdefinestyle{prompt}{
  basicstyle=\ttfamily\scriptsize,
  breaklines=true,
  breakatwhitespace=false,
  columns=flexible,
  keepspaces=true,
  showstringspaces=false,
  frame=none,
  aboveskip=0pt,
  belowskip=0pt,
}

\graphicspath{{figures/}}

\author{Gregor Schubert \\ UCLA Anderson
\thanks{\scriptsize I would like to thank participants at the SF Fed Micro Macro Labor Economics Conference, the UNC CREDA Real Estate Research Symposium, the UCLA Global Economics and Management Brown Bag Lunch, the UCLA Finance Brown Bag Seminar, the Hong Kong University Finance Brown Bag, the Stanford Remote Work Conference, the Wharton Business and Generative AI Conference 2025, the USC AI and Economics Conference, the UEA North America Meeting 2025, as well as Lindsey Raymond (discussant), Alexander Bick (discussant), Nick Bloom (discussant), Alan Kwan, Emma Harrington, Natalia Emanuel, and Anna Stansbury for helpful comments and discussion. I thank the UCLA Ziman Center for Real Estate’s Rosalinde and Arthur Gilbert Program in Real Estate, Finance and Urban Economics for generous funding. Email: gregor.schubert@anderson.ucla.edu. Website: \url{https://sites.google.com/view/gregorschubert}.} }
\title{Organizational Technology Ladders: Remote Work and Generative AI Adoption \\ } 
\date{\vspace{-2ex} \today \vspace{1ex} \\  \vspace{-2ex} }

\begin{document}

\begin{titlepage}
\maketitle
\thispagestyle{empty}
\begin{abstract}
\begin{spacing}{1.1}
This study proposes that firms move along an ``organizational technology ladder'': adopting one technology transforms hiring and work processes and builds skills and organizational capital that change the cost of adopting subsequent technologies. I study how firms' adoption of remote work technology during the COVID-19 period shaped later uptake of generative AI. Using U.S. job-posting data and an instrumental-variables strategy based on predicted differences in labor-market pressure to offer remote work, I estimate that a 10 percentage point increase in remote hiring in 2021–-2022 increases the share of job postings mentioning generative AI in 2023–-2024 by 0.4 percentage points across firms and 0.7 percentage points across occupations within firms. I provide evidence on mechanisms consistent with a technology-ladder channel: remote work adoption shifts hiring toward technical and managerial capabilities that predict faster conversion of generative AI exposure into adoption. Firms with return-to-office mandates--interpreted as revealing low remote productivity--exhibit a substantially larger response of generative AI adoption to remote work, consistent with an organizational frictions channel. I formalize the technology-ladder mechanism using a task-based model of sequential technology investments where remote work changes the costs and benefits of generative AI automation.
\end{spacing}
\end{abstract}

\end{titlepage}


In November 2022, just as firms were settling into post-pandemic work arrangements, ChatGPT's release  triggered one of the fastest waves of technology adoption in history: Within months, generative AI tools spread through organizations at unprecedented speed \citep{bick2024}, spurred by the potential to automate a large share of tasks \citep{eloundou2023} and by field experiments documenting large productivity gains  \citep{noy2023, brynjolfsson2023}. Yet adoption has been strikingly uneven. Some firms rapidly integrated these tools while others lag behind, creating early productivity gaps that may lead to persistent advantages \citep{bloom2021diffusion}. Understanding why requires looking beyond generative AI in isolation to examine how firms' prior technological choices shaped their readiness for this technological transformation.

This paper uncovers a fundamental dynamic in how organizations navigate technological change: firms move along an ``organizational technology ladder'' where each adopted innovation transforms their capacity to embrace the next. The massive shift to remote work during COVID-19 didn't just change where people worked---it also fundamentally altered firms' technological capabilities, hiring patterns, and organizational structures in ways that determined their ability to embrace generative AI when it arrived. Organizations that climbed the earlier rungs of the ladder, for example, by successfully adapting to remote work, positioned themselves to ascend to the next rung when generative AI opportunities became salient. 

My findings show a strong pattern of technological interdependence: Firms induced to accommodate more remote work during the pandemic subsequently adopt generative AI at significantly higher rates. I show that a plausibly exogenous 10 percentage point increase in remote hiring causes a 0.4-0.7 percentage point increase in hiring for generative AI capabilities.

I argue that these empirical patterns can be explained by the organizational technology ladder operating, on the one hand, through infrastructure complementarity, where remote work forces investments in digital systems that generative AI later builds upon; on the other  hand, it is supported by skill accumulation: adapting to remote work requires hiring technical and managerial talent that then enables AI implementation. 

My empirical strategy exploits exogenous variation in \emph{how much} remote work different firms were induced to accommodate during the pandemic. Specifically, I exploit exogenous variation in firms' exposure to labor market pressure to provide remote work options as a benefit to workers: The main instrument in my IV estimations is based on the firm's exposure to labor markets with competition from \textit{other firms} that are more likely to be offering remote work as a perk for their workers. An alternative instrument uses a firm's pre-pandemic exposure to labor markets with long commuting times as a proxy for greater employee demand for remote options during the pandemic. I also identify exogenous within-firm variation in remote work across different occupations by interacting the firm's labor market exposure with the occupation-level remote work feasibility (``teleworkability''). Identification in this approach comes from the effects of  differential labor-market pressure across remote-capable firms, controlling for both the independent effect on generative AI use of observable firm characteristics, pre-pandemic hiring characteristics, and the effect of labor market pressure on remote work among firms more generally.

Using these instruments, I provide causal estimates of the effects of a higher rate of remote work at a firm on its organizational investment in generative AI capability. The main empirical analysis uses data on job postings by firms from Jan. 2017 to Sep. 2024 and labels for which skills (e.g. generative AI) or work modalities (e.g. remote work) are mentioned in them.  I provide descriptive evidence that greater remote hiring during 2021--2022 is associated with higher generative AI adoption. Throughout, I measure organizational generative AI adoption using mentions of generative AI tools or skills in job posting text, which captures formalized hiring and stated work practices rather than informal employee experimentation.  Using the IV approach,  I find that, across firms, a 10 pp higher remote work share causes firms to increase their hiring for generative AI skills by at least 0.4 pp by 2024. Within firms, a 10 pp difference in remote work across occupations is associated with a 0.7 pp increase in generative AI mentions when hiring.  These effects are robust to controlling for a flexible set of control variables that capture firms' fundamental suitability for remote work and generative AI automation, their technology capabilities before the pandemic, as well as industry sector fixed effects; and also to using the alternative IV strategy based on variation in commuting time. Moreover, within-firm estimates control for both firm and occupation fixed effects, and results are robust to geographically disaggregating the data and including MSA fixed effects. The effects are especially pronounced in information- and decision-intensive sectors (including finance and business services), consistent with a mechanism where remote work changes coordination frictions and raises the value of automation in some sectors.

To understand why remote work and generative AI complement each other, I develop a model that rationalizes the observed patterns and suggests further tests to understand the proposed technology ladder mechanism. This model builds on the task-based approach to modeling automation \citep{acemoglu2022tasks, autor2024} but incorporates the organizational effects of the two technologies of interest: I model the effect of remote work as an increase in productivity by the worker that comes at the cost of lower decision-making quality. Generative AI is modeled as an automation technology for which the adoption cost depends on previous investments in information technology at the firm. Thus, if remote work leads firms to invest in technology infrastructure and skill, and these investments are partially reusable, it thereby lowers the incremental investment needed to implement generative AI. As long as these complementarities in technical infrastructure are large enough, a \textit{technology ladder} exists: remote work adoption makes generative AI easier to implement.

The mechanisms driving these effects are supported by the data: the importance of technology skill investments and organizational adaptation to  managing these new technologies is supported by findings that remote work adoption during the pandemic  causes firms to increase hiring for roles that are associated with higher decision-making skills, social skills and managerial responsibility; and also causes greater investments in technology skills. I also show that these skills are associated with a greater adoption of generative AI for a given exposure to the technology's benefits. As a result, technology ladder effects are stronger for firms that previously hired for more technology skills,  for more complex roles and for more managerial roles. As a case study for the importance of adaptation, I show that firms that issued return-to-office mandates---as a proxy for perceiving remote work to have relatively low productivity---have a significantly higher likelihood of investing in generative AI skills in response to higher remote work shares, suggesting they use automation to escape the productivity penalty of distributed teams.

To further explore the idea that sequential technology adoption leads to complementary investments, I also analyze earnings call transcripts  from Capital IQ using a multi-stage LLM classification pipeline. Evidence from discussions of remote work and generative AI in transcripts  highlights the enabling investments that link remote work to later generative AI adoption: for example, communication with investors highlights that remote work is associated with data infrastructure, cloud/remote-access capabilities, and security infrastructure investments, which are then also mentioned in the context of technology investments associated with generative AI adoption. Moreover, prior data infrastructure, as well as existing machine learning and analytics capabilities are highlighted as enablers for generative AI adoption.

These results confirm that firms that adopt one technology---remote work in this case---then benefit more from subsequent technology waves---here, generative AI---that build on the same investments in digital infrastructure and require similar changes in work processes.  This path dependency can further exacerbate differences in productivity among firms as the effects of the two biggest technological changes in firms' work processes during the last decade amplify one another in some organizations, and are dampened in others.

These findings also reveal how technological transformations cascade through organizations, creating persistent differences not just due to the diffusion of a single innovation but due to firms interacting with sequences of interlocking technology waves. Firms that successfully adopted remote work gained compound advantages in the form of better infrastructure and technical teams, and perhaps also a better managerial capacity for adaptation.  As generative AI capabilities expand and new technologies emerge, these diverging trajectories may lock in persistent productivity gaps between organizational technology leaders and laggards.

\textbf{Related Literature.} The proposed notion of a ``technology ladder'' that operates through organizational adaptation builds on earlier studies which showed that technology can affect the organization of firms and thereby change wage inequality \citep{garicano2006}, and that technologies differ in their effect on the centralization of decision-making within firms \citep{bloom2014}. Moreover, there is existing evidence that the effect of technologies on productivity depends on the degree to which complementary organizational changes are made \citep{bresnahan1996, brynjolfsson2000}. For example, \cite{brynjolfsson2016} show that greater IT investments and more educated workers are associated with greater adoption of data-driven decision-making by firms in the   2010s. Similarly, \cite{brynjolfsson2021productivity} argue that complementary firm investments in reorganization and implementing new decision processes are critical for unlocking the productivity benefits of new technologies. \cite{bresnahan2002} argue that this organizational adaptation can reinforce the effect of technologies on labor demand. The contribution in this paper is to propose a link between the effects of adopting one technology and the likelihood and effectiveness of adopting another, and also to provide detailed empirical evidence on the firm investments in skills that accompany technology adoption.

While this paper is, to my knowledge, the first study to draw a direct link between remote work and generative AI adoption, previous studies have found that  remote work benefits from existing investments in communication technology \citep{boeri2024work} in line with evidence that remote work creates more communication overhead \citep{bao2022, gibbs2023}. My finding that greater remote work may be associated with greater automation is also consistent with a literature that finds that remote work changes the attachment between firms and their employees, leading to a decline in the quality of employee engagement with co-workers \citep{gibbs2023, akan2024, emanuel2024}. 

My findings regarding the changes in hiring that enable generative AI adoption relate to an emerging literature that studies the effect of generative AI adoption on firms:  \cite{gulati2025} show that generative AI adoption in particular roles within firms is associated with higher demand for cognitive and social skills in job postings for those roles. \cite{alekseeva2024} show that non-generative AI adoption was also associated with greater hiring for managerial roles and demand for managers with more interpersonal and cognitive skills. \cite{humlum2025small} show, using survey data,  that the introduction of generative AI changes the task structure of the affected jobs.

This paper has the following structure: the next section discusses the data that I use. Section \ref{sec:ea} presents the empirical approach to estimating the causal effects of remote work adoption on generative AI. Then, I provide  a set of new stylized facts regarding the relation between remote work and generative AI in Section \ref{sec:facts}; and estimate the causal effect of remote work on generative AI adoption in Section \ref{sec:r2g}. To interpret the empirical findings, Section \ref{sec:theory} provides a simple conceptual framework. Section \ref{sec:mechanism} provides evidence on heterogeneity and mechanisms, including investments in technical skills, differential effects for firms with return-to-office mandates and an earnings call transcript analysis of technology investment narratives. Finally, Section \ref{sec:discussion} draws out the implications of these results and concludes.

\section{Data} \label{sec:data}

\subsection{Key variables}
Both hiring activity and investments in Generative AI skills and other characteristics of new jobs are measured using job postings data from Lightcast.\footnote{The Lightcast data vintage is as of October 18th, 2024.} This data consists of the near-universe of online job postings in the U.S. from Jan. 2010 to Sep. 2024. I filter job postings in the following way: (1) Drop job postings by staffing companies. (2) Drop all jobs flagged as internships. (3) Retain only full-time jobs, dropping part-time jobs. I crosswalk all occupational data to 6-digit SOC 2010 codes and refer to these as ``occupations'' throughout, unless otherwise noted.

\textbf{Remote work share.} The job posting data allows for the identification of both jobs that are fully remote and jobs that involve a hybrid format with some presence in an office. The classification of \textit{remote} jobs used in most of the analyses below represents all jobs that indicate either of these possibilities. For robustness checks, I also construct versions of the remote job definition where only \textit{hybrid} or only \textit{fully remote} jobs are included, but the results are qualitatively similar. When job postings are aggregated into firm or occupation or firm-by-occupation level panels, the counts reflect the number of job postings \textit{posted} in each category in the time period, which does not necessarily reflect actual hiring, but rather proxies for hiring demand.

\textbf{Generative AI adoption.} While it is difficult to observe the \textit{usage} of Generative AI within firms, the job posting data allow us to see when firms are either requiring Generative AI-related skills from new workers, or describing the work activities in a job as involving Generative AI tools. As a result, job posting data provides a window into whether and where in a firm the technology is used. While not all instances of generative AI use by individual workers will be captured in this way, this measure is likely to reflect a firm's official policy, which is what we are interested in when considering the degree to which organizational characteristics impact adoption.  Therefore, I will use mentions of Generative AI in job postings as a proxy for the degree to which those jobs involve using Generative AI-related tools. Labels for "Generative AI" skills mentioned in a job posting are supplied by Lightcast based on the job posting text. Section \ref{sec:calls} provides a  validation of this measure using Capital IQ earnings call transcripts, where  Figure \ref{fig:calltech}, panel A,  shows strong industry-level correspondence between generative AI use shares measured in earnings calls and in job postings.

\textbf{Generative AI exposure.} As a proxy for generative AI exposure based on the composition of the tasks done in a firm, I use the \cite{eisfeldt2023} measure of occupation exposure to generative AI. This measure uses the approach in \cite{eloundou2023} to first evaluate the task-level ability to potentially deploy generative AI productively and then aggregates from the set of tasks in an occupation to an occupation-level exposure score that represents the weighted share of tasks that could be made more productive with generative AI. I compute the exposure of a firm's hiring in a given year as a job posting weighted average over the exposure of the occupations that it is hiring for.

\subsection{Other data}

\textbf{Occupational wage and employment data.} I obtain data on employment and average wages by occupation and MSA from the Bureau of Labor Statistics' Occupational Employment Statistics. I crosswalk all occupational data to SOC 2010 codes and Core-Based Statistical Area codes (CBSAs).

\textbf{Occupation characteristics.} To evaluate whether the estimated effects differ between occupations with different characteristics, I use O*Net data on the characteristics of different occupations to construct summary measures that assess the prevalence of different skills in different occupations.   I construct the following measures of occupation characteristics:  To capture the importance of ``decision-making'' in a job, I average the level of related work activities involved in an occupation according to O*Net, based on \cite{deming2021}.\footnote{These activities are: Making decisions and solving problems; Developing objectives and strategies; Organizing, Planning and prioritizing work.} An indicator of social skills required for a job is constructed as described in \cite{deming2017}. I also construct a measure of job inflexibility based on \cite{goldin2014},and a measure of leadership skills based on \cite{schubert2019}, which is an average of the work activities for which a high value best predicts managerial leadership positions.  I also use a proxy for the remote work suitability of an occupation---the ``teleworkability'' measure constructed by \cite{dingel2020}.

\textbf{Skill demand.} Firm- and occupation-by-firm measures of skill demand are constructed directly from the job posting data by using tags provided by Lightcast for job postings that require a college degree, an advanced degree, or mention different skills. I focus on communication, decision-making, data management, data science, machine learning, and deep learning skills. I also construct an indicators for jobs that require ``high'' experience when at least 5 years of experience are required.

\textbf{Commuting cost.} To measure differences in commuting cost across different MSAs, I aggregate county-level data (using population weights) on average travel time to work by county from IPUMS NHGIS \citep{manson2024ipums}, which is based on the 2015-2019 American Community Survey data.

\textbf{Return-to-Office policies.} I obtain information on firms' return-to-office (RTO) policies by collecting crowdsourced public information from the website Flex Index. I match companies listed on Flex Index to companies in Compustat using company names and locations. The final dataset includes data on 1,336 companies and their office presence requirements. I classify a company as having provided a  ``Return-to-Office'' mandate, if its RTO policy requires  either a full-time presence, or specifies a non-zero minimum number of days in the office.\footnote{See Appendix \ref{sec:rtodata} for more information on the Flex Index data cleaning and the method for matching to Compustat.}

\textbf{Earnings call transcripts} for the Jan. 2019-Sep. 2025 period from Capital IQ/WRDS are used to provide evidence on technology mentions in investor communication Section \ref{sec:calls}.

\subsection{Sample description}

\textbf{Unit of observation.} Depending on the particular dimension of the variation in generative AI adoption that we are interested in, a different unit of observation is appropriate. One question is whether the \textit{same firms} are seeing higher remote work adoption and generative AI adoption, which matters for inequality in productivity across the economy as some firms might see compounding effects of technology, and might also provide evidence about the degree to which there is variation in corporate strategy across firms. This question is best studied using firm-level evidence.  At the same time, there are questions of whether generative AI adoption is more prevalent in the same occupations where remote work was adopted, which speaks to the degree to which these technologies complement or substitute for one another \textit{within} particular jobs, holding a firm's overall hiring and work processes constant. I will use a firm-by-occupation sample to study the latter questions, such that I can identify differences in effects across occupations within the same firm. In a robustness check, I also analyze data at the firm-by-MSA and at the firm-by-occupation-by-MSA level based on the ``location'' of a job posting---however, the location indicator for remote work is, by definition, not necessarily reliable or binding, and the IV approach is therefore not as compelling at this level of granularity, which is why these results are not one of the main specifications.

\textbf{Time periods.} To capture the sequential nature of the dynamics of interest, I use a number of different time periods: The main outcome variables, and a number of control variables are constructed for the 12 months ending with September 2024. The remote work adoption variable is computed over the 2021--2022 period, but excludes Q4 2022 in order to avoid contamination from the release of ChatGPT in November 2022. This period will be referenced as ```21-`22'' at times for ease of notation, but always omits Q4 2022. Moreover, several control variables and independent variables are constructed for data from single years, e.g. the instruments that are measured in 2019, or the skill variables measured in 2022. Unless otherwise specified, these samples include all months from that year.

\textbf{Summary statistics.}  The descriptive statistics of the two key samples in this paper, at the firm-level and at the occupation-by-firm level, are shown in Table \ref{tab:summary}. I limit the analysis in both samples to firms that have at least 10 job postings in the `23-`24 period in which we measure the generative AI adoption outcome variable. The average firm in our sample has 194 job postings overall in this period, of which \~0.1\% mention generative AI. At the occupation-by-firm level, I observe on average about 11 job postings.

The next section provides an empirical approach for estimating the causal relationship of interest with this data.

\section{Empirical Approach} \label{sec:ea}

The goal of the empirical estimation is to test whether there is in fact a causal ``technology ladder effect'', where one technology's adoption impacts the use of the other by firms. In a later section, I will explore the potential mechanisms that can explain such a relationship. I focus on remote work due to the COVID pandemic as a large, salient shock to firms' technology adoption that  preceded the emergence of widespread generative AI adoption with the release of ChatGPT in November 2022. Given the size and importance of both of these technology shocks, their relationship provides an important setting in which to understand the interaction between sequential technology waves in firms' adoption decisions.

Concretely, we first want to estimate the effect ${\beta}$ of having a greater share of remote jobs on firms' investment in generative AI skills. This effect is estimated from regressions of the form
\begin{align}
 \text{GenAIJobShare}_{i} =& \alpha + {\beta}\; \text{RemoteWorkShare}_{i} + \text{Controls}_{i} +\varepsilon_{i}, \label{eq:r2g}
 \end{align}
where the unit of observation $i$ can be a firm or an occupation-by-firm unit in the main analyses. The variables contained in  ${Controls}_{i}$ are detailed below when discussing how I address identification concerns. The main estimation will be cross-sectional, using job posting data for the 12 months ending in, and including, September 2024, to construct the dependent variable, and data for 2021 and the first 3 quarters of 2022 for the remote work share variable. However, some of the control variables are constructed in earlier periods to capture past characteristics of the firm, occupation, or location, and some of the mechanism analyses will use outcomes from earlier periods.

\subsection{Remote work and Generative AI: identification issues} \label{sec:ident}

Why would having  many workers in remote positions \textit{cause} a firm to adopt generative AI at a faster rate? To structure the empirical approach, it is important to first clarify the potential causal channels of interest and empirical challenges in identifying them.

The effects of remote work on generative AI adoption are potentially ambiguous: generative AI tools could complement remote workers more than in-office workers, for example building on their existing familiarity with digital workflows, or enabling better remote work tools. However, if generative AI is more likely to automate remote work tasks entirely, the use of these tools can also substitute for remote workers.  It is therefore an empirical question whether a greater use of remote workers is likely to lead to more or less investment in generative AI skills by firms.

 The organizational technology ladder mechanism where remote work adoption \textit{causally} affects generative AI adoption can arise from a number of different channels:
\begin{enumerate}
\item \textit{Augmentation:} Remote work jobs might benefit more or less in their productivity from applying generative AI tools, and this differential benefit of adoption can drive higher or lower demand for generative AI use.
\item \textit{Digital infrastructure:} the digital infrastructure, data processing, and other technical capacity improvements that result from implementing a remote work-friendly organization, may make it easier to adopt generative AI technologies.
\item \textit{Workflow restructuring:} by making workflows modular, asynchronous, decentralized, and not reliant on real-time in-person input, while incorporating mechanisms for remote quality validation, remote-adapted firms may be more suitable for inserting automated AI tools into the value chain.
\item \textit{Automation:} For some firms, remote workflows are less suitable, but they find themselves unable to enforce a post-pandemic return to in-person work. In that case, automation through generative AI can provide a channel for organizational adjustment that allows firms to reduce the share of their work done remotely.
\end{enumerate}
The main empirical analysis in this paper does not assume that a single mechanism must be driving any observed causal relationship, but I provide evidence that supports the existence of some of these when exploring potential mechanisms.
On the other hand, there are also possible mechanisms through which remote work and generative AI adoption may coincide in the same firms and occupations, but without a causal link from one to the other. For example, the following may be salient concerns:
\begin{enumerate}
\setcounter{enumi}{4}
\item \textit{Task characteristics:} it is possible that the tasks that are amenable to being done remotely also happen be the ones for which generative AI is useful. In that case, greater adoption of both technologies is driven by greater suitability for the technologies being correlated.
\item  \textit{Innovation orientation / tech-savviness:} some firms may have organizational cultures and leadership that are more open to, and capable at, embracing new technologies and being at the leading edge of innovation. This trait may make these firm more likely to experiment with, and adopt, both remote work and generative AI as the newest waves of technological progress.
\end{enumerate}

One key empirical issue that I address below is trying to distinguish these causal and non-causal channels.  To see that this is a relevant concern, I compare a proxy for the remote work suitability of an occupation---the ``teleworkability'' measure by \cite{dingel2020}---to the measure of generative AI exposure at the occupation level from \cite{eisfeldt2023}. Figure \ref{fig:exptele}, Panel A, shows that there is a positive relationship between occupations having tasks that are suitable for remote work and tasks that are exposed to generative AI capabilities.

The ideal setting for estimating the effect of remote work adoption on generative AI adoption would therefore require comparing groups of jobs that have similar \textit{suitability} for both remote work and generative AI deployment, and also are at companies with similar ``tech-savviness'' or innovative capacity. The ideal experiment to identify causal effects would then require one group of these jobs to experience greater prevalence of remote work for an exogenous reason, so that we can compare generative AI adoption rates to see what the magnitude of the causal channel operating \textit{through} remote work adoption is.

I approximate this natural experiment by controlling for  a large set of potential confounders motivated by the discussion above, and instrumenting for remote job prevalence using exogenous variation that is plausibly unrelated to firm-level differences in unobserved characteristics that might be driving generative AI adoption.

\subsection{Remote work instruments} \label{sec:instruments}

 To identify exogenous variation in the remote job share at the firm level, I exploit a novel source of variation in whether a firm adopts remote work at a high rate in a particular location: the interaction between  \textit{labor market pressure} and the \textit{ability} to let employees work remotely. That is, many studies (e.g. \citealp{maestas2023}; \citealp{powell2024}) have found that workers consider the ability to work remotely as a sizable non-monetary benefit.  Other studies have found that offering hybrid work options increases worker retention \citep{bloom2024hybrid}, and that remote workers are more productive as a result of shorter commute times and better work-life balance \citep{choudhury2024}.
 
 As a result, similar to the way that workers' outside options encourage firms to match wages on offer elsewhere \citep{schubert2024}, labor market competition induces firms to offer remote work options. This pressure should be larger in labor markets where \textit{other} employers are likely offering this perk, or where commuting is more burdensome and workers therefore likely perceive a larger benefit of working from home \citep{flynn2024}. Based on this intuition, I construct two novel instruments for remote work.

\textbf{Main instrument.} While the actual adoption of remote work by other employers in particular labor markets is both difficult to observe (as remote job locations are, by definition, not well defined in job postings), and might be simultaneously determined with post-pandemic choices by the firm of interest, the \textit{average pre-pandemic ability} to work remotely in the labor markets that a firm hires in is both observable and unlikely to be driven by the focal firm's ex post adoption behavior. Thus, I construct a firm-level instrument for remote work adoption based on labor market competition as
\begin{align*}
Z^{T}_{f} & = \left(\sum_m \phi_{fm, 2019} T_{m, 2019}\right) \times T_{f, 2019}\\
&= \underbrace{{\text{Avg. Labor Market Teleworkability}_{f,2019}}}_{\substack{\text{Remote work adoption potential} \\ \text{of the firm's labor market pre-Covid} } } \times \underbrace{{\text{Firm Teleworkability}_{f,2019}}}_{\substack{\text{Firm remote work adoption} \\ \text{potential pre-Covid} } },
\end{align*}
where $T_{m, 2019}$ is the average \cite{dingel2020} teleworkability among job postings in a particular MSA based on 2019 job postings, and the MSAs are weighted by the share $\phi_{fm, 2019}$ of all firm $f$ hiring done in each location $m$ as of 2019.  $T_{f, 2019}$ is the average teleworkability among job postings by firm $f$ in 2019. 

\textbf{Alternative instrument.}  To proxy for the pressure to permit remote work during and after the pandemic that arises from workers' perceived cost of commuting, I use estimates of the pre-pandemic average travel time to work in each MSA, based on ACS data from 2015-2019 (similar to the commuting distance instrument for remote work used by \citealp{kwan2024}). I aggregate these geographic proxies to a firm level measure of commuting cost using the firm's share of hiring in different locations in 2019. As the desire by employees to work remotely will only translate into remote jobs if positions at a firm are teleworkable, I again interact this measure of remote work pressure with the firm's teleworkability when constructing the commuting cost instrument:
\begin{align*}
Z^{C}_{f} & = \left(\sum_m \phi_{fm, 2019} C_{m, `15-`19}\right) \times T_{f, 2019}\\
&= \underbrace{{\text{Avg. Labor Market Commuting Time}_{f,2019}}}_{\substack{\text{Travel time to work} \\ \text{in the firm's labor market pre-Covid} } } \times \underbrace{{\text{Firm Teleworkability}_{f,2019}}}_{\substack{\text{Firm remote work adoption} \\ \text{potential pre-Covid} } },
\end{align*}
where $C_{m, `15-`19}$ is the average travel time by MSA, and the MSAs are weighted by the share $\phi_{fm, 2019}$ of all firm $f$ hiring done in each location $m$ as of 2019.

\textbf{Occupation-by-firm instruments.} As I also want to be able to estimate exogenous variation across occupations \textit{within firms}, I construct  occupation-by-firm-level instruments based on a similar intuition. For occupations with greater teleworkability, employers likely face more labor market pressure to grant this perceived perk relative to workers in the same firm whose jobs are not easily done remotely. Thus, adoption of remote work in an occupation-by-firm cell can vary independent of overall firm incentives to adopt remote work. To be specific, I construct occupation-by-firm instruments by interacting the firm-level instruments with the teleworkability $T_o$ in the occupation:
\begin{align*}
Z^{T}_{fo} & = \left(\sum_m \phi_{fm, 2019} T_{m, 2019}\right)  \times T_{o} \\
Z^{C}_{fo} & = \left(\sum_m \phi_{fm, 2019} C_{m, `15-`19}\right) \times T_{o}
\end{align*}
 For some robustness checks, I also use versions of the main instrument that induce exogenous variation at the firm-by-MSA or occupation-by-MSA level, which are constructed as
 \begin{align*}
Z^{T}_{fm} & =  T_{m, 2019} \times T_{f, 2019} \\
Z^{T}_{mo} & = T_{m, 2019}  \times T_{o}.
\end{align*}
 The uninteracted versions of the terms that are used in constructing the instruments are consistently included as control versions in the IV regressions. That is, the baseline IV estimation includes, for instance, a firm's average labor market teleworkability, and the firm's own teleworkability as separate control variables, and the identification relies on the variation in remote work induced by the interaction between these terms, while controlling for their direct linear effect.

\textbf{Control variables} The baseline firm-level control variables include characteristics of the firm's 12 months ending Sep. 2024 hiring that consist of the firm's teleworkability and generative AI exposure,  the share of job postings at the firm that require at least a college education or an advanced degree, and also industry sector (NAICS 2-digit) fixed effects. They also include  the firm's pre-pandemic 2019 hiring characteristics: separate variables that each capture the share of the firm's 2019 job postings that mention each of a list of technology skills (data management, machine learning, deep learning, artificial intelligence); the firm's 2019 remote hiring share; and the teleworkability and generative AI exposure of the firm's 2019 hiring. Moreover, baseline controls also include the characteristics of the firm's hiring markets before the pandemic--which are constructed as firm hiring-weighted 2019 averages across MSA characteristics--specifically, exposure to MSA remote work shares and MSA teleworkability. Together, these control variables are meant to capture the firm's overall technology orientation and tech-savviness, as well as the omitted variable bias from the observed overlap between tasks that tend to be teleworkable and those that have potential applications for generative AI tools.

\textbf{Exclusion restriction.} The shift-share instruments rely only on pre-Covid job compositions in different locations, firms and occupations, and are therefore unlikely to be correlated with later endogenous generative AI adoption choices by a firm, other than through the remote work channel, conditional on control variables that capture firm characteristics driving technology adoption in general. Formally,  the relevance condition in this setting requires that the \textit{interaction} between labor market suitability for remote work (due to competition or commuting costs) and a firm's ability to let workers go remote drives remote work adoption in a way that is not explained by either overall labor market suitability for remote work, or firm suitability on their own---as the latter uninteracted variables are included as control variables in the estimation. 

 The exclusion restriction then requires, for example, that if we compare two firms that have similar teleworkability, similar remote shares pre-pandemic, similar exposure to generative AI, are in the same industry, and have similar observable hiring for technology skills before the pandemic, then the one that finds itself in a labor market with more other firms that have high teleworkability does not systematically differ from the other firm in a way that drives generative AI adoption \textit{except} through greater remote work adoption during the pandemic. 

At the occupation-by-firm level, the exclusion restriction requires that greater pressure to adopt remote work, arising from a high teleworkability of the occupation and a firm's presence in labor markets with high teleworkability rates, eventually leads to differential generative AI adoption for some occupations within a firm, and that this effect operates only through a remote work channel. Importantly, in this setting we are able to condition on a firm's overall tendency to adopt generative AI, as well as an occupation's overall tendency to use generative AI, by including the corresponding fixed effects in the estimation.

\section{Remote Work and  Generative AI Adoption} \label{sec:r2hiring}

In this section, I first provide descriptive facts about the relation between remote work and generative AI that motivate the causal analyses. Second, I explore what firm characteristics are associated with greater adoption of remote work. Then, I provide evidence of a ``technology ladder effect:'' remote work adoption has a positive causal effect on the subsequent investment in generative AI skills.

\subsection{Stylized facts: remote work and generative AI} \label{sec:facts}

\begin{figure}[h]
\caption[.]{\\\textbf{Remote work and generative AI adoption by industry sector}}

\vspace{-0.1cm}  \small \label{fig:corrscatterintro} This figure plots the share of job postings in the 12 months ending Sep. 2024 that are for jobs that mention generative AI relative to those that are for remote jobs. The job postings data are from Lightcast and are aggregated into 2-digit NAICS industry sectors The red line indicates a linear best fit.

\centering
\includegraphics[width=0.75\textwidth]{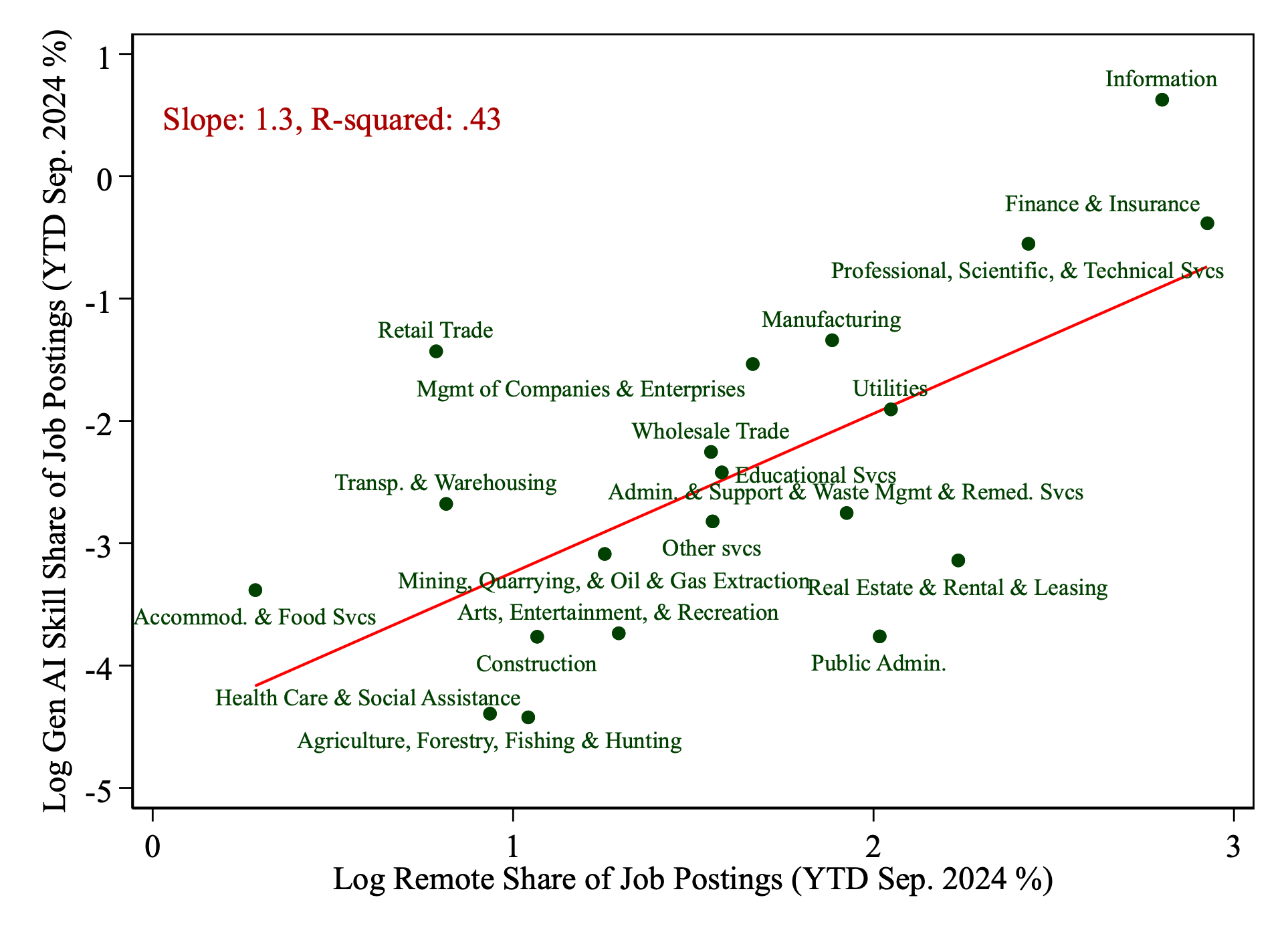} 
\end{figure}

 To understand whether firms are more or less likely to change their investments in hiring for generative AI skills if they already have a remote workforce, I start by providing descriptive evidence that these two technologies are likely to be connected. To motivate the causal estimation that follows, I first document a number of novel facts about remote work and generative AI skill demand:
 \begin{enumerate}
 \item \textbf{Industries with higher remote work prevalence also have higher generative AI adoption.} Figure \ref{fig:corrscatterintro} shows that the log of the remote share explains 43\% of the log generative AI adoption rate at the industry sector level
 \item \textbf{Firms that hire more remotely also hire more for generative AI skills.} Figure \ref{fig:corrscatterfirm} shows in job postings data for the 12 months ending Sep. 2024 that the log of the remote share explains 11\% of the cross-sectional variation in the log generative AI adoption rate across firms.
 \item \textbf{Generative AI adoption rates are higher in remote occupations:} 
Appendix Table \ref{tab:occlist} ranks 6-digit occupations by their prevalence of remote jobs (panel A) and generative AI (panel B) in job postings, as of the 12 months ending Sep. 2024, and shows the top 20 occupations. Occupations that adopt generative AI at high rates are similar between remote and non-remote workers, but generative AI adoption rates are higher among remote workers.  For example, Technical Writer job postings mention generative AI in 6.1 percent of all job postings, but in 16 percent of remote job postings.
\item \textbf{Higher remote hiring during the pandemic predicts rapid generative AI adoption.} 
Appendix Figure \ref{fig:rwgaiscatter}, panel A shows that occupations that adopted remote work at a faster rate from 2019 to 2022 also increased hiring more for generative AI-related skills from 2022-2024. 
 \end{enumerate}

 Next, I will implement the empirical approach discussed in Section \ref{sec:ea} to see whether this association is causal.

\subsection{Remote Work Impact on Generative AI Adoption} \label{sec:r2g}

To test the key hypothesis that there exists a  ``technology ladder effect,'' this section estimates the causal effect of remote work on Generative AI adoption both across firms, and across occupations within firms.

\begin{figure}
\caption[.]{\\\textbf{Firm-level and occupation-by-firm level IV first stage}}\label{fig:first}

\vspace{-0.1cm} \small  This figure shows binscatters of the first-stage relationships underlying the IV results in Table \ref{tab:firmocc_competition}. The first panel shows the effect of the interaction between a firm's exposure to MSA level teleworkability and firm-level teleworkability (both measured in 2019) on the horizontal axis, and the firm's job posting remote work prevalence in 2021-2022 (excluding Q4 2022) on the vertical axis.The second panel shows the effect of the interaction between a firm's exposure to MSA level teleworkability, and occupation-level teleworkability on the horizontal axis, and the occupation-by-firm's job posting remote work prevalence in 2021-2022 on the vertical axis. The values on both axes in both graphs are residualized with regard to the same baseline control variables as the IV results in Table \ref{tab:firmocc_competition}.

\centering
\includegraphics[width=0.65\textwidth]{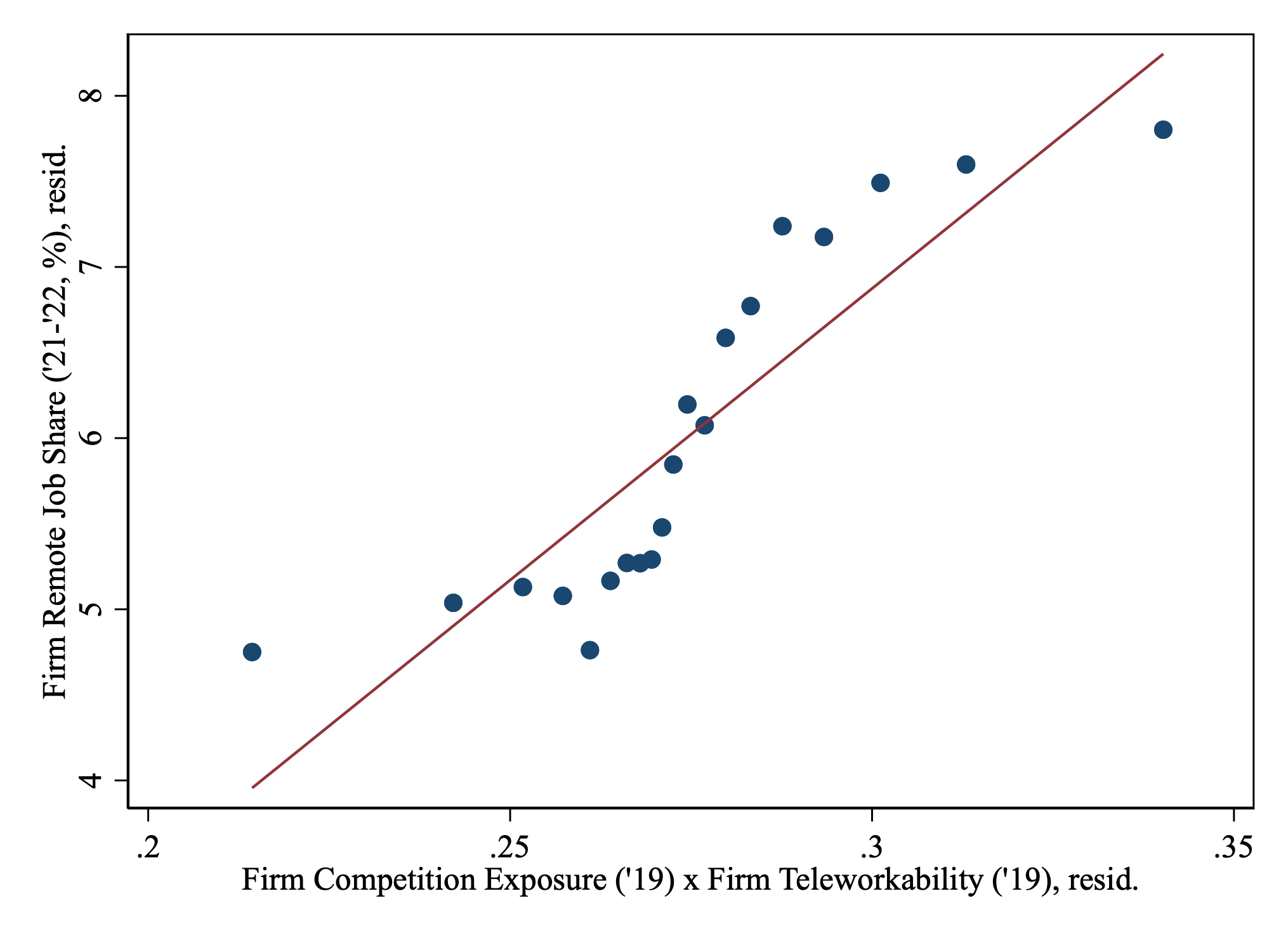} \\
\includegraphics[width=0.65\textwidth]{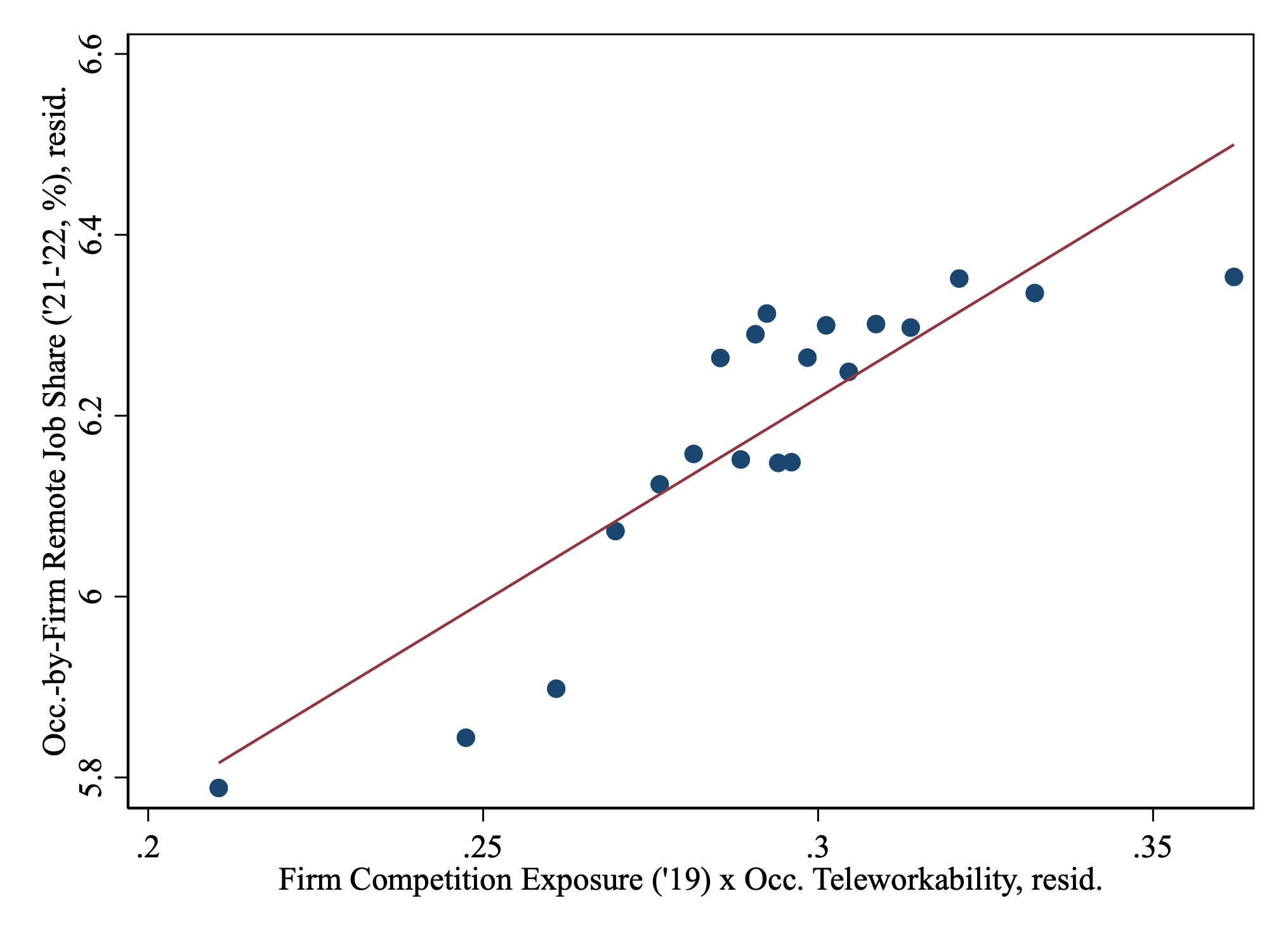} 
\end{figure}

\textbf{First-stage.} To visualize the relationship that underlies the variation in the instrument, Figure \ref{fig:first} plots the relationship between the competition IV instruments and the remote work shares at the firm and occupation-by-firm level, including a full set of control variables. As the figure shows, the instruments have a strongly positive relationship with the endogenous remote work prevalence. Moreover, while the slope of the relationship varies in steepness, the relationship does not show strong evidence of reversals or non-monotonicity.

\textbf{Firm-level results.} To what degree is the correlation between remote work shares at firms and their tendency to invest in hiring workers with generative AI skills causal? I use the IV approach described in Section \ref{sec:instruments} to estimate the equation \begin{align*}
100 \times \text{GenAIJobShare(Oct `23--Sep. `24)}_{i} = &  {\beta} \text{ RemoteWorkShare(`21-`22)}_{i}  + FEs  + \text{Controls}_{i}  +\varepsilon_{i}
\end{align*}
for the generative AI share for the 12 months ending Sep. 2024 period and the remote share during 2021-2022 (excl. Q4 2022). The results using the competition instrument are shown in Table \ref{tab:firmocc_competition}. Column (1) shows the firm-level OLS results, and column (2) the firm-level  IV results. The coefficient corresponds to the percentage point generative AI skill share effect of a 100 pp change in the remote share in the firm's job postings, such that the IV specification in column (2) suggests that a 10 pp higher remote share causes about a 0.4 pp higher generative AI skill share.  The baseline controls mean that the effect is unlikely to be driven by a sorting of jobs that are more suitable for both technologies into the labor markets that are affected  by the exogenous remote work variation induced by labor market competition. Note that the IV estimates are substantially larger than the OLS estimates, suggesting that the bias removed by the IV approach is that the firms which endogenously choose to offer remote work tend to be less likely to adopt generative AI.

\textbf{Within-firm results.} Is the coincidence between remote work and generative AI driven only by firm-level dynamics, or is there also a causal effect of one on the other across occupations \textit{within} firms?
Table \ref{tab:firmocc_competition}, column (4)  shows the results for this within-firm effect using the competition IV, additionally controlling for firm fixed effects and occupation fixed effects. This means that the effect is only identified off within-firm variation across occupations. The estimated effect suggests that higher remote work prevalence has a large positive causal effect across occupations within a firm (and across firms within an occupation), with a  10 pp higher remote share causing about a 0.7 pp higher generative AI skill share. This provides additional confirmation that the across-firm results are not driven by some unobserved firm-level attitude towards innovation---as such an effect would be eliminated by the firm fixed effects included here. 

These results also mean  that, while differences in corporate processes and capabilities as a result of firm-level changes in response to remote work can perhaps explain part of the puzzle of why the adoption of these two technologies is correlated, this is not the full story. The inclusion of occupation and firm fixed effects also means that the effects cannot be explained purely by a change in the composition of hiring by firms towards occupations that tend to use generative AI. The large within-firm causal effects suggest that  \textit{particular jobs} become more likely to use generative AI if those jobs themselves adopted higher rates of remote work at that firm.

\textbf{Results by sector.} In order to provide guidance to managers and other decision-makers, it is important to know whether the size of the technology ladder effect varies between different sectors. To answer these questions, I estimate the baseline IV regressions separately for broad industry sectors, which are defined thematically by aggregating 2-digit NAICS codes until there are at least about 10,000 firms in each broad sector.\footnote{I omit public administration firms (NAICS 92) from this analysis, which do not have a sufficient sample size and do not match thematically with other sectors.} While this sample splitting substantially reduces the power of the IV estimation, it reveals suggestive patterns about the relative size across sectors. The results for the firm-level IV analysis by broad sector are shown in Figure \ref{fig:r2genaibyind}: I find that the technology ladder effects are largest in the technology -related industries (NAICS 51 \& 54)\footnote{I use the definition from \cite{acemoglu2022} and label the information sector (NAICS 51) and the professional, scientific, and technical services sector (NAICS 54) jointly as the AI-related ``tech sector.''} and in the ``Financial Activities \& Business Services'' sectors (NAICS 52, 53, 55, 56), which contain finance and management industries.\footnote{This super-sector aggregates NAICS 52 (Finance and Insurance) with closely related business-services categories (NAICS 53, 55, 56) to maintain sample size and power in the split-sample IV estimates.}
While the results are not significant for individual sectors in the within-firm analysis due to the much smaller sample sizes, the results shown in Appendix Figure \ref{fig:r2genaibyind_within} show the largest (but not significant) coefficient for ``Financial Activities \& Business Services.''  These results confirm that the technology ladder mechanism has particular relevance for firms in the financial sector and in the technology industry. This differential response may reflect organizational adaptation to technological innovation---and I explore the drivers of this heterogeneity further in Section \ref{sec:mechanism}.

\textbf{Extensive margin effects.} As Table \ref{tab:summary} shows, the distribution of generative AI skill shares in job postings is very skewed, with most firms not mentioning the technology at all, while other firms mention it in more than 10\% of all job postings. This raises the question whether the remote work effect is driven only by the intensive margin of greater use of the technology or whether remote work also causes firms to mention generative AI at all. I repeat the analysis from Table \ref{tab:firmocc_competition}, but define the dependent variable as a dummy for whether the firm or firm-by-occupation cell has \textit{any} mentions of generative AI---and the results are shown in Appendix Table \ref{tab:firmocc_extensive}: I find that a 10 pp increase in remote work causes a \~0.9 pp increase in the likelihood of mentioning generative AI at all, both across firms and within firms, which is large relative to the baseline probability of mentioning generative AI in any job posting in the sample of 3.6\% at the firm level and of 0.6\% at the firm-by-occupation level.

\subsection{Robustness checks}

This section explores whether the finding of a causal effect of remote work on generative AI adoption is robust to different identification concerns and empirical specifications.

\textbf{Weak IV tests.} Do we need to be worried about ``weak instrument'' bias in these estimations? \cite{olea2013} suggest constructing their ``effective F'' statistic, which is robust to heteroscedasticity, autocorrelation, and clustering, and to reject the hypothesis that weak instrument bias exceeds 10\% of a worst-case benchmark at a 5\% significance level when this effective-F statistic  for the first stage exceeds an appropriate value. They show that this critical value is below 23.1, making 23.1 a conservative rule-of-thumb cutoff. In the setting that I study, with one endogenous regressor and one instrument, the effective F-statistic is identical to the Kleibergen Paap F-statistic (KPF), which I report in all IV regression tables. I find that the KPF is greater than 23.1 in all main analyses (see Table \ref{tab:firmocc_competition}), which means that it is unlikely that there is weak instrument bias.

\textbf{Fully-remote and hybrid-only jobs.} While the remote share measure in the baseline estimation includes both fully remote and hybrid jobs, the Lightcast data allows me to confirm whether the causal relation between remote work and generative AI is robust to using alternative measures of remote work. Columns (1)-(4) of Table \ref{tab:robust_rwgai} show results analogous to the baseline IV estimates in Table \ref{tab:firmocc_competition}, but with remote work shares defined using only fully remote or only hybrid jobs. As the results show, the estimates are economically and statistically significant even when using these alternative definitions of ``remote'' work, suggesting that this finding is not sensitive to distinguishing precisely which form of remote work an employer offers.

\textbf{Exclude tech sectors.} One potential concern with the estimation is that there may be broader concurrent economic trends that happen to cause an increase in generative AI hiring precisely for those firms that also happened to be more inclined towards remote work for exogenous reasons. The baseline estimations already include 2-digit industry or firm fixed effects, eliminating the possibility that industry-level trends are driving the results. However, it could still be the case that the results are driven by the dynamics \textit{within} only particular sectors. In particular, technology companies faced a dramatic rise and decline in hiring during and after the pandemic, which may lead their hiring responses to be distorted. Moreover, technology companies may be more responsive to a new technology trend---generative AI---than companies in other sectors, as they supply AI services and software and hardware used for AI products themselves.   In columns (5) and (6) of Table \ref{tab:robust_rwgai}, I exclude the tech sector from the analysis and find that, while this halves the estimated effect sizes across firms in column (5), the effect of remote work on generative AI across jobs within firms in column (6) is very similar to that in the corresponding baseline estimation in column (4) of Table \ref{tab:firmocc_competition}. This means that the technology ladder effect is by no means limited to technology firms.

\textbf{Alternative instrument.} In the estimation in Table \ref{tab:firmocc_competition}, I used the instrument based on labor market competition from other teleworkable firms. However, there may be residual concerns that this instrument selects for firms that are unobservably different due to their sorting into labor markets with particular types of competitors. To allay this concern, I repeat the same estimation with the instrument based on average commuting time in a firm's hiring labor markets. If one believes that this source of variation is more plausibly excluded from the estimation equation, conditional on the other controls, then the relevant IV estimates are shown in columns (7) and (8) of  Table \ref{tab:robust_rwgai}: the effect estimates are also significant and positive and of a comparable magnitude to those in Table \ref{tab:firmocc_competition}, albeit with a large coefficient estimate for the within-firm effects. The commuting instrument operates through geographic commuting costs that raise the relative benefit of remote arrangements, which differs from the labor-market pressure from other employers driving the variation in the competition instrument. As a result, any violation of the exclusion restriction would likely differ across these channels, and the fact that I find very similar estimates narrows the set of plausible omitted variables that could explain the results in both. Thus, the alternative IV results provide additional support for the exclusion restriction of the baseline IV.  However, the first-stage Kleibergen-Paap F-statistics in columns (7) and (8) of  Table \ref{tab:robust_rwgai} also suggest that the commuting time IV is a weaker instrument than the competition IV, which means that the latter yields more precise estimates. While the key results in this paper are robust to using either of these instruments, I therefore focus on the competition IV as the main source of variation.

\textbf{Geographic variation.} An additional concern may be that the instrument based on labor market competition captures the association of particular geographic areas (e.g. the Bay Area / Silicon Valley) with both remote work and generative AI adoption. To allay this concern, I construct versions of my data that are disaggregated at the firm-by-MSA and firm-by-occupation-by-MSA level.\footnote{An important caveat here is that ``location'' for fully remote workers is somewhat ill-defined in job postings, as the location listed in the job posting will reflect some reference work location that the worker does not necessarily need to spend time at. However, to the degree that this hiring location still captures key aspects of differential labor market dynamics, we can use it to control for differences in technology use that are associated with being affiliated with a particular branch location (even if that is not where the worker is physically located).} Then, I repeat the baseline IV estimation, but  including MSA fixed effects.\footnote{This also requires using instruments with MSA-level variation, which are discussed in Section \ref{sec:instruments}.} Column (9) shows that the firm-level results are robust to including not just the baseline 2-digit industry fixed effects but also MSA fixed effects. Column (10) shows that the estimated causal effect is also significant and large when including MSA and firm-level fixed effects such that effects use only within-firm variation across locations. Last, column (11) shows that there is even a large and significant effect of remote work on generative AI adoption across MSAs within a firm when we are able to control for firm-level, MSA-level, and occupation-level fixed effects.

\textbf{Tech hubs.} One identification concern is that firms that sort into particular geographic technology hubs face unobserved shocks to AI capability, talent, or complementary investments that could both increase the incentive to offer remote work and also to accelerate generative AI adoption, potentially violating the exclusion restriction. By excluding any data from the most generative AI-oriented tech hub cities, we can remove the subset of observations where these ``ecosystem'' factors may be most concentrated. As shown in Appendix Figure \ref{fig:genaicities}, generative AI hiring shares are distinctly higher in San Jose, San Francisco, and Seattle relative to other large U.S. cities. Column (12) of  Table \ref{tab:robust_rwgai} shows that the remote work effect on generative AI adoption is significant and of a similar magnitude as in the baseline estimation when I exclude these top 3  technology hubs for generative AI from the analysis. As tech hubs are where omitted variables issues are most plausibly distorting the results, the stability of the IV estimates is reassuring with regard to the plausibility of the exclusion restriction. 

\textbf{Placebo test and pre-trends.} The most likely exclusion restriction violation is that the  instrument is correlated with an unobserved firm or occupation-by-firm trait like ``tech-savviness'' or ``innovation capacity'' that drives both remote work adoption and generative AI adoption and is also correlated with the labor market pressure due to teleworkability of competitive employers. One implication of such a trait would be that it should also predict the pre-COVID trend in hiring characteristics in technology-relevant dimensions. As a falsification test for the exclusion restriction, we can therefore conduct a ``placebo analysis'' for the IV approach, by estimating the effect of instrumented remote work during COVID on the characteristics of pre-COVID job postings.

I focus on the ability to predict the pre-trend in hiring for advanced degrees, (non-generative) AI skills, and remote positions as proxies for the degree to which the pandemic period instrument captures  a broader technology orientation of firms. Appendix Table \ref{tab:r2gplacebo} shows that the instrumented remote work share 2021-2022 does not predict changes in remote hiring, AI skill demand or advanced degree hiring for 2017-2019 across firms (columns 1-3). The same is true for within-firm results (columns 4-6), but the sample size for this longer placebo analysis sample is too limited for the IV to have a strong first stage in the within-firm regressions. These placebo analysis results does suggest that the across-firm findings are unlikely to simply reflect a broader firm-level innovation orientation that correlates with the instrument and persistently drives firm-level technology adoption trends.

\section{A Simple Model of Organizational Technology Ladders} \label{sec:theory}

Why would having  many workers in remote positions \textit{cause} a firm to adopt generative AI at a faster rate? In this section, I provide a parsimonious model that rationalizes the empirical finding that remote work adoption during the pandemic causally increases subsequent generative AI adoption. The model builds on task-based approaches to modeling automation \citep{acemoglu2022tasks, autor2024} but emphasizes two organizational features that are central in this setting: (i) remote work trades off effective labor input against coordination and decision quality, and (ii) technology adoption is path dependent because investments in ICT infrastructure and complementary capabilities are partially reusable across technology waves. The framework is intentionally stylized and partial equilibrium. Its goal is not to fit all features of remote work or generative AI adoption, but to isolate the minimal forces that generate an ``organizational technology ladder,'' where adoption of one technology changes the incentives for adopting the next. Moreover, it generates predictions that discipline the mechanism evidence in Section \ref{sec:mechanism}.

\subsection{Production with remote time savings and decision quality}

Consider a firm $f$ employing a production worker in occupation $j$.
Production consists of a continuum of tasks indexed by $x\in[0,j]$, ordered by automatability, where lower-$x$ tasks are easier to automate. Let $k_{fj}\in[0,j]$ denote the current automation frontier, so tasks $x\in[0,k_{fj}]$ are automated and tasks $x\in(k_{fj},j]$ require human labor.

\paragraph{Decision quality and coordination.} Human time spent on tasks produces additional output as a result  of good decision-making, which varies across different human tasks depending on how sensitive they are to effective managerial guidance and coordination.
Let $D_{fj}\ge 0$ denote the decision sensitivity of human tasks. Then, the additional output above baseline from good decision-making is given by $D_{fj} \cdot q_{f}(M_f)$, where $q_{f}(M_f)$ is the decision-making quality at a job, which depends on  the level of managerial support $M_f$ at the firm. Technology in this setting can affect output through three channels: (i) changing the productivity of human labor in human tasks; (ii) changing the decision-making quality; or (iii) changing the automation frontier and shifting tasks away from humans.

\paragraph{Remote work effects.}
Let $R_f\in\{0,1\}$ indicate whether a job is performed remotely. I assume that remote work yields a productivity or time-savings benefit $r_j$ on \emph{human-executed} tasks,  motivated by empirical observations that remote workers save time on grooming and commuting, and that some of these time savings are used to work more \citep{barrero2020, pabilonia2022}. This productivity boost can vary across occupations $j$, such that there might be more remote-suitable jobs with a high $r_j$ and other jobs where $r_j$ is small or even negative \citep{dingel2020}.

Remote work also affects the quality of decision-making. Let the effect of remote work on decision support be summarized by
\[
q_f(M_f,R_f) \equiv \ln M_f - R_f \rho_f,
\]
where $M_f>0$, and  $\rho_f \ge 0$ captures a coordination discount. The restriction $\rho_f \ge 0$ reflects that co-location facilitates coordination, while distributed work can reduce effective managerial input \citep{choudhury2025}.

Log output in this setting is represented in reduced form as
\begin{equation}
\ln y_{fj}(R_f,k_{fj})
=
A_{fj}(k_{fj})
+
(j-k_{fj})\left(\,r_j\,R_f
+
D_{fj}\cdot q_f(M_f, R_f)\right)
\label{eq:output_reduced}
\end{equation}
where $A_{fj}(k_{fj})$ collects terms that do not depend on remote status (including the direct output contribution of automated tasks). Equation~\eqref{eq:output_reduced} isolates the key organizational trade-off: remote work raises effective labor input on human tasks through $r_j$ but reduces decision quality through $\frac{\partial q_f}{\partial R_f} <0$.

\subsection{Digital capability and technology implementation costs}

Firm's have a baseline \emph{digital capability stock} (ICT infrastructure, data assets, data management capabilities, etc.) as a result of investments in technologies $\chi\in \mathbbm{T}$ previously implemented by the firm.
Consider a new technology $\tau$ that the firm wants to implement, which requires a level of technology-specific capabilities $S^\tau$.  Crucially, previous technology investments have \textit{reusability} $\omega_{\tau}$, such that the implementation cost of the new technology is lower if the firm has invested in other advanced technologies (high $S^{\chi}$), which required similar digital capabilities (high reusability $\omega_{\tau}$).

Let $H_f$ denote a firm's tech-skill intensity (e.g., pre-pandemic hiring shares for technology roles). I allow the reusability of past technology investments to be firm-specific and to depend on the firm's technology skill:
$\omega_f\equiv \omega(H_f)\in[0,1]$ with $\omega'(H_f)>0$.

 Therefore, implementation costs for reaching the required capability level for the new technology depend on the capability gap:
\begin{equation}
c_f(\tau) = \big(S^\tau - \sum_{\chi \in \mathbbm{T}} \omega_f S^{\chi}\big)_+,
\label{eq:cost}
\end{equation}
where $(x)_+\equiv \max\{x,0\}$. Therefore, a new technology, such as generative AI, is cheaper to implement if the firm has existing technologies with some overlap in capabilities, and high technology skills to enable the adaptation of the existing technology for generative AI.

\subsection{Remote work, generative AI, and the ladder mechanism}

\paragraph{Reusability of remote work technology.}
While this framework can be applied more generally to think about sequential technology investments, in the setting of interest two technology waves arrive: remote work (first) and generative AI (later). Let $S^R$ denote the capability requirement for remote work, and $S^G$ the capability requirement for generative AI. For expositional clarity, assume $S_f^0<S^R<S^G$, where $S_f^0$ is the sum of pre-pandemic investments in technologies other than remote work. Moreover, $\omega_f$ is the fraction of these past technology investments that is reusable for generative AI.\footnote{Here, 
$\omega=1$ corresponds to full reuse (e.g., shared cloud/data infrastructure and similar engineering talent required);
$\omega=0$ corresponds to no reuse (purely remote-specific tools that do not carry over to generative AI use cases).
} Accordingly, the generative AI implementation cost depends on whether the firm previously adopted remote work
\begin{equation}
c_f(G\mid R_f)
=
\Big(S^G - \omega_f \big(S_f^0 + S^R\,R_f\big)\Big)_+
\label{eq:genai_cost}
\end{equation}

\paragraph{Reusable capability channel.}
From \eqref{eq:genai_cost}, the \emph{reusability benefit}---the reduction in generative AI implementation cost induced by prior remote adoption---is
\[
\Delta c_f^G
=
\Big(S^G-\omega_f S_f^0\Big)_+ - \Big(S^G-\omega_f(S_f^0+S^R)\Big)_+ \;\ge\; 0.
\]
In the interior case where $(\cdot)_+$ does not bind, $\Delta c_f^G=\omega(H_f)S^R$, so the cost complementarity from remote work is increasing in the firm's tech-skill intensity $H_f$ when $\omega'(H_f)>0$. This makes one mechanism for a technology ladder explicit:  generative AI adoption is cheaper if reusability $\omega_f$ is higher due to a firm having high tech skills, and if firms invested substantially in remote work technology.

\paragraph{Generative AI adoption.} Generative AI is modeled as expanding the automation frontier by an occupation-specific amount $g_j\ge 0$ if the firm adopts the technology:
\begin{equation}
k_{fj}^G = k_{fj} + g_j\cdot \mathbf{1}\{\text{GenAIAdopt}_f=1\}.
\label{eq:kG}
\end{equation}
The parameter $g_j$ captures the occupation's generative AI automation potential (empirically proxied by occupation-level generative AI exposure measures).

Given remote status $R_f$, the firm adopts GenAI when the gain in log output exceeds the implementation cost:
\begin{equation}
\ln y_{fj}(R_f,k_{fj}^G) - \ln y_{fj}(R_f,k_{fj})
\;\ge\;
c_f(G\mid R_f).
\label{eq:adopt_genai}
\end{equation}

The model's implications can be summarized in two propositions (derivations in Appendix \ref{sec:deriv}):

\paragraph{Proposition 1 (Technology ladder):} \textit{Firms that adopt remote work have a higher incentive to also adopt generative AI.}

A sufficient condition for a positive ladder effect where remote work adoption increases the incentive to adopt generative AI is
\begin{equation}
\Delta c_f^G +  g_j\,\rho_f \, D_{fj} \;>\; g_j r_j.
\label{eq:ladder_condition}
\end{equation}
That is, the ladder effect occurs if the work-from-home productivity boost for the automated tasks is not too large relative  to the  technology investment cost savings for generative AI adoption when already having implemented remote work and the decision-making output penalty among remote workers.

\medskip
Proposition~1 corresponds to the baseline ``technology ladder effect'' estimated in Section \ref{sec:r2g} using IV variation in remote work intensity.

\paragraph{Proposition~2 (Heterogeneity of the ladder effect):} 
\textit{The technology ladder effect is stronger when: (i) remote coordination is less effective (higher penalty $\rho_f$); 
(ii) the set of tasks automated by generative AI is more decision-sensitive (larger $D_{fj}$);
 (iii) remote work productivity benefits $r_j$ are smaller; and (iv)  firm technology skill intensity is higher (higher $H_f$ and hence higher $\omega_f=\omega(H_f)$).}

Decision-critical environments---such as financial and business decision-making settings---naturally correspond to higher decision sensitivity. Moreover, interpreting return-to-office (RTO) mandates as revealing low perceived remote effectiveness (a higher penalty $\rho_f$), Proposition~2 motivates the RTO heterogeneity tests.

\subsection{Mapping to empirical tests}

The empirical estimates in Section \ref{sec:r2g} directly test Proposition~1 by estimating the causal effect of remote work intensity in 2021--2022 on generative AI-related hiring in 2023--2024.

Section \ref{sec:r2skill} provides evidence for one mechanism behind Proposition~1 by documenting remote-induced changes in hiring toward technology and managerial capabilities. This evidence combines with evidence in Section \ref{sec:g2g} and Section \ref{sec:heterogeneity} to show that technology skills then enable greater generative AI adoption, which together supports Proposition 2 (iv). Proposition~2 (ii) can be mapped onto the heterogeneity in the ladder effect results by sector. Section \ref{sec:rto} tests Proposition~2 (i) and (iii) by interacting remote work with return-to-office mandates as a proxy for low remote effectiveness (a higher penalty $\rho_f$ or low benefits $r_j$). Finally, section \ref{sec:calls} provides direct evidence from investor communications that firms associate remote work adoption and generative AI adoption with investments in particular technologies, and that these technologies overlap, in line with the proposed reusability mechanism.

\section{Mechanism: organizational adaptation} \label{sec:mechanism}

The empirical results in Section \ref{sec:r2g}  showed that greater remote work adoption leads companies to adopt generative AI at higher rates.  In this section, I provide evidence on organizational adaptation, showing how remote work changes skill demand in ways that facilitate subsequent AI deployment.

\paragraph{Finance and decision-critical work.} One piece of evidence for the mechanism is the earlier finding that the technology-ladder effect is not uniform across industries. As shown in Figure \ref{fig:r2genaibyind}, the IV estimates are largest in the technology sector and in ``Financial Activities \& Business Services'' (NAICS 52, 53, 55, 56), a group that includes finance and management industries. Within-firm sector estimates are noisier, but Appendix Figure \ref{fig:r2genaibyind_within} also shows the largest (though statistically insignificant) coefficient for this finance-related super-sector. This sector pattern is consistent with the model's emphasis on decision sensitivity: in decision-critical environments such as finance, coordination frictions and decision quality are first-order, increasing the value of organizational adaptation and the potential appeal of generative AI-based automation, especially if remote work deteriorates the quality of critical activities.

I organize the additional mechanism evidence below around two channels emphasized by the conceptual framework:
(i) \emph{capability accumulation}, where remote work adoption induces investments in reusable digital and organizational capabilities that lower the incremental cost of deploying generative AI;
and (ii) \emph{coordination frictions}, where firms with larger remote-work coordination losses have stronger incentives to substitute toward automation.
Empirically, I first estimate the causal effect of remote work on intermediate hiring and skill investments, and then show that these same capabilities predict whether firms translate a given level of generative-AI exposure into adoption. Then, I use return-to-office mandates as a proxy for low adaptation to remote work to test whether the remote work impact on generative AI adoption is amplified when remote work is perceived to be less effective. Finally, I show that firms highlight investments in particular complementary technologies when discussing investments in remote work or generative AI.

\subsection{Building the organizational technology ladder: remote work and hiring} \label{sec:r2skill}

This section tests how the adoption of remote work changed firm-level hiring patterns with regard to skills. These results help us understand through which mechanism the technology ladder effects on subsequent generative AI adoption might be operating. I estimate specifications of the form 
\begin{align}
 \text{Skill(2022)}_{i} = \alpha + {\beta} \text{ RemoteWorkShare(`21-`22)}_{i} + \gamma  \text{Skill(2019)}_{i} + \text{Controls}_{i} +\varepsilon_{i}, \label{eq:r2hiring}
 \end{align}
where the dependent variable captures different measures of the composition of the firm's hiring with regard to indicators of skill in its job postings, or the average characteristics (derived from O*Net scores) of the occupations that the firm is hiring for. These regressions also control for the past level of the dependent variable pre-pandemic, so the estimated coefficients capture the effect on changes in the skill composition of job postings between the period before and after the lockdown period of the pandemic.

\begin{figure}
\caption[.]{\\\textbf{Effects of remote work adoption on firm hiring characteristics}}\label{fig:rw2skill}

\vspace{-0.1cm} \footnotesize This figure shows coefficients estimated using IV for the effect of remote work prevalence in a firm's  2021/2022 (excl. Q4 2022) job postings on the standardized characteristics of the firm's job postings in 2022 in a regression of the form
$$ \text{Skill(2022)}_{i} = \alpha + {\beta} \text{ RemoteWorkShare(`21-`22)}_{i} + \gamma  \text{Skill(2019)}_{i} + \text{Controls}_{i} +\varepsilon_{i}, $$
where the controls in all regressions include the standardized 2019 average value of the dependent variable as a control variable, so the coefficients can be interpreted as the effect of changes in remote work shares on changes in the composition of hiring from before to after the pandemic. The instrument consists of the interaction between firm-level exposure to MSA teleworkability through its hiring labor markets and firm-level teleworkability (all measured in 2019). All regressions also  include the following control variables: NAICS 2-digit fixed effects; company's remote work share in 2019, the company's uninteracted exposure to MSA teleworkability in 2019, uninteracted firm-level teleworkability in 2019, the company's share of jobs requiring a college education and the share requiring an advanced degree in 2019, the share of the company's job postings in 2019 that was for computer occupations or manager positions; the log of total job postings in 2019 and in 2022; the company's labor market exposure to MSA remote shares in 2019. The 95\% confidence intervals shown are based on heteroskedasticity-robust standard errors clustered at the firm level.

\centering
\includegraphics[width=0.9\textwidth]{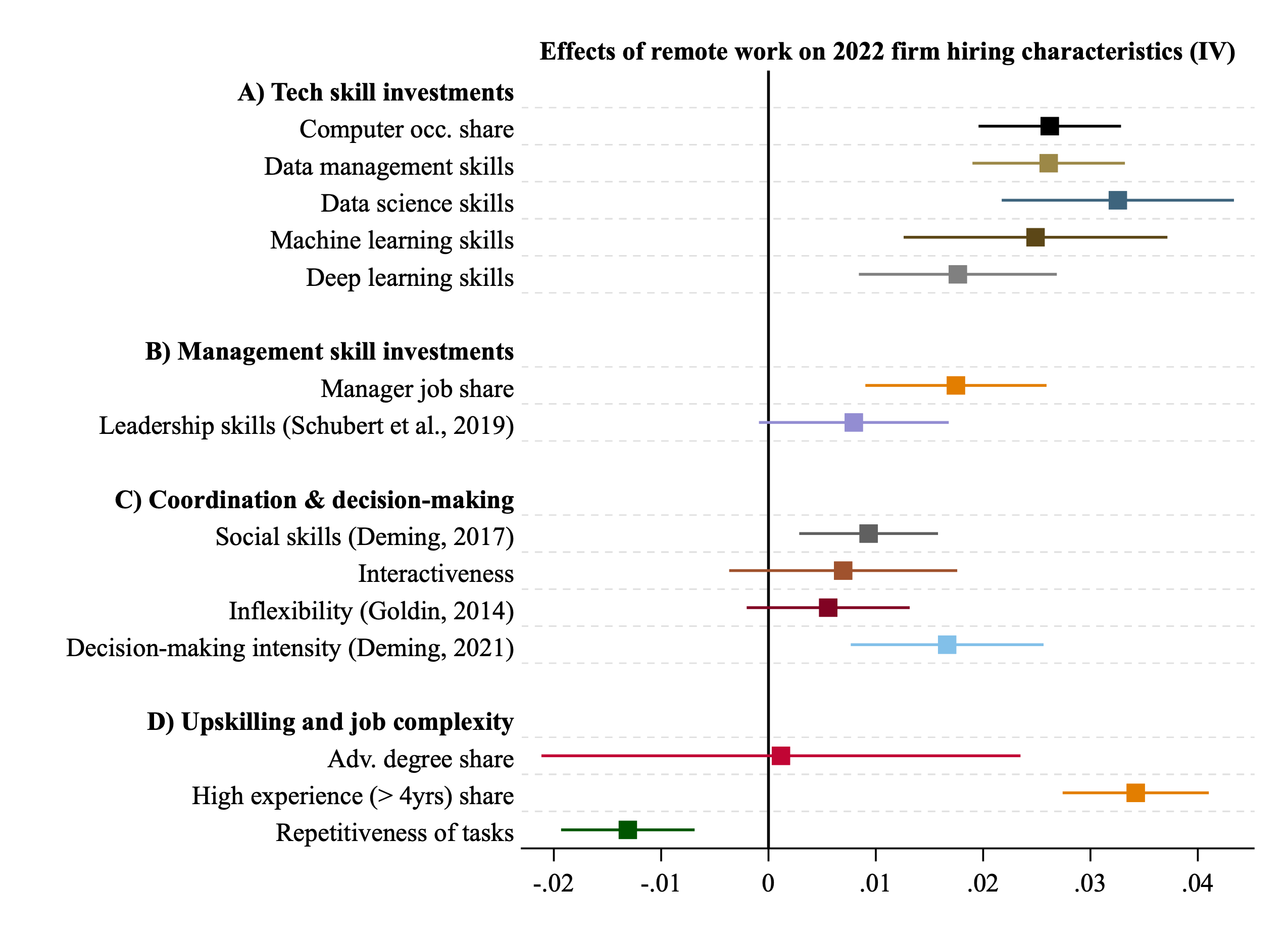}
\end{figure}

Figure \ref{fig:rw2skill} shows the results of the IV estimation, which additionally controls for a rich set of pre-pandemic hiring characteristics to capture potential confounders in hiring skill trends.\footnote{The firm-level regressions include the following control variables: NAICS 2-digit fixed effects; company's remote work share in 2019, the company's uninteracted exposure to MSA teleworkability in 2019, uninteracted firm-level teleworkability in 2019, the company's share of jobs requiring a college education and the share requiring an advanced degree in 2019, the share of the company's job postings in 2019 that was for computer occupations or manager positions; the log of total job postings in 2019 and in 2022; the company's labor market exposure to MSA remote shares in 2019.} The sets of skills considered map onto intuitive categories of characteristics and skills that are relevant for technology adoption and organizational adaptation.\footnote{See Section \ref{sec:data} for references for skill measures constructed based on other papers.} 

The results suggest three key patterns: First, remote work adoption induces sizable investments in \emph{technical and managerial capability}. The share of computer occupations and job postings mentioning data management, data science, machine learning, or deep learning increases significantly in firms where remote work rises (panel A). This is consistent with remote-work implementation requiring upgrades to digital workflows, data infrastructure, and investment in complementary human capital---capabilities that are plausibly reusable when deploying generative AI (section \ref{sec:calls} and Figure \ref{sec:calls}, panel B provide additional evidence of this). Moreover, panel B shows that adopting remote work causes firms to skew hiring towards more managerial roles (SOC code group 11) and towards roles that tend to have greater leadership skills, as defined by \cite{schubert2019}. This suggests that firms may increase their demand for centralized managerial skills to deal with the increased difficulty of supervising remote workflows. The standardized skill variables mean that the coefficients are comparable across skills and can be interpreted as the standard deviation change in the skill that is caused by a 1 pp change in remote work. 

Second, remote work adoption shifts hiring toward \emph{coordination and decision-related capabilities}. Firms become more likely to hire for roles requiring social skills  \citep{deming2017} and for occupations with higher decision-making intensity \citep{deming2021}, as shown in panel C of Figure \ref{fig:rw2skill}. This pattern is consistent with distributed work raising the marginal value of roles that coordinate information flows and support decentralized decision-making.

Third, remote work adoption is associated with \emph{skill upgrading and changes in job complexity}. Firms increase hiring for positions requiring at least five years of experience and hire less into repetitive roles, but do not significantly increase advanced-degree requirements (panel D).

\subsection{Generative AI adoption and hiring characteristics} \label{sec:g2g}

A technology ladder requires not only that remote work shifts capabilities, but also that these capabilities increase the probability that a firm converts potential generative AI value into actual adoption. In this section, I show that firm-level generative AI exposure is more likely to translate into generative AI adoption in firms that, as of 2022, have higher technical capability, more managerial capacity, higher decision-making intensity, and more complex roles.

I build on the idea that some job tasks are more exposed to the benefits from generative AI deployment and that this can be used to quantify differences in exposure across different occupations \citep{eloundou2023} and to predict firm-level exposure to generative AI \citep{eisfeldt2023}. This exposure to generative AI strongly predicts actual adoption: Appendix Figure \ref{fig:genaiscatter} shows for the 12 months ending Sep. 2024 that increasing deciles of firms' exposure to generative AI are monotonically associated with greater hiring for generative AI skills in the same time period, which validates these exposure measures as predictors of actual use of generative AI. To the degree that exposure only indicates potential benefits that not all firms will be able to realize, I use the heterogeneity in which firms translate exposure into higher generative AI skill demand as a proxy for the degree to which particular firm characteristics help in adopting generative AI.

I split my sample based on the characteristics of a firm's hiring in 2022 to see if particular hiring patterns lead to a stronger adoption of generative AI when a firm (or an occupation within a firm) has generative AI exposure. I estimate cross-sectional specifications of the form
\begin{align*}
100 \times \text{GenAIJobShare(Oct `23--Sep. `24)}_{i} = & {\beta} \text{GenAIExp(`21-`22)}_{i} + {\gamma} \text{GenAIExp(`21-`22)}_{i}\times \mathbbm{1}[\text{High Skill(2019)}_i] \\
& FEs + \text{Controls}_{i} +\varepsilon_{i},
\end{align*}
where the unit of observation $i$ is either a firm, or an occupation-by-firm. Here, $\mathbbm{1}[\text{High Skill(2019)}_i]$ indicates whether a firm is above the hiring-weighted median (relative to the 2023-2024 regression sample) in the characteristics of its hiring in 2019. Some firm  characteristics are based on the 2019 firm-level average of time-invariant 2018 O*Net occupation characteristics, which is the case for repetitiveness, social skills, interactiveness, inflexibility, decision-making intensity, computer occupation status, manager occupation status, and leadership skills. The remaining characteristic variables are means of skill mentions in job postings in the 2019 Lightcast data. Moreover, $\text{GenAIExp(`21-`22)}_{i}$ is the mean generative AI exposure in a firm's 2021--2022 (excl. Q4 2022) hiring, using the occupation-level exposure scores from \citep{eisfeldt2023}.

The results are shown in Figure \ref{fig:g2gbyfirmskill}, where the top panel shows estimates of $\hat{\gamma}$ that correspond to the difference in effects between the high and the low group, and the bottom panel shows the baseline effect $\hat{\beta}$ for the low group.
 
Focusing on the graph at the top, I find that firms that had higher technical and managerial capability pre-pandemic (panels A and B) are more likely to adopt generative AI if they have exposure to the productivity potential from the technology. The most pronounced difference in adoption for a given exposure occurs between firms that have high vs. low technical skills with regard to data science, machine learning or deep learning, where firms that hired more for these skills are substantially more likely to adopt generative AI. 

Panel C shows that firms with more roles involving social skills or decision-making are also more likely to adopt generative AI for a given exposure, and panel D shows that firms with more workers with advanced degrees and with experience, as well as less repetitive tasks, are more likely to mention generative AI in hiring. These findings suggest that coordination impacts the ability to use generative AI technology, and confirms the importance of judgment and experience in using generative AI successfully, in line with the finding in the literature that the employment impacts of generative AI favor more senior workers \citep{brynjolfsson2025canaries, lichtinger2025}.
 	
Taken together with Figure \ref{fig:rw2skill}, these patterns support a capability accumulation ladder: remote work induces investments in technical skills and complementary capabilities that are predictive of faster subsequent generative AI deployment.

\subsection{Heterogeneity of remote work effects on generative AI adoption} \label{sec:heterogeneity}

In this section, I estimate whether particular firm characteristics are associated with higher or lower impacts of remote work adoption on generative AI adoption  (in line with the conceptual framework's comparative statics in Proposition~2 in Section \ref{sec:theory}). I estimate interacted regression models of the form
\begin{align*}
100 \times \text{GenAIJobShare(Oct `23--Sep. `24)}_{i} = &  {\beta} \text{ RemoteWorkShare(`21-`22)}_{i}  + \alpha_{ind}  + \text{Controls}_{i} \\
& +  {\gamma} \mathbbm{1}[\text{High Skill(2019)}_f] \times \text{ RemoteWorkShare(`21-`22)}_{i}  +\varepsilon_{i}
\end{align*}
where the high skill indicator is computed as before for different firm characteristics and indicates that the firm is above the median across firms for the measure as of 2019. The IV estimations use the same instruments as in the baseline estimates in Table \ref{tab:firmocc_competition}, as well as an interaction between the instrument and the group indicator.

\begin{figure}
\caption[.]{\\\textbf{Heterogeneity by firm characteristics in remote work effects on  Gen. AI adoption}}\label{fig:r2genaibyskill}

\vspace{-0.1cm} \scriptsize This figure shows coefficients estimated using IV for the effect of remote work prevalence in a firm's  2021/2022 (excl. Q4 2022) job postings on the prevalence of generative AI mentions in a firm's job postings in the 12 months ending Sep. 2024 in a regression of the form:
\begin{center}$100 \times \text{GenAIJobShare(Oct `23-Sep. `24)}_{i} = \alpha_{ind} + {\beta} \text{RWS(`21-`22)}_{i} + {\gamma} \text{RWS(`21-`22)}_{i}\times \mathbbm{1}[\text{High Skill(2019)}_f] + \text{Controls}_{i} +\varepsilon_{i}, $\end{center}
where $RWS$ is the $\text{RemoteWorkShare}$ and the dependent variable has been scaled by 100 for better readability. So, a coefficient of 10 indicates that a 10 pp change in remote work causes a 1pp change in generative AI adoption. $\mathbbm{1}[\text{High Skill(2019)}_f]$ indicates whether a firm is above median in the characteristic. The characteristics of jobs are  measured in 2019 in Lightcast data, or or are time-invariant characteristics based on 2018 O*Net data, and are computed from 2019 averages over a firm's hiring composition. Panel A shows the difference in effects between high and low group firms (corresponding to allowing for heterogeneity in the effect in column 2 of Table \ref{tab:firmocc_competition}), and panel B shows the difference in the within-firm effects across occupations between high and low group firms (corresponding to allowing for heterogeneity in the effect in column 4 of Table \ref{tab:firmocc_competition}). The control variables are the same as in Table \ref{tab:firmocc_competition}, except for adding a dummy for level differences between the low and high groups. The instruments are also the same, except for adding an interaction between the $\mathbbm{1}[\text{High Skill(2019)}_f]$ indicator and the respective instrument.   The 95\% confidence intervals shown are based on heteroskedasticity-robust standard errors clustered at the firm level (panel A) or double-clustered at the occupation and firm level (panel B).

\centering
\begin{subfigure}{.7\textwidth}
  \centering
\includegraphics[width=\textwidth]{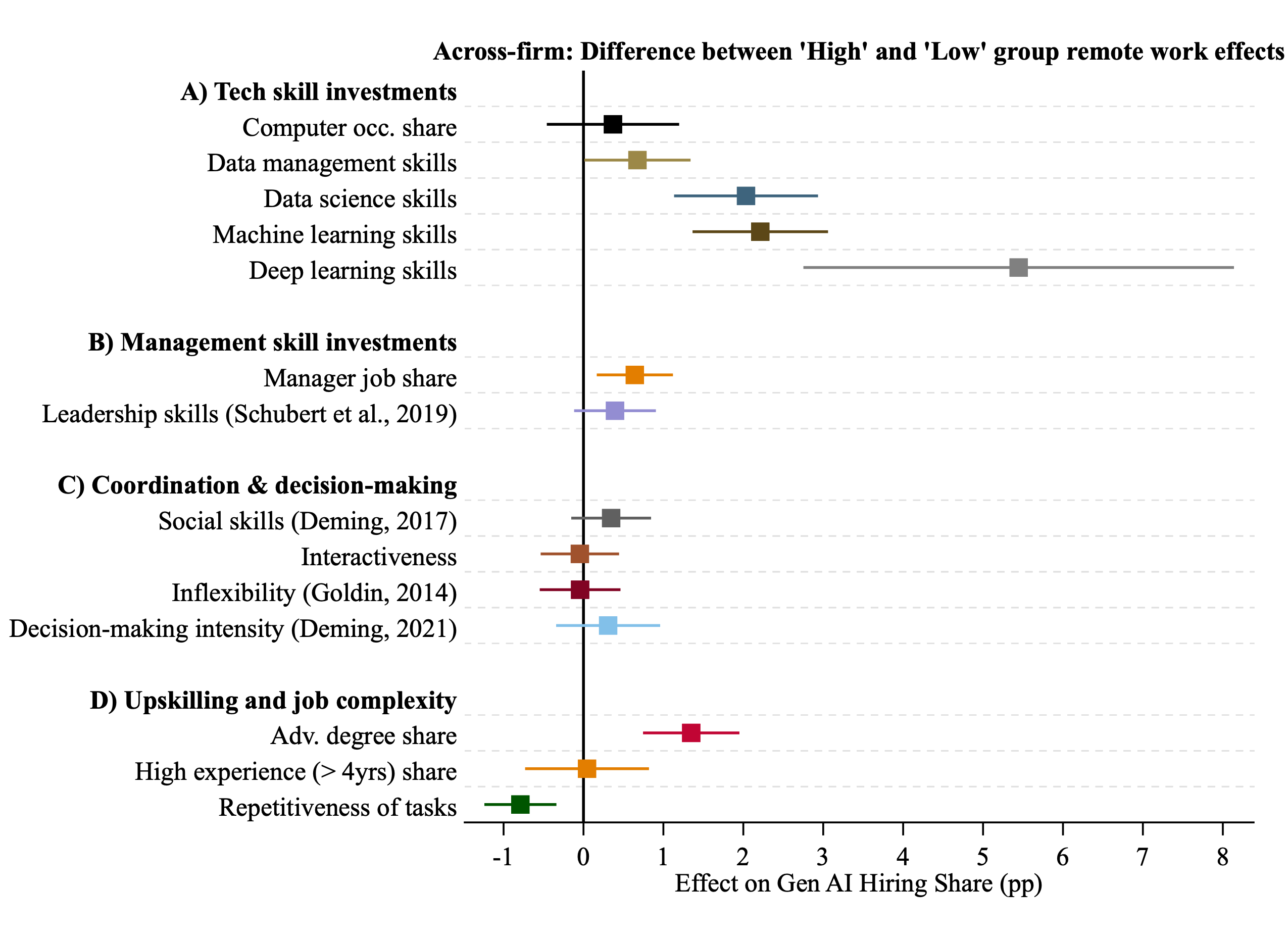}\\
  \caption{Across-firm estimate of difference in effects ($\gamma$)}
  \label{fig:sub1}
\end{subfigure} 
\begin{subfigure}{.7\textwidth}
  \centering
\includegraphics[width=\textwidth]{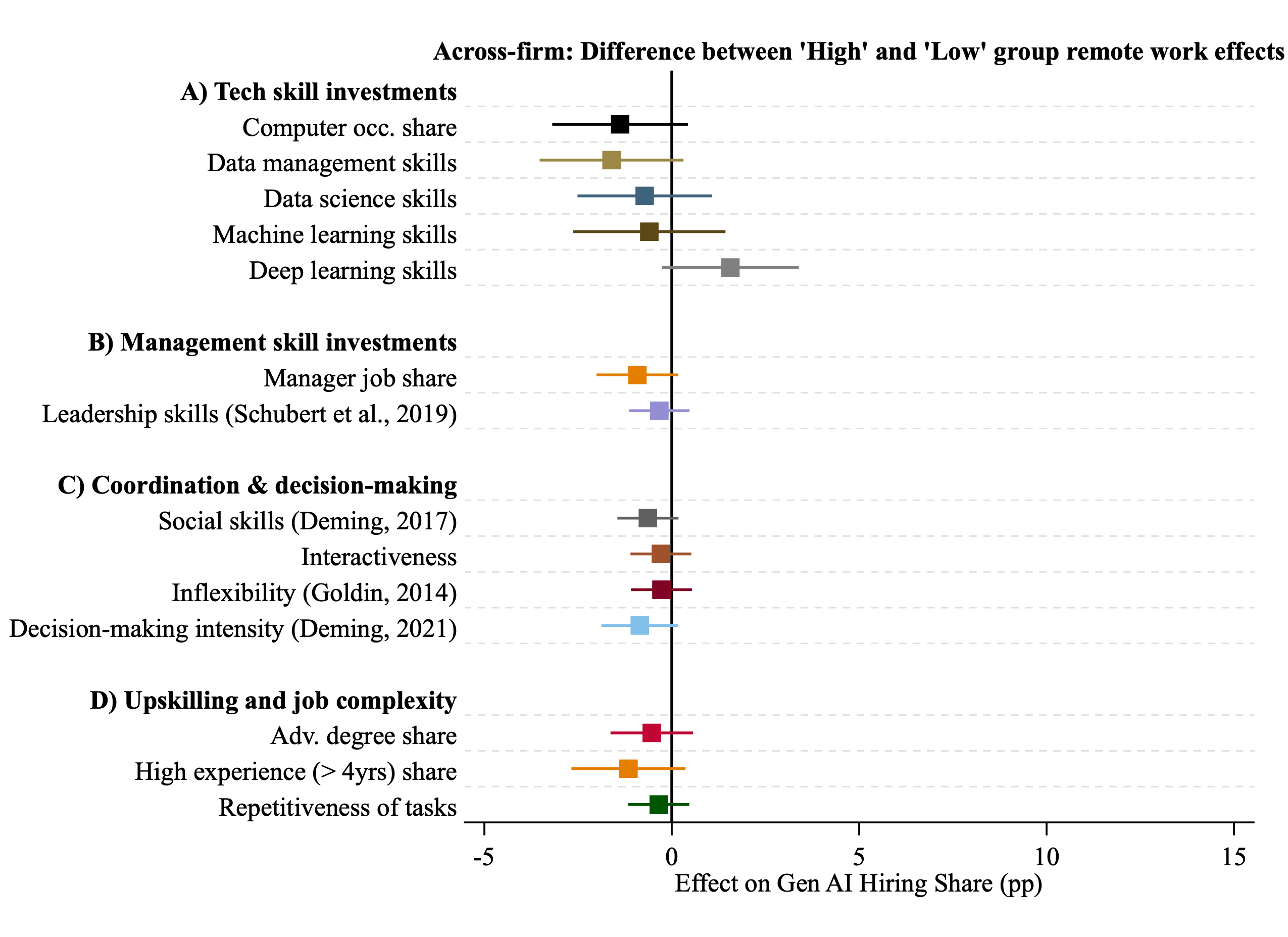}
  \caption{Within-firm estimate of difference in effects ($\gamma$)}
  \label{fig:sub2}
\end{subfigure}
\end{figure}

The results of estimating these interacted models are shown in Figure \ref{fig:r2genaibyskill}, where I show only the coefficients on the interaction term that captures the difference in remote work effects between the `high' and the `low' group for each characteristic. Panel A shows the estimates for the across-firm estimation and panel B for the within-firm estimation, corresponding to heterogeneity in the baseline estimates in columns 2 and 4 of Table \ref{tab:firmocc_competition}, respectively. Thus, panel~A identifies which \emph{firms} become faster generative AI adopters overall after remote work use, while panel~B identifies in which firms generative AI adoption is targeted more toward more-remote occupations. Appendix Figure \ref{fig:r2genaibyskill_base} shows the corresponding `low'-group effects, which are consistently positive, showing that the baseline technology ladder effect is not driven by a particular subset of the data based on the characteristics considered.

The heterogeneity patterns are consistent with two distinct margins: 

\emph{(i) Capability amplification.}
Across firms, the remote-to-generative-AI effect is significantly larger in firms with stronger pre-existing technical capability---especially data science, machine learning, and deep learning skill hiring---and in firms with greater managerial capacity. These characteristics plausibly lower implementation frictions after remote work, by enabling firms to build on / reuse existing remote work technology for generative AI use cases.

\emph{(ii) Task targeting.} Generative AI adoption is more likely in more-remote firms when the firm's job mix is more complex in the sense of hiring more for advanced-degree holders and for less repetitive tasks. This pattern is consistent with generative AI being more likely to be used by firms that may face more difficulty in conducting complex work in a decentralized manner, but also with greater job complexity making firms better at building on existing technologies when incorporating new technologies.

These findings directly support the notion that organizational adaptation matters for the ability to deploy technologies sequentially, as firms with more capabilities are more likely to translate remote work use into adopting generative AI. This mechanism seems to operate at the level of the entire firm's capabilities: Panel B shows that, if anything, firms with more technology skills or managerial capacity are \textit{less} likely to internally target generative AI adoption at the more remote jobs, albeit none of the within-firm heterogeneity results are statistically significant.

Taken together with the sector heterogeneity in Figure \ref{fig:r2genaibyind}, this heterogeneity in responses suggest that the ladder mechanism is likely to be especially relevant in finance-related industries, where decision-making intensity and information-processing capabilities are key inputs to production.

\subsection{Return-to-office policies and technology frictions} \label{sec:rto}

The previous results suggest that greater generative AI adoption in response to remote work may be associated with particular organizational characteristics, such as complex tasks, that may make remote work harder to sustain, or with characteristics, such as tech skills, that make it easier to implement the new technology.

In this section, I further explore the idea that not just the technological capability to adopt generative AI matters, but also the adaptation to the previous remote work wave---and the firm's productivity while working remotely. To identify this variation in adaptation, I consider firms that have publicly shown that remote work is not working out for them by issuing ``return-to-office'' mandates. For example, JPMorgan Chase CEO Jamie Dimon when asked about the firm's return-to-office policy has reportedly said at an internal company town hall that ``[remote work] simply doesn't work. It doesn’t work for creativity. It slows down decision-making.''\footnote{See Barron's (Feb 13, 2025). ``Leaked Audio:JPMorgan CEO Dimon on Hiring \& Remote Work.''  URL: \url{https://www.barrons.com/articles/jamie-dimon-leaked-audio-jpmorgan-return-to-office-7064ee64}} Given that employees tend to value working from home, firms likely need to perceive substantially negative effects of remote work on firm productivity  before forcing employees back into the office.

I use data on firms that have stated RTO policies, collected from the crowd-sourced Flex Index website.  I create an indicator for whether a firm appears in the Flex Index data as requiring a minimum presence in the office. All other firms (whether they appear in Flex Index or not) are labeled as not having an RTO mandate.\footnote{See Appendix \ref{sec:rtodata} for details on the Flex Index data.}

Then, I estimate IV regression specifications of the form  
\begin{align*}
100 \times \text{GenAIJobShare(Oct `23--Sep. `24)}_{i} = &  {\beta} \text{ RemoteWorkShare(`21-`22)}_{i}  + FEs  + \text{Controls}_{i} \\
& +  {\gamma} \mathbbm{1}[\text{RTOMandate}_i] \times \text{ RemoteWorkShare(`21-`22)}_{i}  +\varepsilon_{i}
\end{align*}
both at the firm and at the occupation-by-firm level. I interpret RTO mandates as a proxy for revealed implementation frictions: conditional on observables (which are the same as in the baseline effect estimations in Table \ref{tab:firmocc_competition}) and industry fixed effects, firms that publicly commit to require an office presence are likely those that perceive larger coordination costs from remote work. Accordingly, the interacted IV estimates should be read as suggestive evidence on heterogeneity by perceived remote effectiveness rather than as a causal effect of the RTO policy itself.

The results are shown in Table \ref{tab:rto}: I find that firms that have an RTO mandate are more than twice as likely to adopt generative AI for each percentage point of remote hiring during the pandemic (column 1). Moreover, in line with the idea that this remote work aversion is based on a firm-level lack of adaptation to remote work leading to a firm-wide investment in generative AI, rather than a difference in how targeted adoption is to remote jobs, column (2) shows that there is no evidence that the within-firm effect of remote work on generative AI is different in RTO firms. While the within-firm instrument is somewhat weaker in this setting due to the interaction term, the zero effect on the interaction terms is precisely estimated.

These results suggest that firms which are likely to have low productivity when working remotely are more likely to embrace generative AI, in line with generative AI investments being influenced by the degree of organizational adaptation to the technologies and work processes that the new technology replaces.

\subsection{Investor communication about technology investments} \label{sec:calls}

One drawback of using job postings as a source of information about technology investments is that they tend to focus on the relevance of new technologies for employees. However, firms may communicate a different, and potentially richer, set of information to other stakeholders, such as investors and other financial market participants.  In this section, I therefore explore how firms talk about generative AI and remote work in their communication with financial investors. I analyze the text of earnings call transcripts using a multi-stage natural language processing pipeline that uses LLM classification to extract mentions of the relevant technologies.

\paragraph{Earnings call data.} I focus on firm earnings calls held in Jan. 2019- Sep. 2025, which are sourced from Capital IQ via WRDS. The raw data includes transcript text segmented by speaker turn (e.g., CEO remarks, analyst questions), along with identifiers allowing me to link each transcript to Compustat via GVKEY. After removing duplicate transcripts and calls with missing text, the sample includes transcripts from \~15K unique firms.  I identify a subset of all the text segments in all transcripts for detailed classification by whether they contain any relevant keywords indicating that they may be discussing either generative AI or remote work (see Appendix \ref{sec:methodology_calls} for methodology details).

\paragraph{LLM classification of technology investments.} The filtered segments of the earnings calls are processed through a three-step LLM pipeline designed to classify whether each segment provides evidence of technology adoption, investment, or reported outcomes. I use separate prompts submitted to GPT 5.1-nano via the Azure OpenAI API for each of the following classification steps.
\begin{enumerate}
\item \textbf{Relevance:} The first step filters for segments that genuinely discuss the firm's own adoption of the technology, excluding segments that mention the technology only in the context of market trends, customer behavior, or products sold to others. For example, a software company discussing its ``AI-powered product for customers'' would be filtered out, while discussion of ``deploying Copilot to our engineering team'' would pass.
\item \textbf{Technology investment classification.} Segments passing the first step are evaluated against three hypotheses:
\begin{itemize}[nosep]
    \item \textit{H\textsubscript{invest}}: The firm is investing in the technology (spending, deploying, training employees).
    \item \textit{H\textsubscript{experience positive}}: The firm reports positive outcomes from adoption (productivity gains, cost savings, etc.).
    \item \textit{H\textsubscript{experience negative}}: The firm reports negative outcomes or obstacles (implementation challenges, costs, quality concerns).
\end{itemize}
For each hypothesis, the LLM determines whether the text ``entails,'' ``contradicts,'' or ``does not address'' the claim. Crucially, the prompt instructs the model to distinguish between mere mentions (``we are excited about AI'') and substantive evidence of investment or outcomes (``we deployed GitHub Copilot to 5,000 developers and reduced code review time by 40\%''). The LLM also extracts verbatim quotes supporting each determination and classifies details, such as the type of investment that the firm is making or whether any existing technologies are mentioned as enabling the technology.
\item \textbf{``Skeptic Audit'' (LLM-as-a-judge):} To reduce false positives from LLM overconfidence or hallucinations, a third step acts as a ``hostile auditor'' reviewing the Step 2 output. This step checks whether quoted evidence actually appears in the source text, whether mere vague mentions were incorrectly treated as evidence of investment or outcomes, and whether the segment discusses the firm's own adoption rather than products or market trends. Labels are revised downward when the audit identifies overreach. This type of ``LLM-as-a-judge'' \citep{zheng2023} design is meant to reduce the LLM's tendency to hallucinate responses or over-interpret available evidence in order to be helpful. 
\end{enumerate}

I aggregate segment-level classifications to the firm-call level by taking the maximum across segments within a call (and across calls within a year to aggregate to the firm-year level). A call is coded as showing technology adoption if any audited segment indicates adoption, and as showing technology investment if any audited segment supports the investment proposition. This conservative approach prioritizes precision over recall, reducing the risk that general commentary on a technology is miscoded as evidence of adoption.

\begin{figure}
\caption[.]{\\\textbf{Earnings call communication about technology investments}}\label{fig:calltech}

\vspace{-0.1cm} \scriptsize Panel A of the figure shows the relationship at the 2-digit industry level between the share of firms having a nonzero prevalence of generative AI mentions in their job postings in the 12 months ending Sep. 2024, and the share of firms suggesting that they have adopted generative AI in an earnings call during 2024. The scatter plot omits two outlier industries (NAICS 81 \& 11) for better visibility but includes them when computing the job-posting count weighted line of best fit and R-squared shown in the graph. Panel B shows the share of earnings calls that mention investments to support remote work, in which specific categories of investments are mentioned. Panel C shows the share of earnings calls that mention investments to support generative AI, in which specific categories of investments are mentioned. Panel D shows the share of earnings calls that show evidence of the firm having adopted generative AI, in which specific existing capabilities that make AI adoption easier (``enablers'') are mentioned.

\centering
\begin{subfigure}{.51\textwidth}
  \centering
\includegraphics[width=\textwidth]{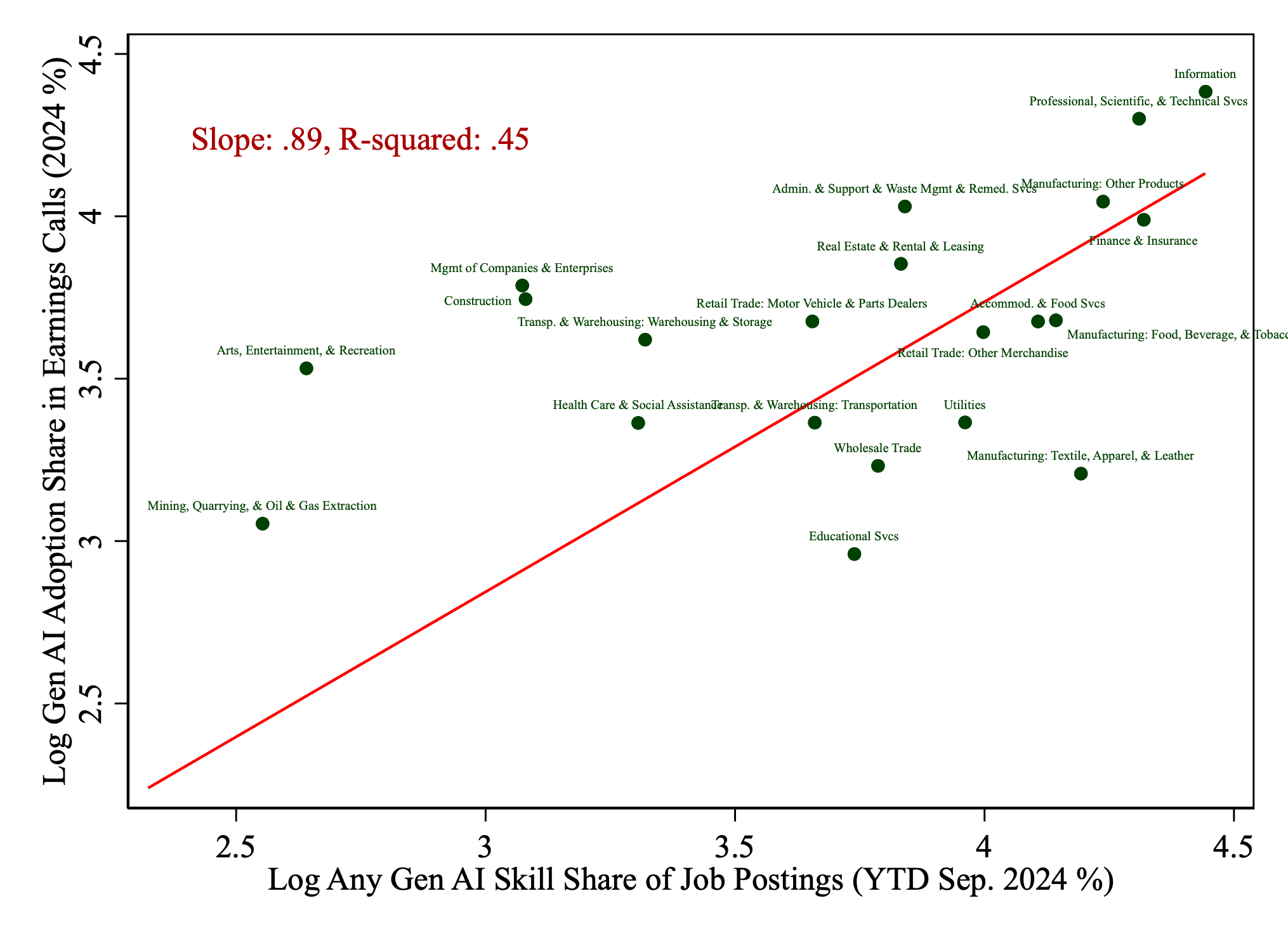}\\
  \caption{Gen. AI earnings call mentions and job posting mentions by industry}
  \label{fig:sub1}
\end{subfigure} 
\begin{subfigure}{.48\textwidth}
  \centering
\includegraphics[width=\textwidth]{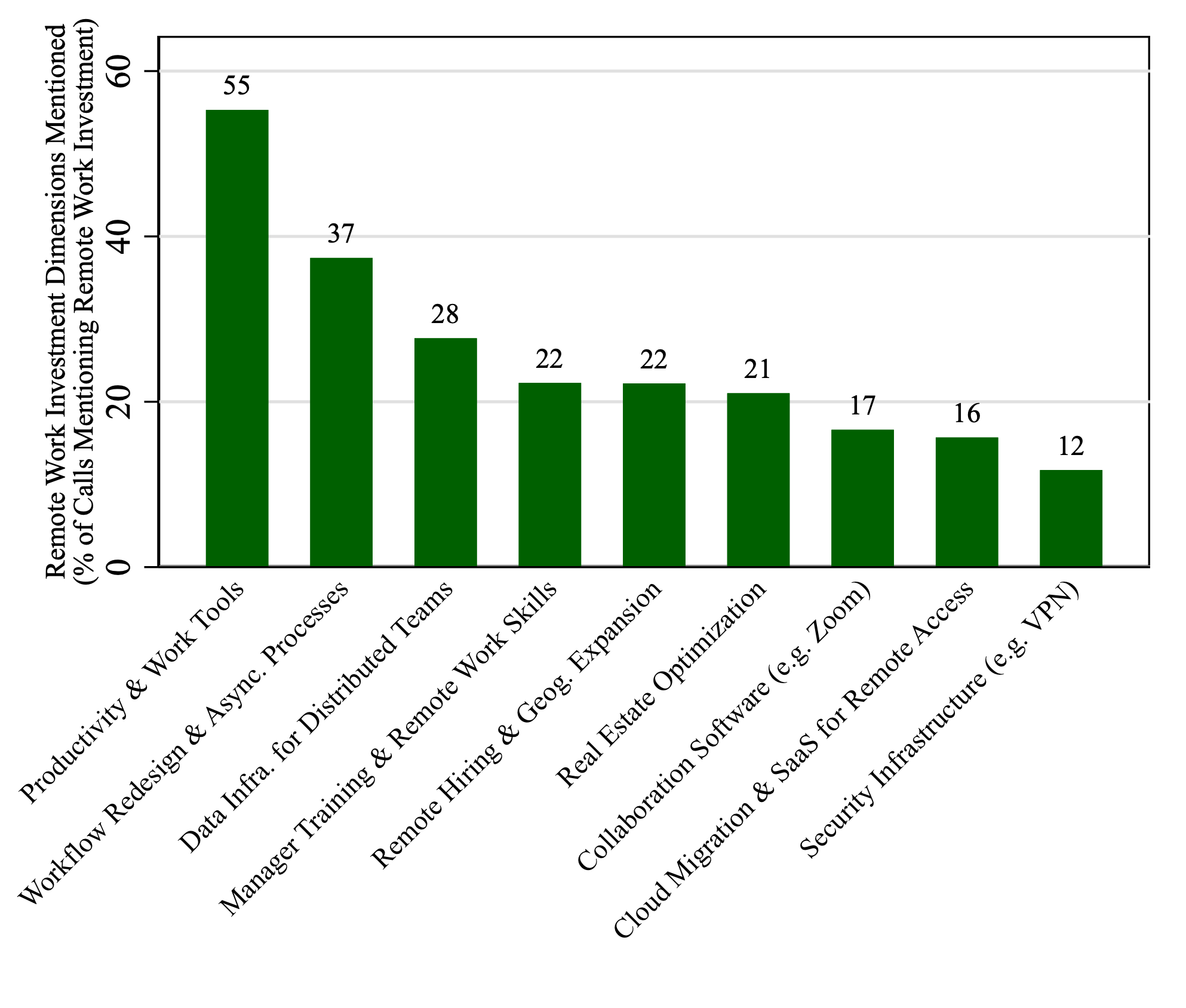}
  \caption{Remote work technology investments in earnings calls}
  \label{fig:sub2}
\end{subfigure}
\begin{subfigure}{.49\textwidth}
  \centering
\includegraphics[width=\textwidth]{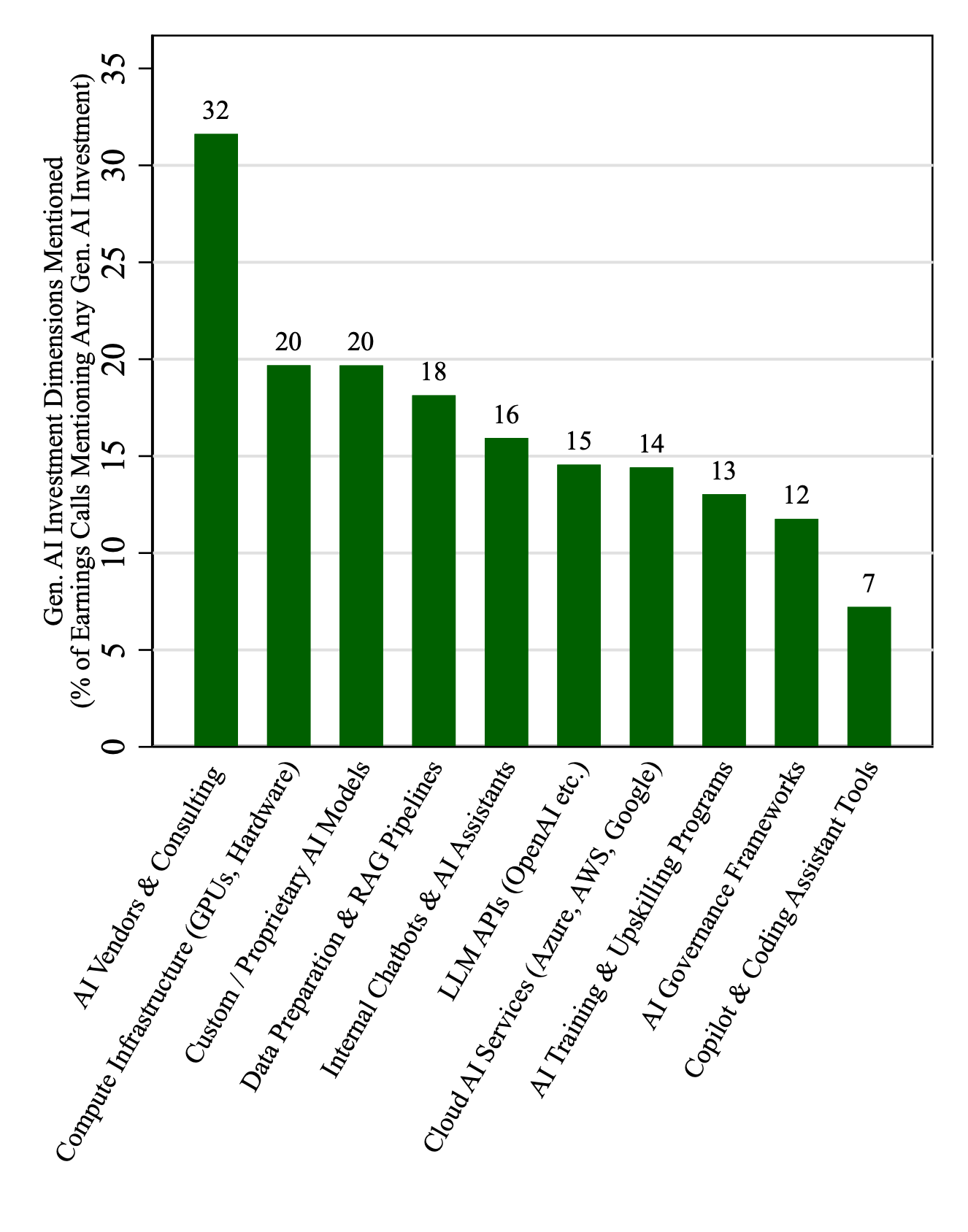}
  \caption{Gen. AI technology investments in earnings calls}
  \label{fig:sub2}
\end{subfigure}
\begin{subfigure}{.49\textwidth}
  \centering
\includegraphics[width=\textwidth]{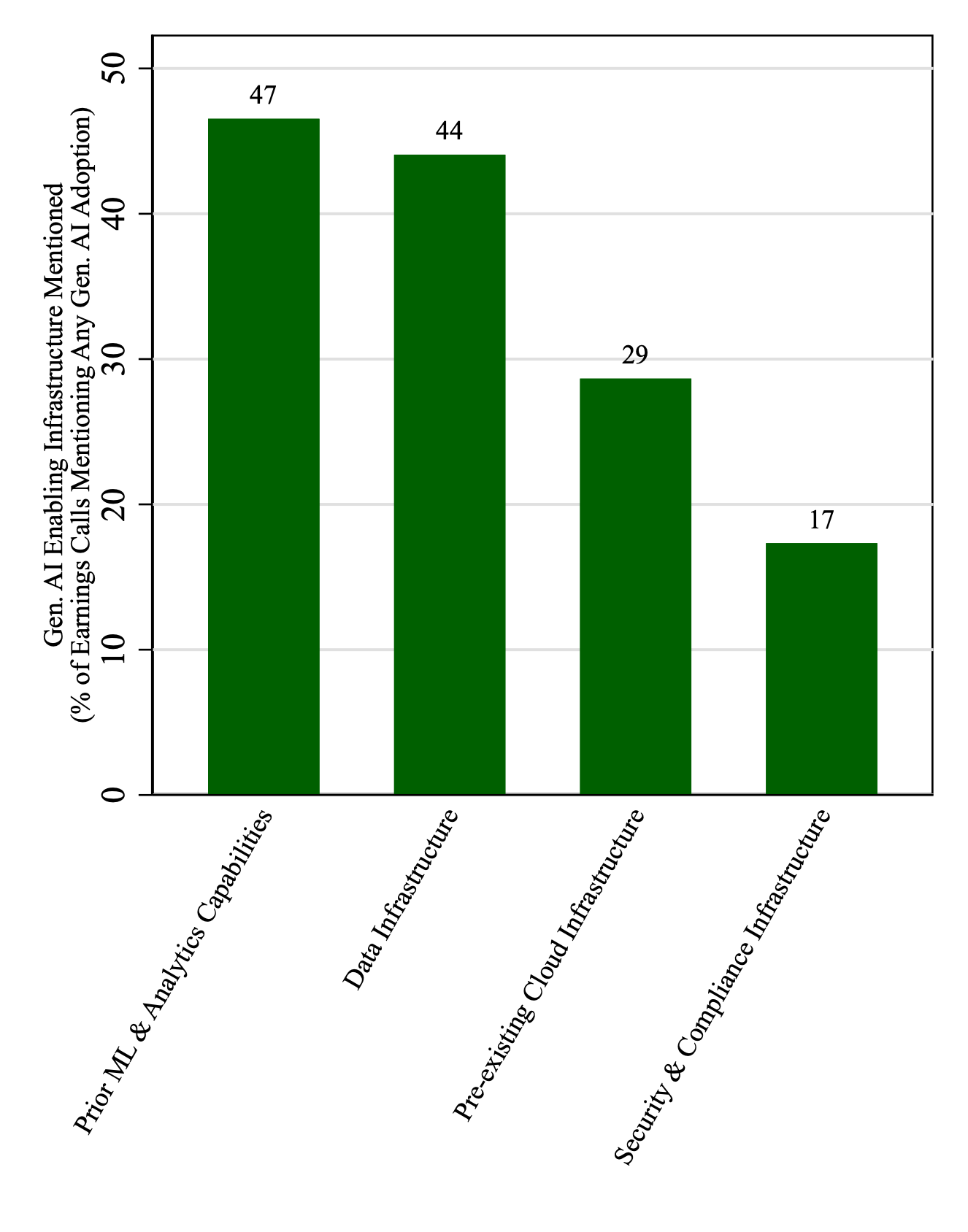}
  \caption{Previous technology investments enabling Gen. AI in earnings calls}
  \label{fig:sub2}
\end{subfigure}
\end{figure}

\textbf{Validation.} To validate the LLM classifications, I manually inspected and labeled a random sample of 25 segments for each technology and year and compared the results against the pipeline's output. This validation surfaced edge cases and errors that were then used to refine the LLM prompts. I also verify that aggregate adoption patterns are consistent with other measures: Figure \ref{fig:calltech}, panel (A) shows the relationship between NAICS 2-digit level shares of firms labeled as having adopted generative AI based on the earnings call LLM classification for 2024, compared to whether any of the firm's job postings in the 12 months ending Sep. 2024 mention generative AI. The figure shows a reassuringly strong correlation ($R^2 = 45\%$), given that the two methodologies use different data sources, different classification methodologies, and potentially cover different dimensions of technology use at the underlying firms. This suggests that the technology investment labels likely capture the relative likelihood of technology use across firms, even if the LLM classifier is excessively strict or lenient with regard to accurately capturing the levels of technology investment.

\paragraph{Earnings call discussion of technology investments.} Which technologies do firms invest in when they adopt remote work---according to their communication with financial investors? Figure \ref{fig:calltech}, panel B, shows the share of earnings calls that mention any investments to support remote work, in which specific categories of investments are mentioned.\footnote{One remote work investment discussion can mention multiple types of investments, or none, so the shares shown in the graph can add to more or less than 100\%.}  More than half of discussions of remote work-supporting investments mention productivity \& work tools provided, and more than a third mention workflow and process changes; earnings calls also discuss training for managers and remote-work skills (22\%) and real estate optimization (21\%)---all of which supports the idea that remote work requires substantial organizational adaptation, which may vary in effectiveness across firms. More than a quarter of remote work investment discussions specifically mention data infrastructure investments, and 17\% focus on collaboration software, 16\% on cloud and remote access infrastructure, and 12\% on security infrastructure. These findings suggest that remote work induces substantial investments in technical capabilities, consistent with the finding of greater hiring for technical skills in Section \ref{sec:r2skill}. 

How do technical infrastructure investments support generative AI? Panel C shows the share of earnings calls that mention investments to support generative AI, in which specific categories of investments are mentioned. About a third of these discussions reference external vendors, but many discussions reference internal technology investments in computing hardware (20\%), customizing AI models (20\%), data preparation and storage pipelines (18\%), or internal chatbot tools (16\%). This shows that, far from being an off-the-shelf solution available to all firms equally, internal technical capabilities play an important role in deploying generative AI. This is also supported by the finding that 13\% of firm explicitly mention AI upskilling programs, and 12\% talk about AI governance programs when discussing generative AI investments, which highlights the advantage of having related skill sets among a firm's employees. Additionally, firms discuss setting up API connections (15\%), cloud services (14\%), and coding assistants (7\%).

Finally, panel D shows the share among earnings calls that show any evidence of the firm having adopted generative AI, in which specific existing capabilities that make AI adoption easier (``enablers'') are mentioned. Almost half of firms mention that prior ML or analytical capabilities enabled generative AI use (47\%) or that existing data infrastructure was useful (44\%). Moreover, a substantial number of firms that adopt generative AI also highlight the importance of pre-existing cloud infrastructure (29\%) and security \& compliance infrastructure (17\%).
These findings provide a direct narrative link in investor communications between the technology investments resulting from remote work and the technologies that enable generative AI deployments, in line with the organizational technology ladder mechanism.

\section{Discussion} \label{sec:discussion}

\paragraph{The mechanisms for the organizational technology ladder.}
The evidence in this paper supports a simple view of an ``organizational technology ladder'': adopting one technology changes both (i) the \emph{firm capabilities} that make subsequent technologies cheaper or more feasible to deploy, and (ii) the \emph{incentives} to substitute across organizational forms when coordination frictions change.

\begin{figure}
\caption[.]{\\\textbf{Gen. AI exposure effects on Gen. AI adoption by firm characteristics}}\label{fig:g2gbyfirmskill}

\vspace{-0.1cm} \scriptsize This figure shows the the cross-sectional estimates of the differential effect of generative AI exposure in 2021/2022 on generative AI adoption in 2023/24 based on the hiring characteristics of a firm in 2019. Each row in the figures shows coefficient estimates corresponding to a different specification of the form
\begin{center}$100 \times \text{GenAIJobShare(Oct `23--Sep. `24)}_{i} = \alpha_{ind} + {\beta} \text{GenAIExp(`21-`22)}_{i} + {\gamma} \text{GenAIExp(`21-`22)}_{i}\times \mathbbm{1}[\text{High Skill(2019)}_i] + \text{Controls}_{i} +\varepsilon_{i}, $\end{center}
where $i$ is a firm unit and the dependent variable has been scaled by 100 for better readability. So, a coefficient of 10 indicates that a 10 pp change in generative AI exposure causes a 1pp change in generative AI adoption. $\mathbbm{1}[\text{High Skill(2019)}_i]$ indicates whether a firm is above the hiring-weighted median (relative to the 2023-2024 regression sample) in the characteristics of its hiring in 2019. Some firm  characteristics are based on the 2019 firm-level average of time-invariant 2018 O*Net occupation characteristics, which is the case for repetitiveness, social skills, interactiveness, inflexibility, decision-making intensity, computer occupation status, manager occupation status, and leadership skills. The remaining characteristic variables are means of skill mentions in job postings in the 2019 Lightcast data. The top panel shows estimates of $\hat{\gamma}$ that correspond to the difference in effects between the high and the low group, and the bottom panel shows the baseline effect $\hat{\beta}$ for the low group. The 95\% confidence intervals shown are based on heteroskedasticity-robust standard errors clustered at the firm level.

\centering
\includegraphics[width=0.65\textwidth]{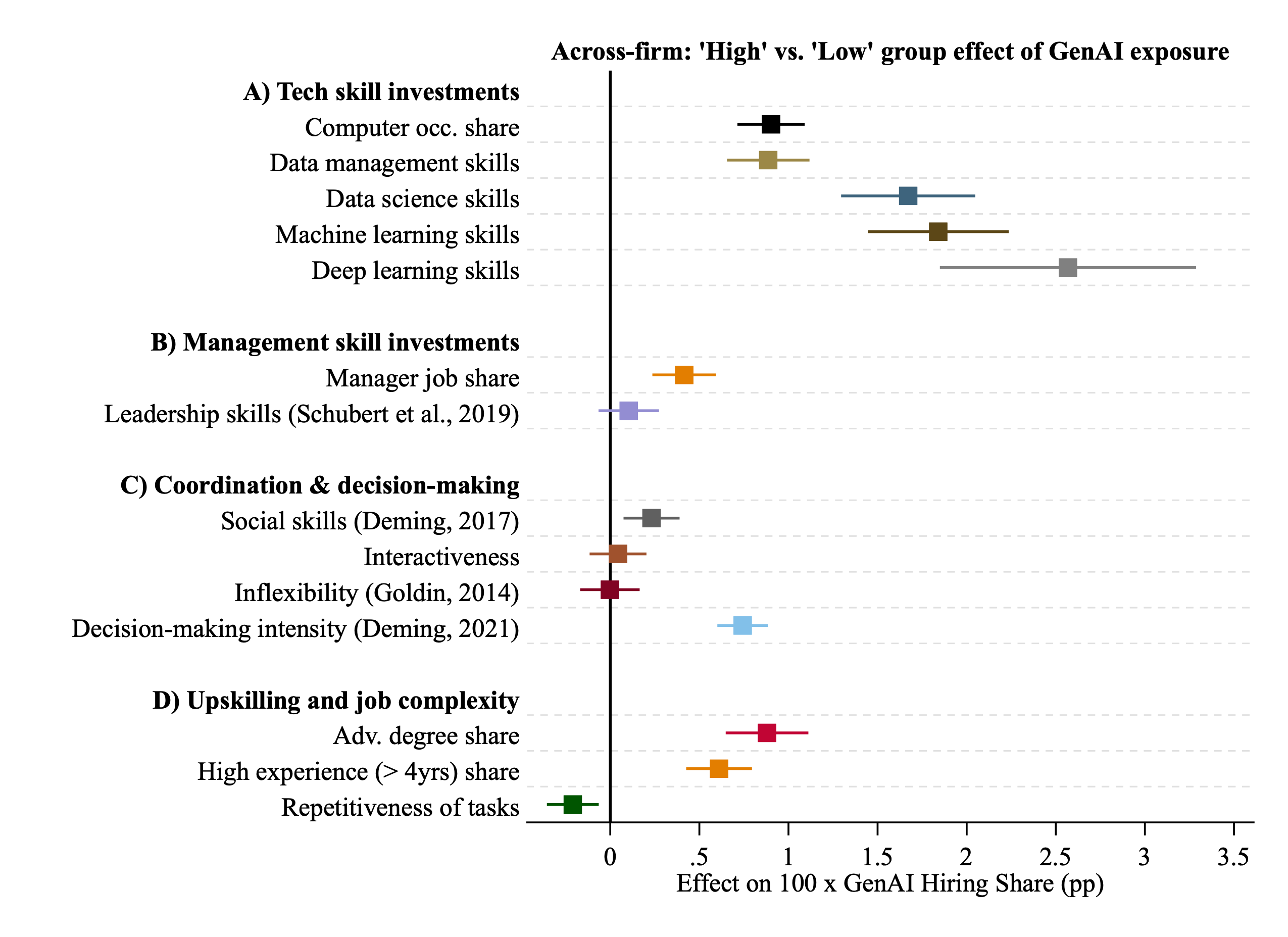}\\
\includegraphics[width=0.65\textwidth]{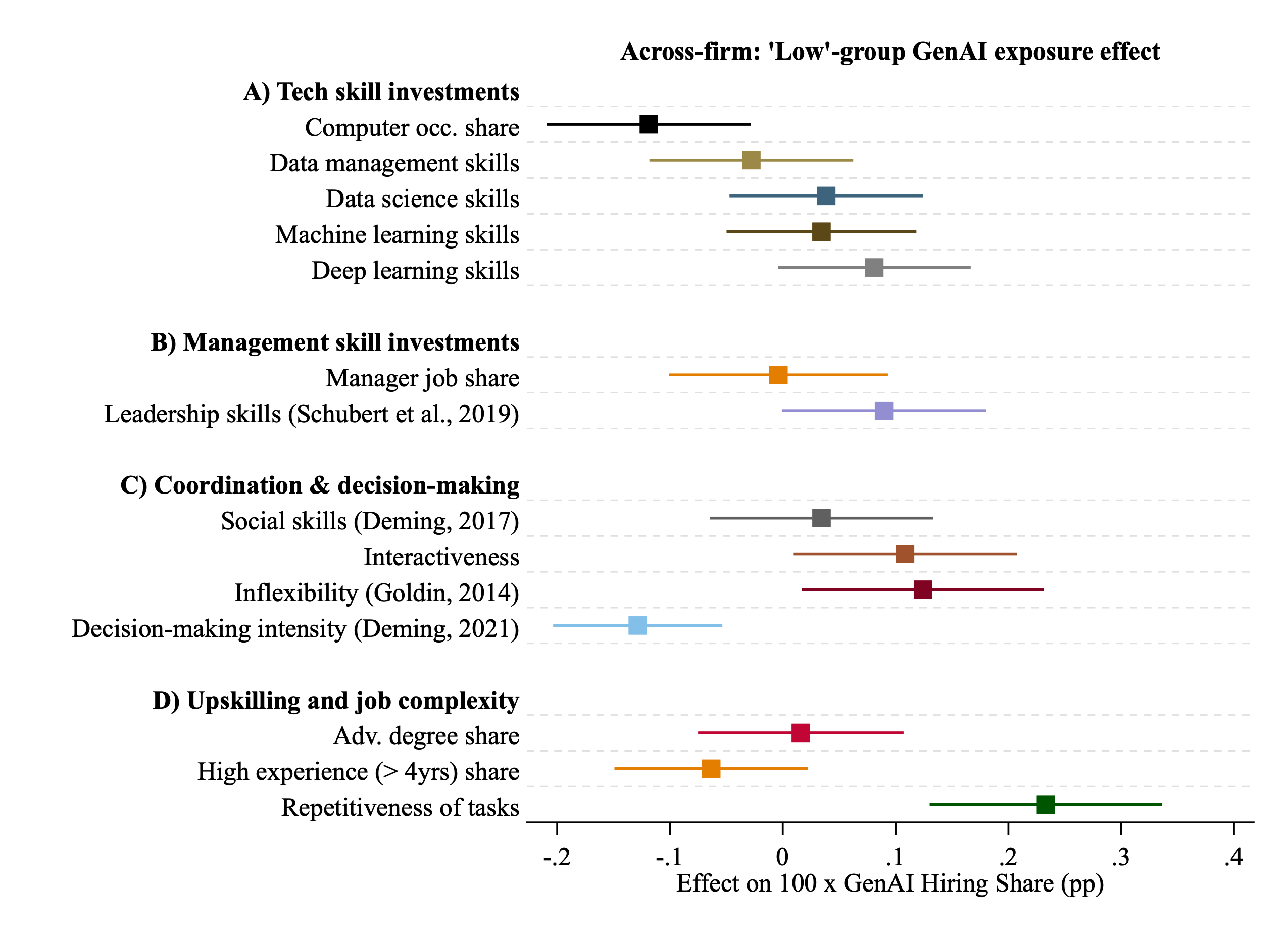}
\end{figure}

In the capability channel, remote work adoption induces investment in technical skills (such as data science and machine learning) and complementary managerial/coordination capacity (Figure \ref{fig:rw2skill}).  This is in line with other studies that have found that firms increased the share of investments going to IT equipment and computers during the pandemic \citep{barrero2021}, and that remote work increases the time and effort devoted to communication \citep{gibbs2023}.  These same capabilities predict a higher likelihood of translating generative-AI exposure into adoption (Figure \ref{fig:g2gbyfirmskill}), consistent with reusability of digital infrastructure and implementation skills.

In the friction-driven substitution channel, firms that appear to experience larger coordination losses from remote work---proxied by the adoption of return-to-office mandates---exhibit a substantially larger remote-to-generative-AI adoption effect (Table \ref{tab:rto}). This pattern is consistent with generative AI serving as an organizational adjustment margin when distributed coordination is costly. Both channels are also consistent with the technology-enabling investments highlighted in investor communications (Figure \ref{fig:calltech}).

\begin{figure}
\caption[.]{\\\textbf{Remote work effects on  Gen. AI adoption by industry sector}}\label{fig:r2genaibyind}

\vspace{-0.1cm} \footnotesize This figure shows coefficients estimated using IV for the effect of remote work prevalence in a firm's  2021/2022 (excl. Q4 2022) job postings on the prevalence of generative AI mentions in a firm's job postings in 2023/2024 (excl. Q4 2024). The specification is the same as in column (2) of Table \ref{tab:firmocc_competition}, but restricting the sample of firms to different industry super-sectors. Firms are categorized as follows into broad sectors (using 2-digit NAICS codes): Technology (NAICS 51 \& 54); Financial Activities and Business Services (NAICS 52, 53, 55, 56);  Trade, Transportation and Utilities (NAICS 22, 42, 44, 45, 48, 49); Manufacturing (NAICS 31, 32, 33); Education and Health Services (NAICS 61 \& 62); Natural Resources, Mining, and Construction (NAICS 11, 21, 23); Leisure and Hospitality and Other Services (NAICS 71, 72, 81). The government sector is omitted due to insufficient sample size. The 95\% confidence intervals shown are based on heteroskedasticity-robust standard errors clustered at the firm level.

\centering
\includegraphics[width=0.99\textwidth]{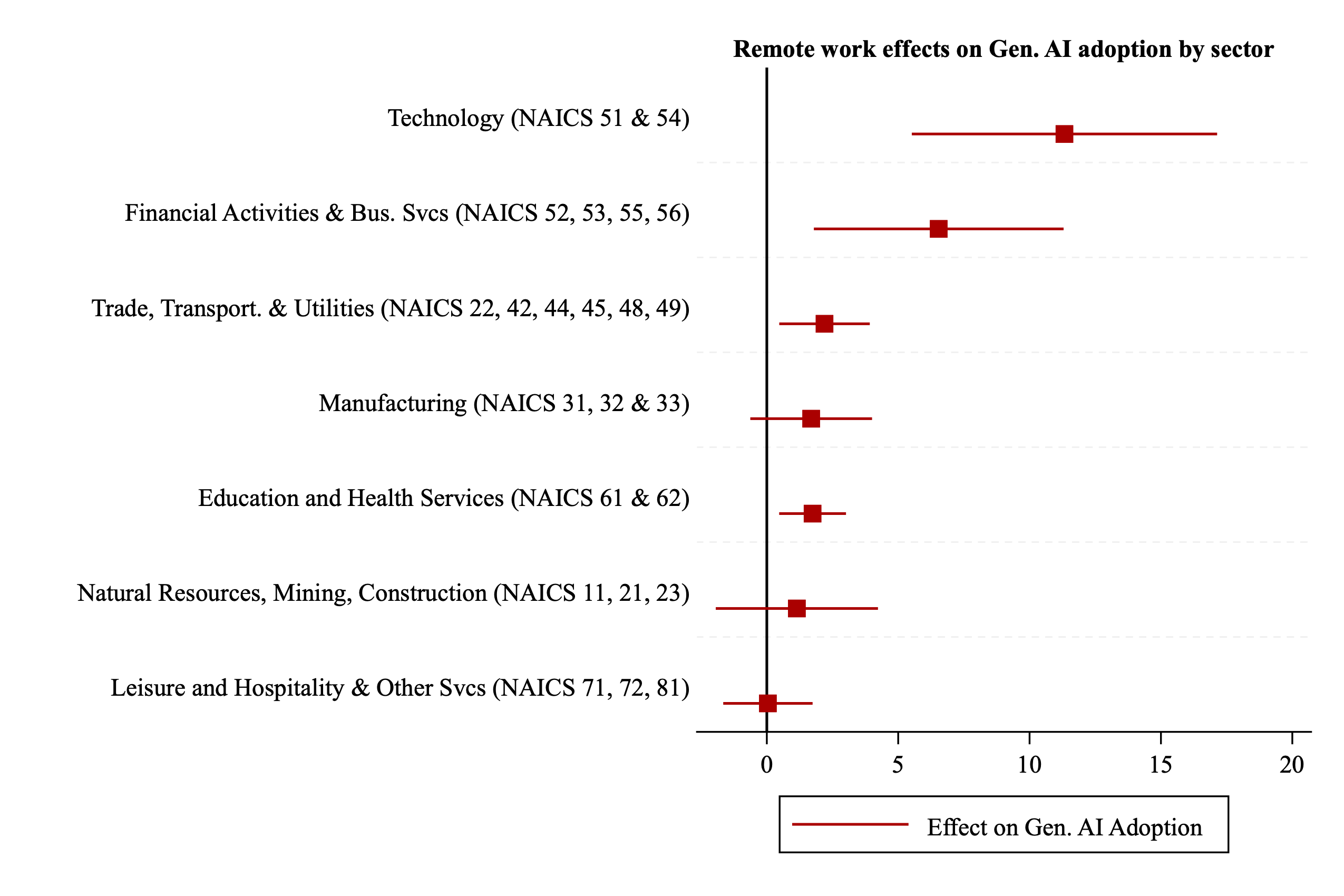}\\
\end{figure}

\paragraph{Implications for AI adoption in finance and business services.}
A key implication for finance and business decision settings is that generative AI adoption is shaped not only by task exposure, but by prior organizational investments that determine implementation feasibility. Consistent with this idea, the estimated remote-to-generative-AI effects are largest in the ``Technology'' and  ``Financial Activities \& Business Services'' sectors (Figure \ref{fig:r2genaibyind}), which are environments where work is information-intensive and decision quality is important. In such settings, investments that improve digital workflows, data access, and managerial coordination can disproportionately increase the set of viable generative AI applications among the many use cases in forecasting, risk management, client communication, and financial research \citep{eisfeldt2024}.

\paragraph{Managerial implications: sequencing AI investments.} More generally, the results imply that AI adoption is an outcome of sequencing complementary investments: digital infrastructure, technical talent, and managerial coordination capacity. Firms that treated remote work and other technology investments as a capability-building investment (rather than a temporary accommodation), are better positioned to deploy generative AI. For business decision-makers, this suggests that measuring ``AI readiness'' requires tracking prior organizational investments and constraints, not only task exposure to the new technology. Moreover, whether an organization should adopt generative AI is not just a function of generative AI capabilities in isolation, but also of the productivity of existing work processes that would be displaced by generative AI tools. This insight reinforces the need for generative AI feasibility assessments to be done in a local context for a specific firms, with one-size-fits-all recommendation unlikely to suit all organizations. Given the novelty of the technology, figuring out which contexts enable greater benefits from its use will require experimentation that may be helped by previous organizational experience with trying out new technologies---as shown in this paper with the example of remote work.

\paragraph{Limitations.}  Two limitations of this study are worth emphasizing. First, generative AI adoption is proxied using job posting mentions of generative AI skills and tools. This captures formalized hiring and organizational intent, but not necessarily informal or decentralized adoption by incumbent workers. Second, the mechanism evidence identifies organizational capability shifts through hiring and occupation composition. It is therefore informative about potential implementation enablers and constraints, but does not directly measure productivity, risk, or decision quality outcomes after adoption. These limitations suggest natural extensions that connect adoption dynamics to eventual firm performance.

\paragraph{Policy implications.}  The dynamic of this ``organizational technology ladder'' also raises important issues for policymakers responding to technological changes. It suggests that differences in the ability to innovate and to transform an organization in response to an initial technology shock can compound into broader competitive advantages over time if the ability to benefit from later technological waves depends on how eagerly an organization embraced the former.  This path dependency means that an initial heterogeneity in technology exposure can cascade into some firms and worker groups being systematically affected in a way that would be difficult to anticipate based on studying technology shocks in isolation. 

Looking forward, as generative AI and related technologies are experiencing rapid innovations and improvements, firms' ability to build on early investments in related technologies may compound into sustained competitive advantages that raise issues for competitive market functioning and market power. Understanding these technological complementarities is crucial for predicting how labor markets and firm productivity will evolve, and for policymakers and researchers to be able to design policies that harness these changes for the benefit of society.  

\renewcommand{\bibfont}{\small}

\bibliography{bibliogeo}
\bibliographystyle{jf}

\clearpage

\section*{Figures and Tables}

\FloatBarrier

\begin{figure}
\caption[.]{\\\textbf{Remote work and occupational generative AI adoption trends.}}

\vspace{-0.1cm}  \small \label{fig:trends} This figure shows the share of job postings in each quarter that are for jobs that mention generative AI. The occupations are aggregated into job posting-weighted quartiles of adoption of remote work: the graph shows generative AI adoption by the occupation's quartile of national remote work adoption in 2021-2022 (excl. Q4 2022). The grey drop line indicates the period (Q4 2022) when ChatGPT was released.

\centering

\includegraphics[width=0.7\textwidth]{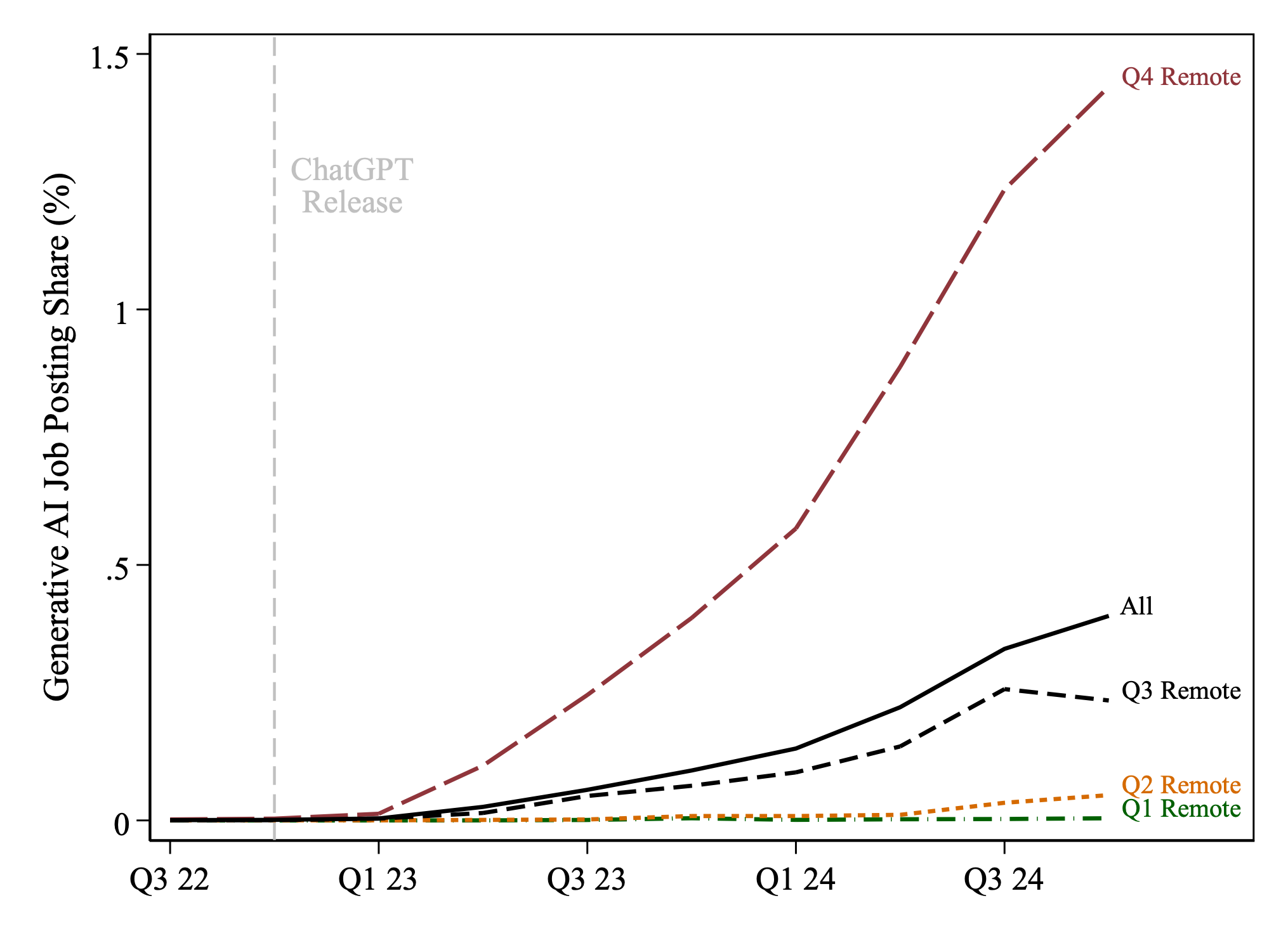} 
  \label{fig:sub1}

\end{figure}

\clearpage

\clearpage

\clearpage

\clearpage


\begin{table}[htbp]
 
  \caption{\\ \textbf{Summary Statistics}}  

  \vspace{-.1cm}  \small  This table reports summary statistics of key variables at the firm-level in Panel A and at the occupation-by-firm level in Panel B. See Section \ref{sec:data} for data sources and variable definitions. \vspace{.2cm}

\centering \footnotesize
 
\label{tab:summary}  
\setlength{\tabcolsep}{2.5pt} 
    \begin{tabular}{lcccccccccc} 
        \toprule
     \ \ \  \    &            \multicolumn{1}{c}{  \  Mean\    }  &    \multicolumn{1}{c}{SD} &  \multicolumn{1}{c}{ 10th } &   \multicolumn{1}{c}{  50th  } &     90th  &  99th       & \multicolumn{1}{c}{     99.9th      } &   \multicolumn{1}{c}{ Obs.  } \\   

    \midrule
                \multicolumn{8}{c}{\textit{Panel A: Firm-level measures}} \\
    \midrule
Remote Work Share '19 (\%)&1.295&7.157&0.00&0.00&1.18&33.33&100.00&87,032\\
Remote Work Share '21-'22 (\%)&6.001&14.178&0.00&0.00&18.87&74.67&100.00&87,032\\
Fully Remote Work Share '21-'22 (\%)&4.533&12.275&0.00&0.00&12.66&67.46&100.00&87,032\\
Hybrid Remote Work Share '21-'22 (\%)&1.468&5.991&0.00&0.00&2.78&30.77&71.43&87,032\\
Generative AI Share YTD Sep. '24 (\%)&0.090&1.090&0.00&0.00&0.00&2.37&11.84&87,032\\
$\mathbbm{1}$[Generative AI Share YTD Sep. '24 > 0]&3.607&18.646&0.00&0.00&0.00&100.00&100.00&87,032\\
Total Job Postings YTD Sep. '24&194.482&1261.407&11.00&30.00&246.00&3088.00&17310.00&87,032\\
Firm Teleworkability YTD Sep. '24&0.500&0.293&0.11&0.48&0.92&1.00&1.00&87,032\\
Firm Teleworkability '19&0.526&0.328&0.00&0.52&1.00&1.00&1.00&87,032\\
Firm Generative AI Exposure YTD Sep. '24&0.336&0.126&0.18&0.33&0.51&0.60&0.72&87,032\\
Firm Generative AI Exposure '19&0.349&0.150&0.16&0.35&0.54&0.73&0.87&87,032\\
$\mathbbm{1}$[Tech sector]&0.163&0.369&0.00&0.00&1.00&1.00&1.00&87,032\\
MSA Teleworkability Exposure '19&0.510&0.080&0.41&0.51&0.61&0.69&0.72&87,032\\
MSA Remote Share Exposure '19&2.381&0.917&1.52&2.31&3.22&5.42&13.80&87,032\\
MSA Commute Time Exposure '19&29.345&4.650&23.68&29.12&34.61&39.68&39.68&87,032\\
$\mathbbm{1}$[Return to Office]&0.009&0.093&0.00&0.00&0.00&0.00&1.00&87,032\\
College Degree Required Share YTD Sep. '24&0.264&0.267&0.00&0.18&0.69&0.95&1.00&87,032\\
Adv. Degree Required Share YTD Sep. '24&0.030&0.082&0.00&0.00&0.09&0.41&0.78&87,032\\
Data Mgmt. Skill Share '19&2.536&7.660&0.00&0.00&7.97&33.33&100.00&87,032\\
Machine Learning Skill Share '19&0.325&2.949&0.00&0.00&0.00&7.02&50.00&87,032\\
AI Skill Share '19&0.183&2.428&0.00&0.00&0.00&3.25&33.36&87,032\\
Deep Learning Skill Share '19&0.046&1.085&0.00&0.00&0.00&0.60&7.69&87,032\\
 \midrule
    \multicolumn{8}{c}{\textit{Panel B: Occupation-by-firm-level measures}} \\
     \midrule
Remote Work Share '21-'22 (\%)&6.156&18.974&0.00&0.00&19.35&100.00&100.00&1,495,607\\
Fully Remote Work Share '21-'22 (\%)&4.979&17.041&0.00&0.00&11.54&100.00&100.00&1,495,607\\
Hybrid Remote Work Share '21-'22 (\%)&1.178&8.234&0.00&0.00&0.00&40.00&100.00&1,495,607\\
Generative AI Share YTD Sep. '24 (\%)&0.125&2.682&0.00&0.00&0.00&0.00&40.00&1,495,607\\
$\mathbbm{1}$[Generative AI Share YTD Sep. '24 > 0]&0.642&7.989&0.00&0.00&0.00&0.00&100.00&1,495,607\\
Total Job Postings YTD Sep. '24&10.872&105.311&1.00&2.00&15.00&132.26&873.00&1,495,607\\
$\mathbbm{1}$[Tech sector]&0.139&0.346&0.00&0.00&1.00&1.00&1.00&1,495,607\\
College Degree Required Share YTD Sep. '24&0.342&0.430&0.00&0.00&1.00&1.00&1.00&1,495,607\\
Adv. Degree Required Share YTD Sep. '24&0.039&0.171&0.00&0.00&0.00&1.00&1.00&1,495,607\\
     \bottomrule
    \end{tabular}
\end{table}%

\begin{table}[htbp]
\caption{\\ \centering \textbf{Remote work effects on generative AI adoption}}  

\vspace{-0.1cm} \small This table shows estimates of the remote share effect at the firm- and  occupation-by-firm level in specifications of the form
\begin{center}$100 \times \text{GenAIJobShare(Oct `23--Sep. `24)}_{i} =   {\beta} \text{ RemoteWorkShare(`21-`22)}_{i}  + FEs  + \text{Controls}_{i}  +\varepsilon_{i}$\end{center}
for the generative AI share for the 12 months ending Sep. 2024 period and the remote share during 2021-2022 (excl. Q4 2022). The instrument consists of a firm's exposure to MSA teleworkability through its hiring labor markets (measured in 2019), interacted with the teleworkability of the firm's hiring in 2019 in column (2), and with the occupation's teleworkability in column (4).  The definition of baseline characteristics included as control can be found in Section \ref{sec:instruments}, and fixed effects are noted in each column. T-test statistics based on heteroskedasticity-robust standard errors clustered at the firm level (columns 1 and 2) or double-clustered at the company and 6-digit occupation level (columns 3 and 4) in parentheses: * p$<$0.10, ** p$<$0.05, *** p$<$0.01. \label{tab:firmocc_competition}  \vspace{.2cm}
 
 \centering
 \begin{adjustbox}{max width=\textwidth}
\begin{tabular}{@{}l*{4}{c}@{}}
\toprule
     \emph{Dependent variable:}                & \multicolumn{4}{c}{100 $\times$ Generative AI Job Share (\%)}   \\   
      \addlinespace
      \cmidrule(lr){2-5}
  \emph{Unit:}                & \multicolumn{2}{c}{Firm}    & \multicolumn{2}{c}{Firm $\times$ Occupation}   \\   
      \addlinespace
      \cmidrule(lr){2-3}   \cmidrule(lr){4-5}      
      
  \textit{Estimation:} & OLS & IV & OLS  & IV  \\
                                        &\multicolumn{1}{c}{(1)}   &\multicolumn{1}{c}{(2)}   &\multicolumn{1}{c}{(3)} &\multicolumn{1}{c}{(4)}      \\ 
                   \midrule
Remote Job Share (\%)&        0.35***&        4.06***&        0.05*  &        7.11***\\
                    &      (4.92)   &      (7.35)   &      (1.65)   &      (2.74)   \\
\midrule \addlinespace Observations&      87,032   &      87,032   &   1,495,607   &   1,314,930   \\
1st-stage KP F-stat.&               &         417   &               &          30   \\
  \midrule
Firm baseline characteristics \ 	&   X 	& X &  &       \\
2-dig. Industry FEs & X & X &        &  \\
Firm $\times$ Occ. adv. educ. requirements &  &   & X  & X \\
Occupation FEs	&  &  &  X &  X     \\
Firm FEs & & &X & X \\
\midrule
\addlinespace
\textit{Instrument} &&  $\substack{\text{MSA Telework. Exposure} \times \\ \text{Firm Teleworkability} }$ && $\substack{\text{MSA Telework. Exposure} \times \\ \text{Occ. Teleworkability} }$ \\
 \bottomrule
\end{tabular}
\end{adjustbox}
\end{table}

\begin{landscape}
\begin{table}[htbp]
\caption{\\ \centering \textbf{Robustness checks: remote work effects on generative AI adoption}}  

\vspace{-0.1cm} \small This table shows estimation results for the following regression specifications: 

The specification in columns 1 and 3 is identical to the one in column 2 of Table \ref{tab:firmocc_competition} and the specifications in columns 2 and 4 is identical to that in column 4 of Table \ref{tab:firmocc_competition}, but with different measures of remote work: the independent variable in columns 1 and 2 are fully remote jobs, and in columns 3 and 4 it measures only hybrid jobs. Columns 5 and 6 omit any data by firms that are in the tech sector (NAICS 51 and 54). Columns 7 and 8 use the alternative instrument that is based on a firm's labor market exposure to different MSA commuting times. Columns 9, 10, and 11 use data that is disaggregated to the MSA-firm and MSA-firm-occupation level and include MSA and industry, MSA and firm, and MSA and firm and occupation fixed effects, respectively. The instruments for columns 9, 10, and 11 interact MSA teleworkability in 2019 with firm or occupation teleworkability. Column 12 repeats the estimation in column 10 but excludes all data points from the 3 tech hubs with the largest share of generative AI adoption, which are San Francisco, San Jose, and Seattle.  T-test statistics based on heteroskedasticity-robust standard errors in parentheses that are: clustered at the firm level (columns 1, 3, 5, 7); or double-clustered at the company and 6-digit occupation level (column 2, 4, 6, 8); or double-clustered at the company and MSA level (column 9, 10, 12); or triple-clustered at the company and 6-digit occupation and MSA level (column 11): * p$<$0.10, ** p$<$0.05, *** p$<$0.01. \label{tab:robust_rwgai}  \vspace{.2cm}
 
 \centering
 \begin{adjustbox}{max width=\linewidth} \setlength{\tabcolsep}{0.2em}
\begin{tabular}{@{}l*{12}{c}@{}}
\toprule
     \emph{Dependent variable:}                & \multicolumn{12}{c}{100 $\times$ Generative AI Job Share (\%)}   \\   
      \addlinespace
      \cmidrule(lr){2-13}
  \textit{Specification:} & \multicolumn{2}{c}{ Fully remote only} & \multicolumn{2}{c}{ Hybrid only} & \multicolumn{2}{c}{Exclude tech sector} & \multicolumn{2}{c}{Alternative IV} & \multicolumn{3}{c}{MSA FEs}  & \multicolumn{1}{c}{Excl. tech hubs} \\
 \cmidrule(lr){2-3}   \cmidrule(lr){4-5}    \cmidrule(lr){6-7}   \cmidrule(lr){8-9}  \cmidrule(lr){10-12} \cmidrule(lr){13-13}    
  
                                        &\multicolumn{1}{c}{(1)}   &\multicolumn{1}{c}{(2)}   & (3) & (4) & (5) & (6) & (7) & (8) & (9) & (10) & (11) & (12)    \\ 
                   \midrule
Remote Job Share (\%)&        5.29***&       10.34***&       17.50***&       22.76***&        2.12***&        7.28***&        4.60***&       10.82** &        5.00***&        8.31***&        7.39***&        5.25***\\
                    &      (7.26)   &      (2.59)   &      (6.07)   &      (2.74)   &      (5.11)   &      (2.88)   &      (6.61)   &      (2.43)   &      (6.38)   &      (3.10)   &      (3.59)   &      (3.54)   \\
\midrule \addlinespace Observations&      87,032   &   1,314,930   &      87,032   &   1,314,930   &      72,860   &   1,126,671   &      87,032   &   1,314,930   &     627,055   &     623,808   &   2,683,275   &     601,495   \\
1st-stage KP F-stat.&         315   &          21   &          99   &          26   &         294   &          29   &         211   &          14   &         113   &          44   &          55   &          51   \\
  \midrule
Baseline firm characteristics	&X  &    	&X  &   	&X  &   &X  &  &X  &  &   \\
Firm $\times$ Occ. adv. educ. requirements   &   & X  &   & X  &   & X &   & X \\
Firm $\times$ MSA adv. educ. requirements   &&&&&&&&&X&X& & X \\
Firm $\times$ Occ. $\times$ MSA  adv. educ. requir.   &&&&&&&&&&&X \\
2-dig. Industry FEs  &      X  &   &      X  &   &      X  & &      X  & &      X  & &  \\
Occupation FEs	 &   &  X    &   &  X    &   &  X    &   &  X   &   &  & X  &   \\
Firm FEs  & & X & & X & & X & & X & & X & X &X \\
MSA FEs    &&&&&&&&&X&X&X &X \\
\midrule
\addlinespace
\textit{Unit:} & Firm &  $\substack{\text{Firm} \\ \times \text{Occ}}$  & Firm &  $\substack{\text{Firm} \\ \times \text{Occ}}$  & Firm &  $\substack{\text{Firm} \\ \times \text{Occ}}$  & Firm &  $\substack{\text{Firm} \\ \times \text{Occ}}$  &  $\substack{\text{Firm} \\ \times \text{MSA}}$  &  $\substack{\text{Firm} \\ \times \text{MSA}}$  & $\substack{\text{Firm} \times  \text{Occ} \\ \times \text{MSA}}$ & $\substack{\text{Firm} \\ \times \text{MSA}}$ \\
\midrule
\addlinespace
\textit{Instrument} &  $\substack{\text{MSA Tele. Exp.} \times \\ \text{Firm Tele.} }$ & $\substack{\text{MSA Tele. Exp.} \times \\ \text{Occ. Tele.}}$  &   $\substack{\text{MSA Tele. Exp.} \times \\ \text{Firm Tele.} }$ & $\substack{\text{MSA Tele. Exp.} \times \\ \text{Occ. Tele.}}$  &  $\substack{\text{MSA Tele. Exp.} \times \\ \text{Firm Tele.} }$ & $\substack{\text{MSA Tele. Exp.} \times \\ \text{Occ. Tele.}}$   &  $\substack{\text{MSA Commute}\\ \text{Time Exp.} \times \\ \text{Firm Tele.} }$ & $\substack{\text{MSA Commute}\\ \text{Time Exp.}  \times \\ \text{Occ. Tele.} }$  
&  $\substack{\text{MSA Tele.} \times \\ \text{Firm Tele.} }$ &  $\substack{\text{MSA Tele.} \times \\ \text{Firm Tele.} }$  &  $\substack{\text{MSA Tele.} \times \\ \text{Occ.Tele.} }$  &  $\substack{\text{MSA Tele.} \times \\ \text{Firm Tele.} }$
 \\
 \bottomrule
\end{tabular}
\end{adjustbox}
\end{table}
\end{landscape}

\begin{table}[htbp]
\caption{\\ \centering \textbf{Return-to-Office firms have higher remote work effects on Gen. AI Adoption}}  

\vspace{-0.1cm} \small This table shows estimation results for the following regression specifications: The specification in column 1 is identical to the one in column 2 of Table \ref{tab:firmocc_competition} and the specification in column 2 is identical to that in column 4 of Table \ref{tab:firmocc_competition}, except for the following differences:  the endogenous remote share is interacted with an indicator for a company that has a return-to-office (RTO) policy as of March 2025, and so is the instrument for the endogenous variable; an indicator for a firm having an RTO policy is added as an additional control variable where this is not collinear with fixed effects.  T-test statistics based on heteroskedasticity-robust standard errors clustered at the firm level (column 1) or double-clustered at the company and 6-digit occupation level (column 2) in parentheses: * p$<$0.10, ** p$<$0.05, *** p$<$0.01. \label{tab:rto}  \vspace{.2cm}
 
 \centering
 \begin{adjustbox}{max width=\textwidth}
\begin{tabular}{@{}l*{4}{c}@{}}
\toprule
     \emph{Dependent variable:}                & \multicolumn{2}{c}{100 $\times$ Generative AI Job Share (\%)}   \\   
      \addlinespace
      \cmidrule(lr){2-3}
  \textit{Estimation:} & IV & IV   \\
                                        &\multicolumn{1}{c}{(1)}   &\multicolumn{1}{c}{(2)}       \\ 
                   \midrule
Remote Job Share (\%)&        3.99***&        7.10***\\
                    &      (7.23)   &      (2.72)   \\
Remote Job Share (\%) x $\mathbbm{1}$[Return-to-Office]&        6.08***&        0.21   \\
                    &      (4.10)   &      (0.21)   \\
\midrule \addlinespace Observations&      87,032   &   1,314,930   \\
1st-stage KP F-stat.&         208   &          15   \\
  \midrule
 ``Firm has RTO Policy'' Indicator  & X &  \\
Baseline firm characteristics \ 	&X  &       \\
2-dig. Industry FEs  &      X  &  \\
Firm $\times$ Occ. adv. educ. requirements   &   & X \\
Occupation FEs	 &   &  X     \\
Firm FEs  & & X \\
 \bottomrule
\end{tabular}
\end{adjustbox}
\end{table}

\FloatBarrier

\newgeometry{top=1in,bottom=1in,left=1in,right=1in}

\appendix

\begin{appendices}
\renewcommand{\appendixname}{}
\setcounter{table}{0}  
\renewcommand{\thetable}{A.\arabic{table}}

\section{Appendix Figures \& Tables}
\setcounter{table}{0}
\setcounter{figure}{0}
\renewcommand{\thetable}{\thesection.\arabic{table}}
\renewcommand{\thefigure}{\thesection.\arabic{figure}}

\FloatBarrier

\begin{figure}
\caption[.]{\\\textbf{Remote work and generative AI at the firm level}}

\vspace{-0.1cm}  \small \label{fig:corrscatterfirm} This figure plots the share of job postings in the 12 months ending Sep. 2024 period that are for jobs that mention generative AI relative to those that are for remote jobs. The job postings data are from Lightcast and are aggregated by company. The red line indicates a linear best fit. The analysis only includes companies that have at least 10 job postings in the sample period shown and have non-zero job postings in all quarters from Q1 2021 to Q3 2024.

\centering
\includegraphics[width=\textwidth]{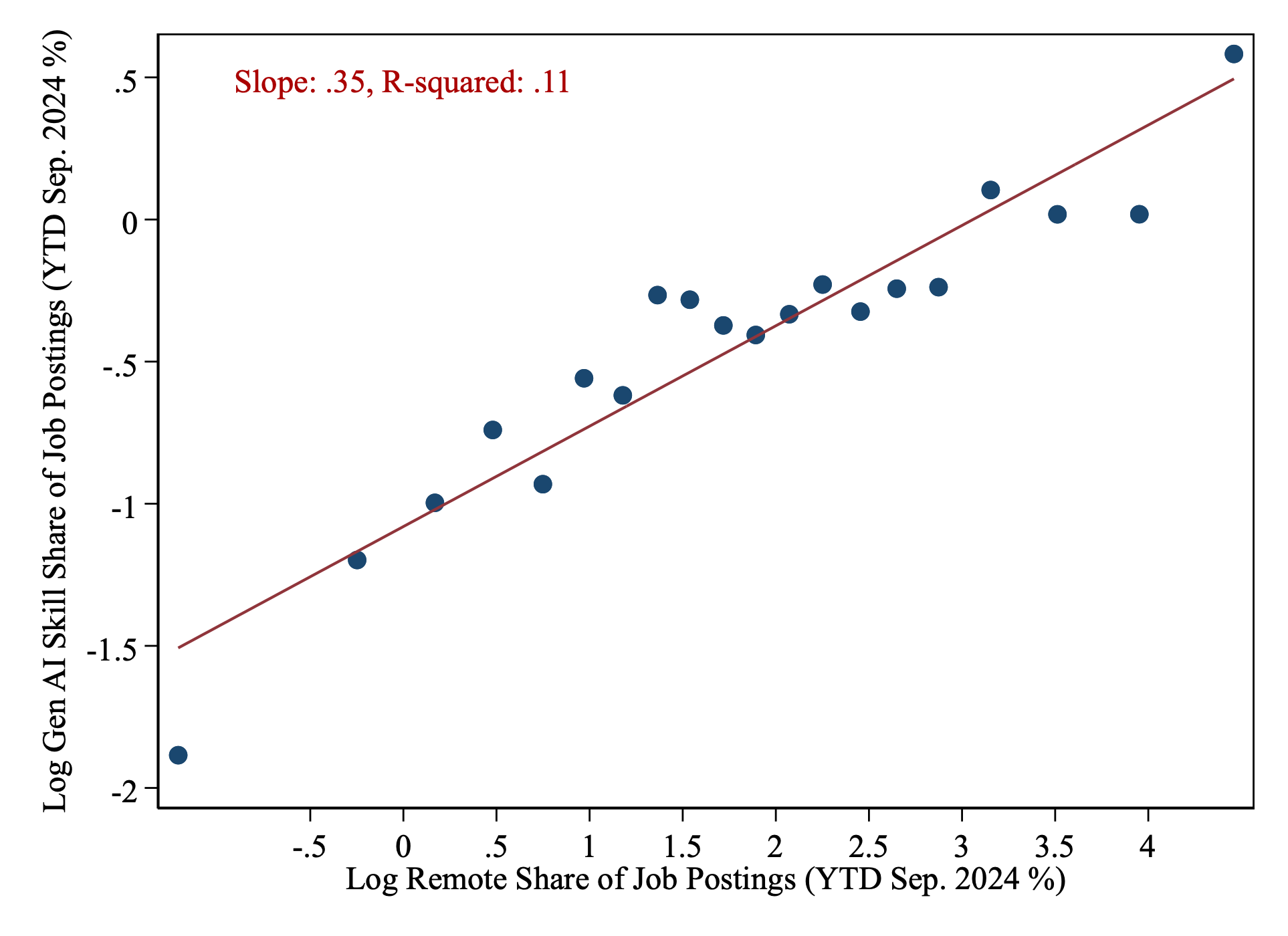} 
\end{figure}

\begin{figure}[t]
\caption[.]{\\\textbf{Pandemic remote hiring and generative AI adoption}}

\vspace{-0.1cm}  \footnotesize \label{fig:rwgaiscatter} This figure focuses on the top quartile of remote work adopting occupations (based on 2021/2022 job postings). It shows the change in hiring for generative AI skills after the release of ChatGPT (Q3 2022-Q3 2024) as a function  of remote work changes during the pandemic (Q3 2019-Q3 2022). The graph only includes occupations with $>1$K job postings in Q3 2022 with some Q3 2024 Gen AI adoption ($>0.1\%$ of jobs). Two outlier occupations, 'Computer Programmers' and 'Writers \& Authors' are not included for better visibility, but follow similar patterns. The data are Lightcast job postings where ``Gen AI share'' is the share of jobs mentioning Gen. AI-related keywords.

\centering
\includegraphics[width=0.7\textwidth]{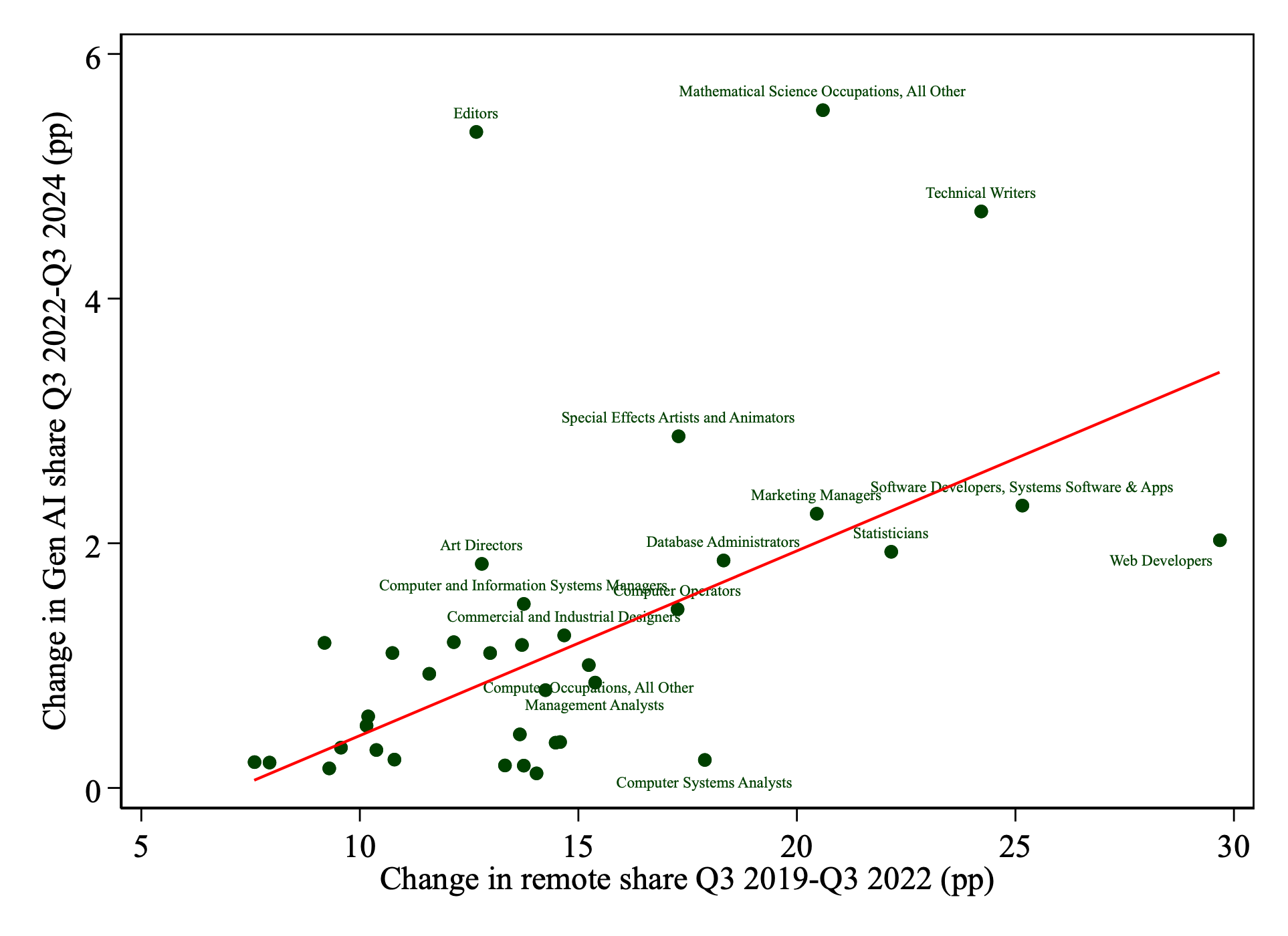} 

\end{figure}

\begin{figure}
\caption[.]{\\\textbf{Remote Work Suitability and Generative AI Exposure by Occupation}}

\vspace{-0.1cm}  \small  \label{fig:exptele} This figure shows the relation between Generative AI exposure and remote work suitability. Generative AI exposure and  teleworkability  are measured at the SOC 2010 6-digit occupation level based on data from \cite{eisfeldt2023} and \cite{dingel2020}, and aggregated across detailed occupations based on employment weights as of 2022 from the Occupational Employment Statistics. The line of best fit in red is estimated with employment weights.

\centering
\includegraphics[width=\textwidth]{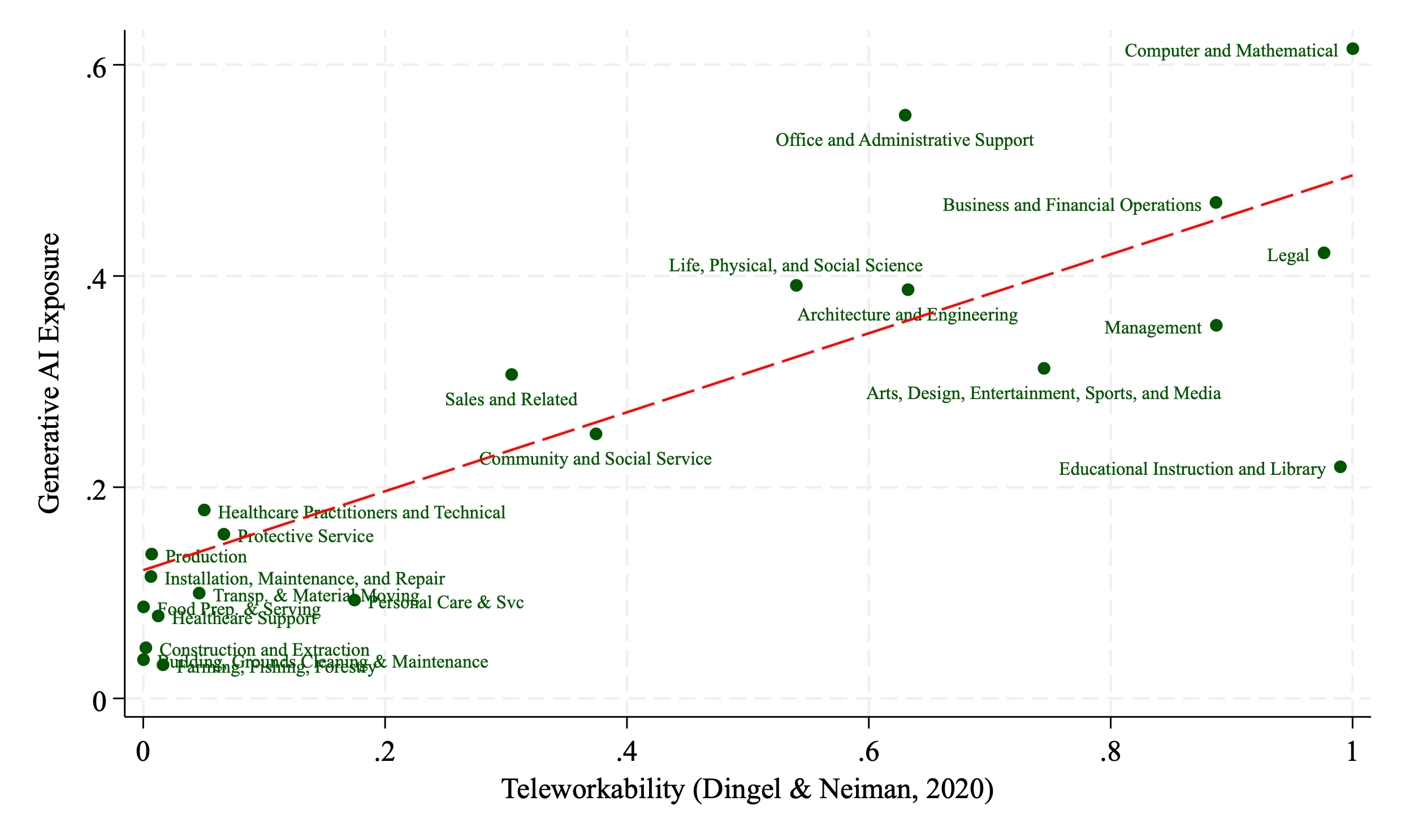} 
\end{figure}

\begin{figure}
\caption[.]{\\\textbf{Within-firm remote work effects on  Gen. AI adoption by industry sector}}\label{fig:r2genaibyind_within}

\vspace{-0.1cm} \footnotesize This figure shows coefficients estimated using IV for the effect of remote work prevalence in a firm's  2021/2022 (excl. Q4 2022) job postings on the prevalence of generative AI mentions in a firm's job postings in 2023/2024 (excl. Q4 2024). The specification is the same as in column (4) of Table \ref{tab:firmocc_competition}, but restricting the sample of firms to different industry super-sectors. Firms are categorized as follows into broad sectors (using 2-digit NAICS codes): Technology (NAICS 51 \& 54); Financial Activities and Business Services (NAICS 52, 53, 55, 56);  Trade, Transportation, and Utilities (NAICS 22, 42, 44, 45, 48, 49); Manufacturing (NAICS 31, 32, 33); Education and Health Services (NAICS 61 \& 62); Natural Resources, Mining, and Construction (NAICS 11, 21, 23); Leisure and Hospitality and Other Services (NAICS 71, 72, 81). The government sector is omitted due to insufficient sample size. The 95\% confidence intervals shown are based on heteroskedasticity-robust standard errors double-clustered at the firm and occupation level.

\centering
\includegraphics[width=0.99\textwidth]{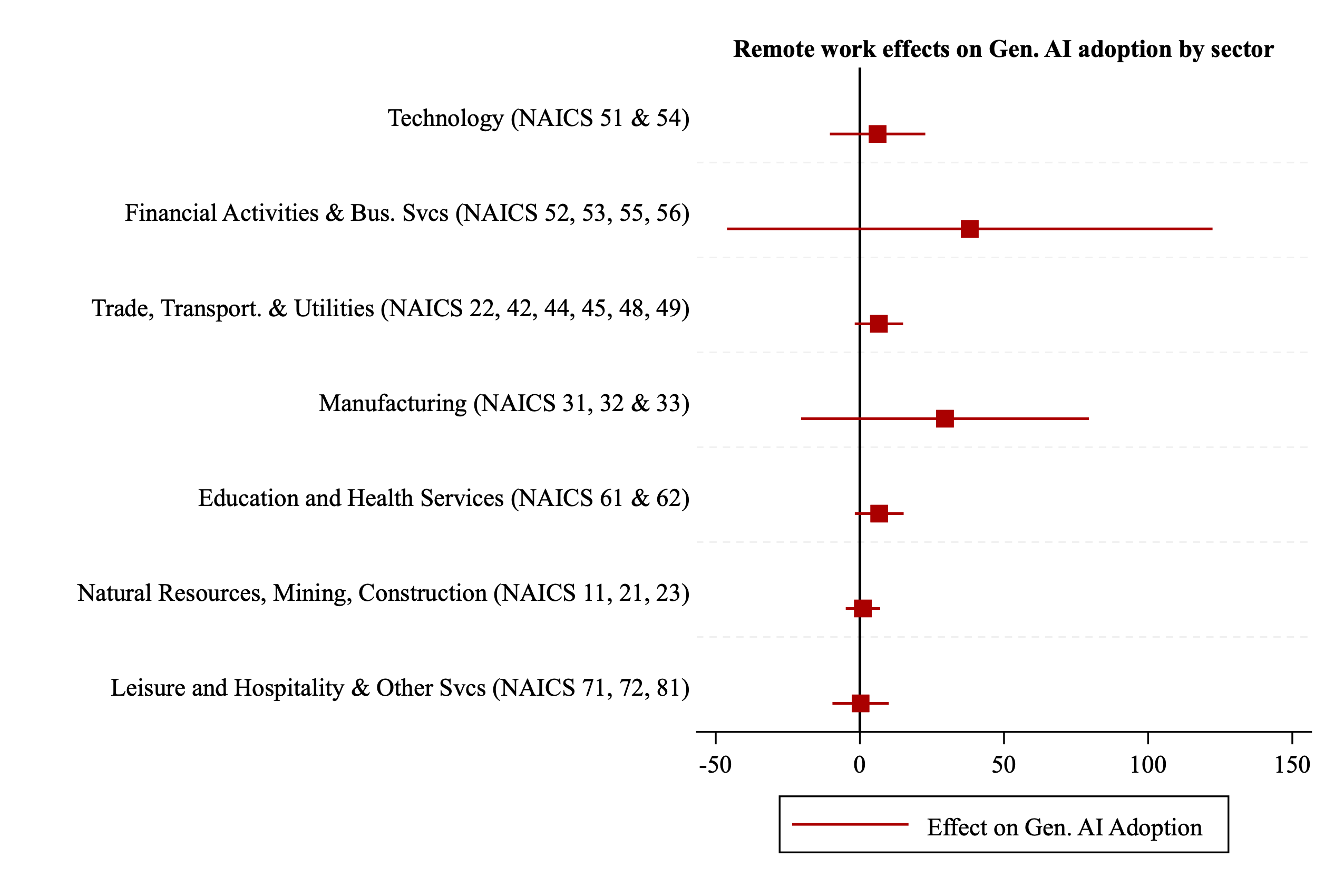}\\
\end{figure}

\clearpage

\begin{figure}
\caption[.]{\\\textbf{Generative AI share of job postings in large CBSAs}}

\vspace{-0.1cm}  \small \label{fig:genaicities} This graph shows the share of job postings at the CBSA level that mention generative AI skills in the first six months of 2024 for all U.S. cities with employment above 1 million.  The vertical axis shows the share of generative AI skill mentions in all job postings. 

  \centering
\includegraphics[width=0.95\textwidth]{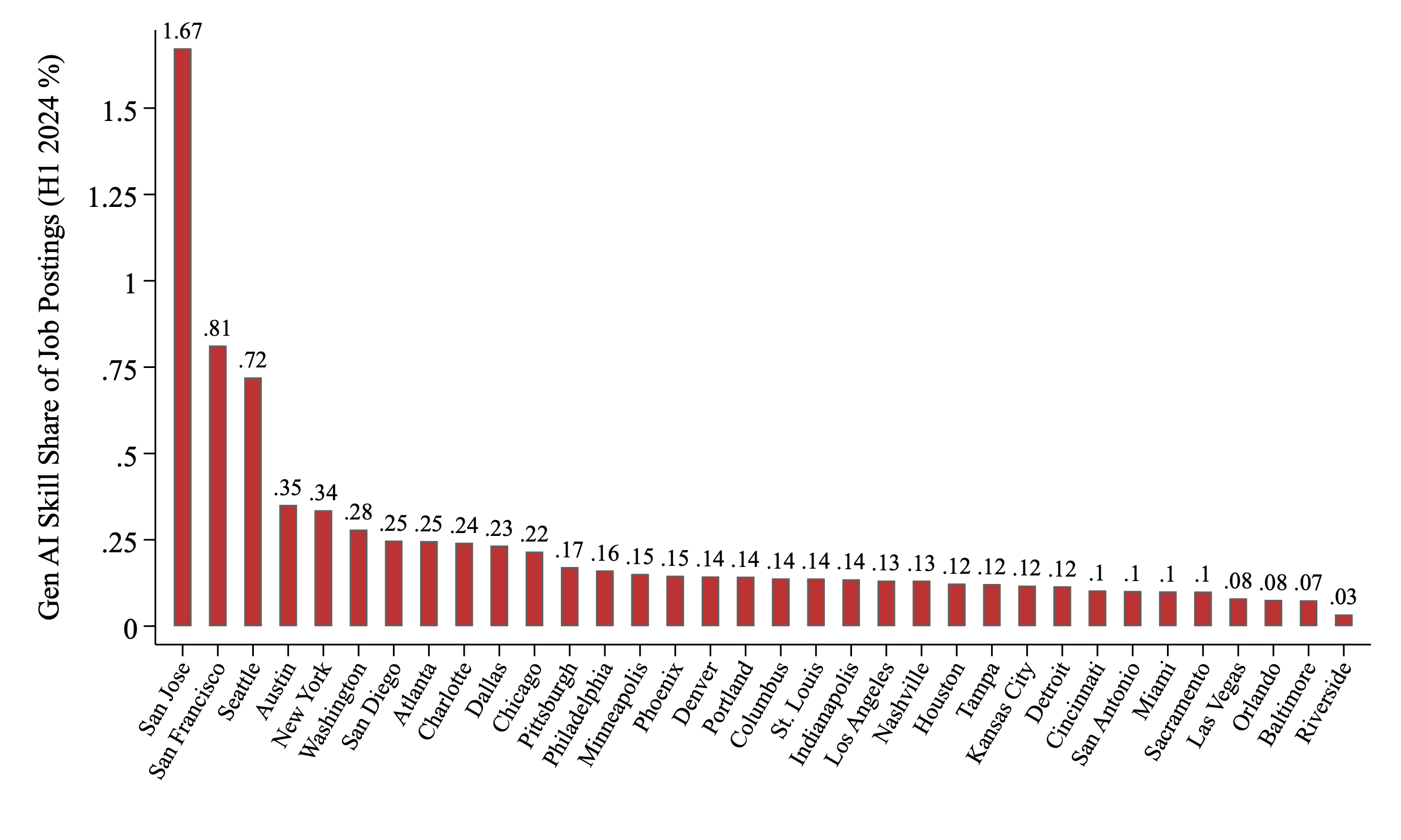}  
\end{figure}

\clearpage

\begin{figure}
\caption[.]{\\\textbf{Generative AI exposure and generative AI adoption by firms}}

\vspace{-0.1cm}  \small \label{fig:genaiscatter} These graphs show the relationship between the share of job postings at the firm level that mention generative AI skills in the 12 months ending Sep. 2024 period, and the \cite{eisfeldt2023} measure of generative AI exposure of the firm's job postings. The vertical axis shows the share of generative AI skill mentions in all job postings. Firms are sorted into deciles by exposure and each bar shows the total job posting-weighted mean share of all posted jobs at firms in that decile of exposure that mention generative AI.

  \centering
\includegraphics[width=0.85\textwidth]{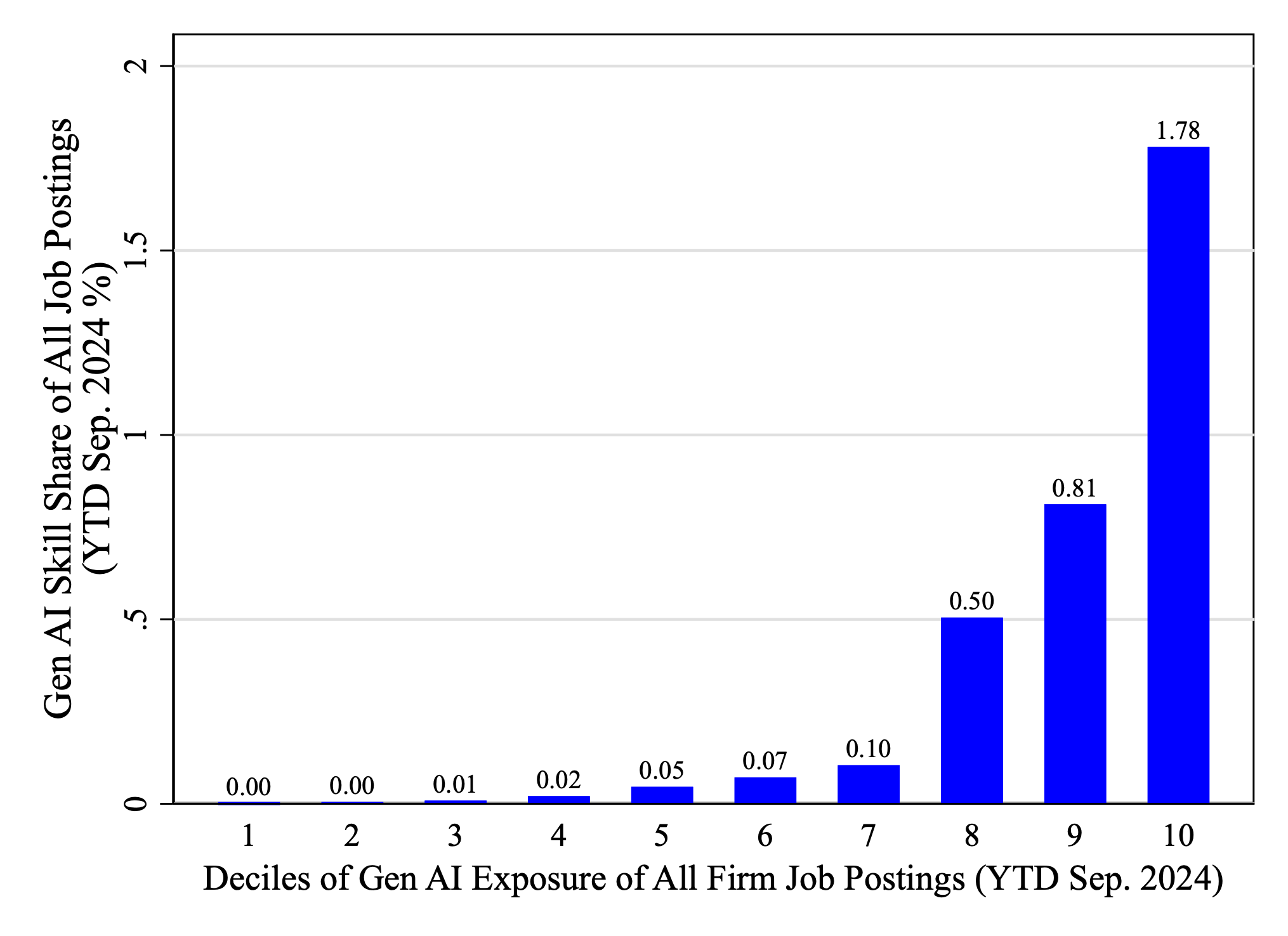}
\end{figure}

\begin{figure}
\caption[.]{\\\textbf{Firm  characteristics and remote work adoption}}\label{fig:skill2rw}

\vspace{-0.1cm} \footnotesize This figure shows OLS estimates of coefficients for the effect of standardized firm characteristics in 2019 on the firm's remote work prevalence in 2021/2022 (excl. Q4 2022) job postings in a regression of the form
$$ \text{RemoteWorkShare(`21-`22)}_{i} = \alpha_{ind} + {\beta} \text{ FirmCharacteristic(2019)}_{i}  + \text{Controls}_{i} +\varepsilon_{i}, $$
where the controls in all regressions include the 2019 value of the dependent variable as a control variable, so the coefficients can be interpreted as the effect of a standardized difference in the characteristic on changes in the remote work share. The control variables also include NAICS 2-digit fixed effects, firm-level teleworkability in 2019 and 2021/2022; and  the company's share of jobs requiring a college education in 2019. The 95\% confidence intervals shown are based on heteroskedasticity-robust standard errors clustered at the NAICS 2-digit sector level.

\centering
\includegraphics[width=0.9\textwidth]{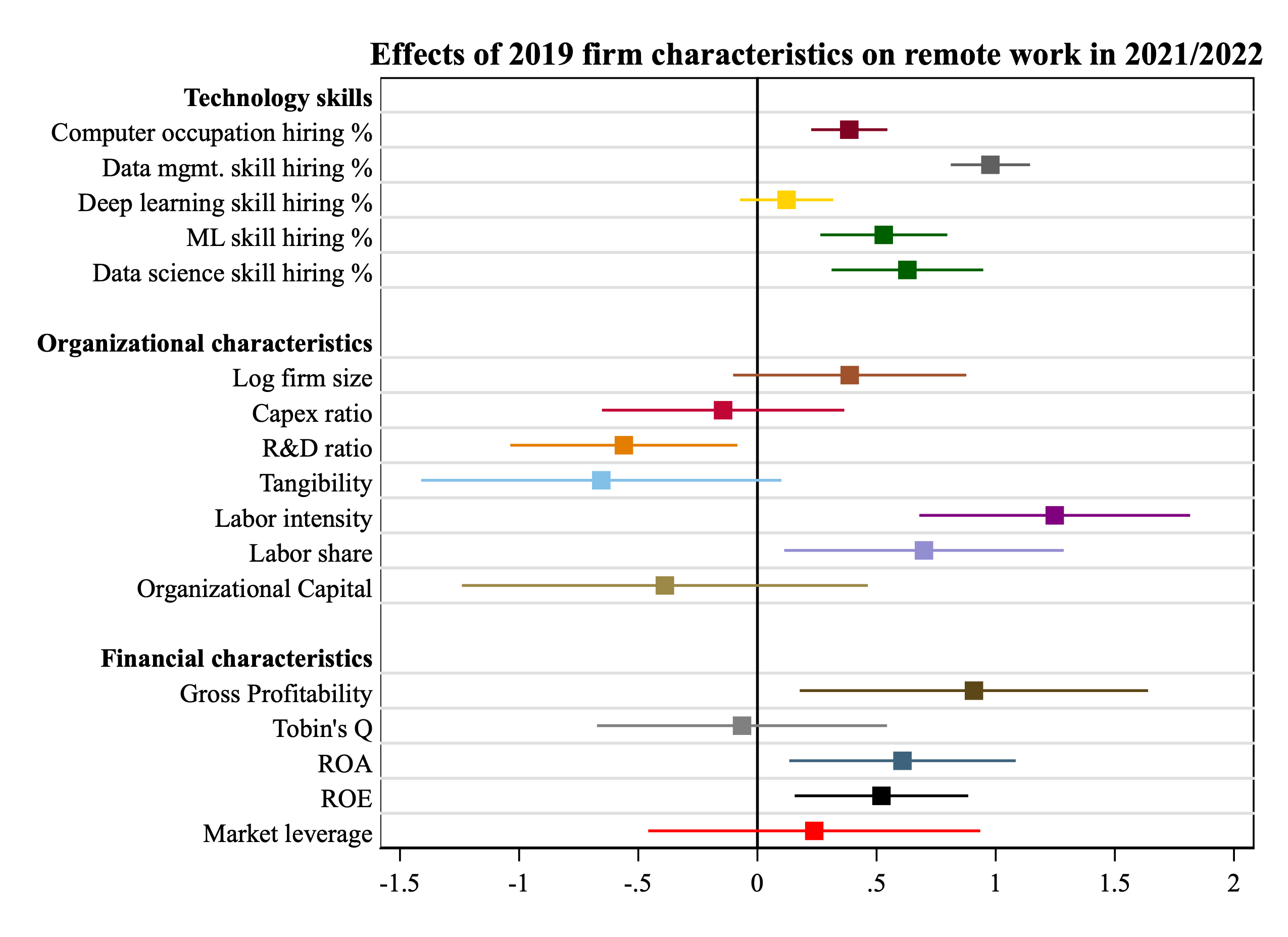}
\end{figure}

\begin{figure}
\caption[.]{\\\textbf{Remote work effects on  Gen. AI adoption: `low'-group effects by firm characteristics}}\label{fig:r2genaibyskill_base}

\vspace{-0.1cm} \scriptsize This figure shows coefficients estimated using IV for the effect of remote work prevalence in a firm's  2021/2022 (excl. Q4 2022) job postings on the prevalence of generative AI mentions in a firm's job postings in 2023/2024 (excl. Q4 2024) in a regression of the form:
\begin{center}$100 \times \text{GenAIJobShare(Oct `23--Sep. `24)}_{i} = \alpha_{ind} + {\beta} \text{RWS(`21-`22)}_{i} + {\gamma} \text{RWS(`21-`22)}_{i}\times \mathbbm{1}[\text{High Skill(2019)}_f] + \text{Controls}_{i} +\varepsilon_{i}, $\end{center}
where $RWS$ is the $\text{RemoteWorkShare}$ and the dependent variable has been scaled by 100 for better readability. So, a coefficient of 10 indicates that a 10 pp change in remote work causes a 1pp change in generative AI adoption. $\mathbbm{1}[\text{High Skill(2019)}_f]$ indicates whether a firm is above median in the characteristic. The characteristics of jobs are  measured in 2019 in Lightcast data, or or are time-invariant characteristics based on 2018 O*Net data, and are computed from 2019 averages over a firm's hiring composition. Panel A shows the `low'-group effects for the across-firm estimation (corresponding to allowing for heterogeneity in the effect in column 2 of Table \ref{tab:firmocc_competition}), and panel B shows the the `low'-group effects for the within-firm estimation (corresponding to allowing for heterogeneity in the effect in column 4 of Table \ref{tab:firmocc_competition}). The control variables are the same as in in Table \ref{tab:firmocc_competition}, except for adding a dummy for level differences between the low and high groups. The instruments are also the same, except for adding an interaction between the $\mathbbm{1}[\text{High Skill(2019)}_f]$ indicator and the respective instrument.    The 95\% confidence intervals shown are based on heteroskedasticity-robust standard errors clustered at the firm level (panel A) or double-clustered at the occupation and firm level (panel B).

\centering
\begin{subfigure}{.7\textwidth}
  \centering
\includegraphics[width=\textwidth]{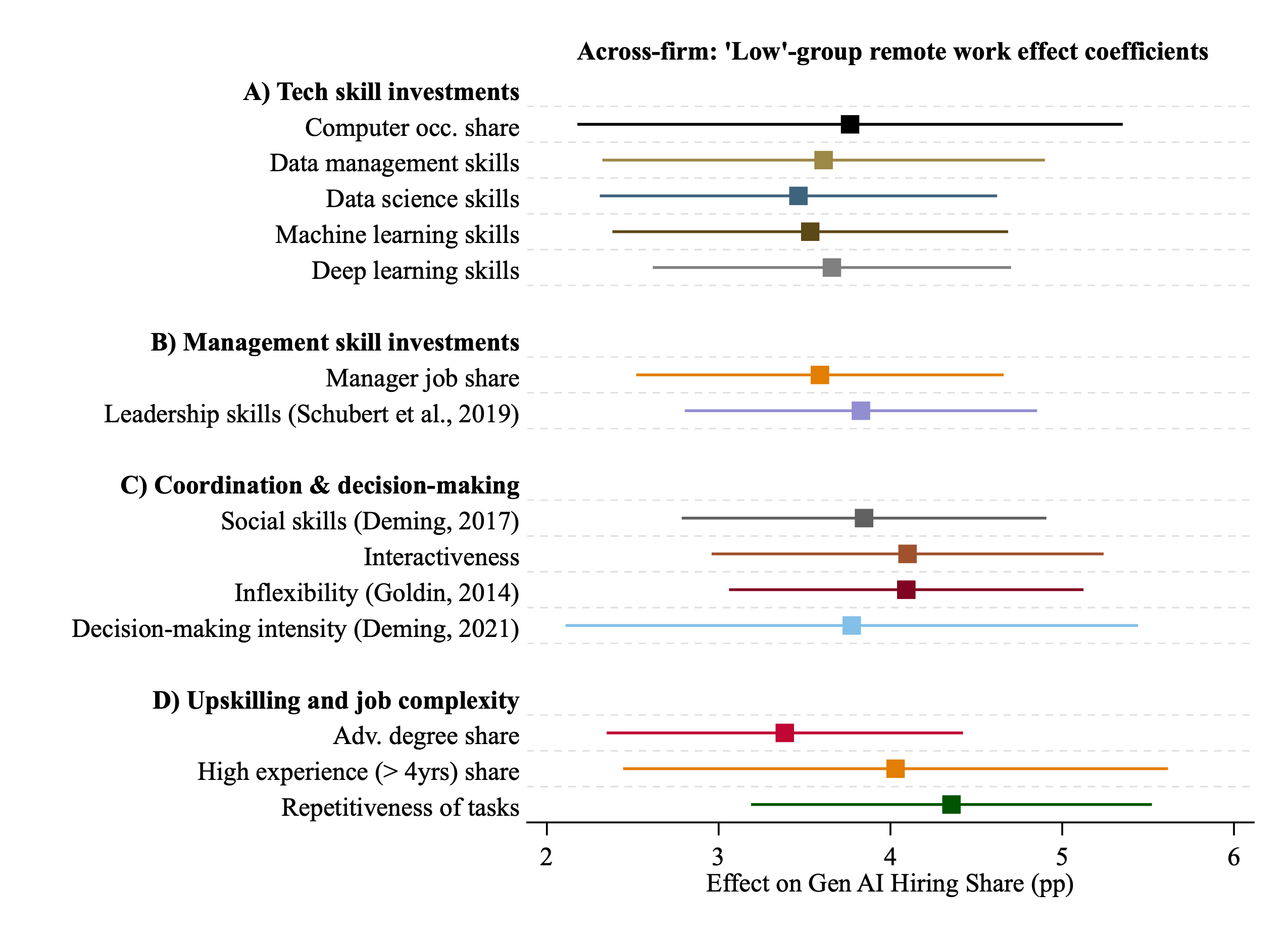}\\
  \caption{Across-firm estimate of `low'-group effects ($\beta$)}
  \label{fig:sub1}
\end{subfigure} 
\begin{subfigure}{.7\textwidth}
  \centering
\includegraphics[width=\textwidth]{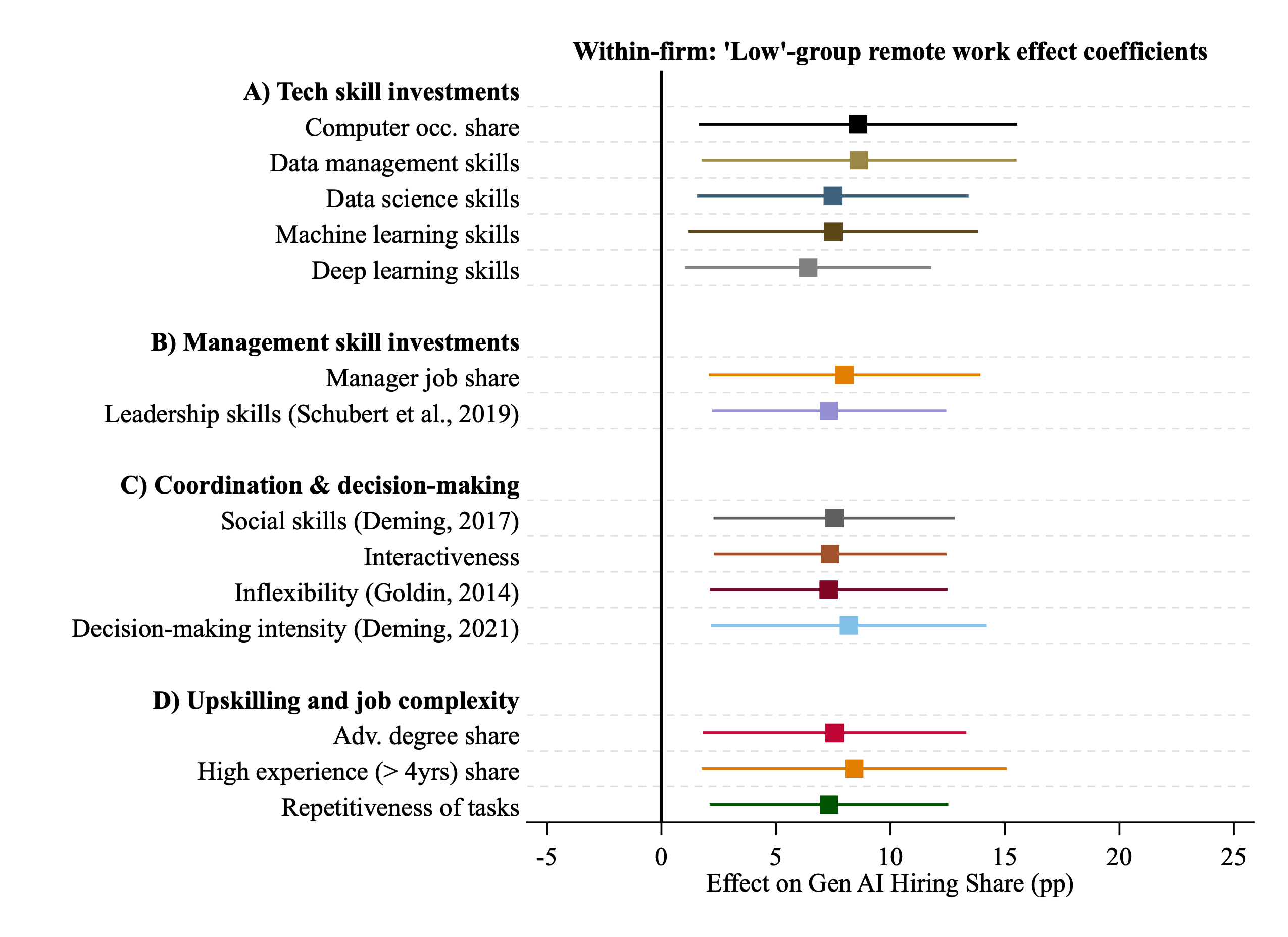}
  \caption{Within-firm estimate of `low'-group effects ($\beta$)}
  \label{fig:sub2}
\end{subfigure}
\end{figure}

\clearpage

\FloatBarrier

\begin{table}[htbp]
\caption{\\ \centering \textbf{Remote work and generative AI prevalence by occupation. }}

\vspace{-0.1cm} \small This table shows the share of all job postings in the 12 months ending Sep. 2024 in each 6-digit SOC 2010 occupation that are for remote jobs (panel A), require generative AI-related skills (panel B), and are both remote and require generative AI skills (panel C). Each panel shows the top 20 occupations, ranked by the measure of interest, which have at least 5,000 job postings during the measurement period. The last column of the table shows the total job postings in the sample for that occupation. \label{tab:occlist}  \vspace{.2cm}
 
 \centering \tiny \addtolength{\tabcolsep}{-0.5em}
\begin{subtable}{.499\textwidth}
      \centering
        \caption{A: Remote shares}
        \begin{tabular}{ll*{7}{c}@{}}
\toprule
SOC & Occupation title & Remote & Jobs \\ 
   \cmidrule(lr){1-1}    \cmidrule(lr){2-2}    \cmidrule(lr){3-3}  \cmidrule(lr){4-4} 
15-2011&Actuaries&43&9,871\\
13-2053&Insurance Underwriters&31&19,385\\
27-3042&Technical Writers&30&19,584\\
41-9041&Telemarketers&30&6,026\\
21-1014&Mental Health Counselors&29&60,334\\
41-3041&Travel Agents&28&5,208\\
13-1031&Claims Adjusters, Examiners, and Investigators&28&58,753\\
43-9041&Insurance Claims and Policy Processing Clerks&26&20,677\\
13-2082&Tax Preparers&25&33,201\\
19-3094&Political Scientists&25&6,688\\
15-1134&Web Developers&23&69,144\\
15-1121&Computer Systems Analysts&23&78,807\\
27-3043&Writers and Authors&23&27,545\\
13-1075&Labor Relations Specialists&23&5,305\\
41-9031&Sales Engineers&23&13,462\\
23-1011&Lawyers&22&103,553\\
13-1111&Management Analysts&22&125,206\\
15-1122&Information Security Analysts&21&36,023\\
15-1132&Software Developers, Applications&21&218,581\\
15-1133&Software Developers, Systems Software&21&218,581\\
 \bottomrule
\end{tabular}
         \end{subtable}
\begin{subtable}{.6\textwidth}
      \centering
        \caption{B: Generative AI shares}
        \begin{tabular}{ll*{7}{c}@{}}
\toprule
SOC & Occupation title & Gen. AI & Jobs \\ 
   \cmidrule(lr){1-1}    \cmidrule(lr){2-2}    \cmidrule(lr){3-3}  \cmidrule(lr){4-4} 
15-1131&Computer Programmers&8.5&19,768\\
27-3042&Technical Writers&6.1&19,584\\
27-3043&Writers and Authors&5.8&27,545\\
15-2099&Mathematical Science Occupations, All Other&4&187,460\\
27-3091&Interpreters and Translators&3&22,998\\
27-3041&Editors&1.9&16,550\\
11-2021&Marketing Managers&1.7&197,468\\
15-1132&Software Developers, Applications&1.7&218,581\\
15-1133&Software Developers, Systems Software&1.7&218,581\\
11-3021&Computer and Information Systems Managers&1.4&17,145\\
15-1134&Web Developers&1.4&69,144\\
15-1141&Database Administrators&1.2&133,219\\
27-1011&Art Directors&1.1&9,699\\
43-9011&Computer Operators&1.1&305,659\\
27-1014&Special Effects Artists and Animators&1.1&6,281\\
25-9031&Instructional Coordinators&.98&28,773\\
25-3099&Teachers and Instructors, All Other&.95&38,229\\
11-9041&Architectural and Engineering Managers&.95&65,851\\
27-1021&Commercial and Industrial Designers&.89&18,390\\
15-2041&Statisticians&.78&6,128\\
 \bottomrule
\end{tabular}
         \end{subtable} \\ \centering   \vspace{0.8em}
         
         \begin{subtable}{.8\textwidth}
      \centering
        \caption{C: Generative AI share in remote jobs}
        \begin{tabular}{@{}llcc@{}}
\toprule
SOC & Occupation title & Gen. AI | Remote & Jobs \\ 
   \cmidrule(lr){1-1}    \cmidrule(lr){2-2}    \cmidrule(lr){3-3}  \cmidrule(lr){4-4} 
27-3091&Interpreters and Translators&20&22,998\\
27-3042&Technical Writers&16&19,584\\
15-1131&Computer Programmers&9.6&19,768\\
27-3043&Writers and Authors&8.9&27,545\\
27-1014&Special Effects Artists and Animators&5.6&6,281\\
25-9021&Farm and Home Management Educators&4&20,921\\
15-2099&Mathematical Science Occupations, All Other&3.9&187,460\\
25-9031&Instructional Coordinators&3.1&28,773\\
11-9041&Architectural and Engineering Managers&3.1&65,851\\
25-2031&Sec. School Teachers (exc. Special/Tech. Ed.)&2.8&131,514\\
15-2041&Statisticians&2.5&6,128\\
15-1134&Web Developers&2.2&69,144\\
17-1011&Architects, Except Landscape and Naval&2&9,934\\
11-9032&Education Administrators, K-12&2&54,092\\
19-4021&Biological Technicians&1.9&7,372\\
17-2011&Aerospace Engineers&1.8&11,394\\
47-2231&Solar Photovoltaic Installers&1.8&5,719\\
29-9011&Occupational Health and Safety Specialists&1.6&47,336\\
11-2021&Marketing Managers&1.6&197,468\\
15-1133&Software Developers, Systems Software&1.5&218,581\\
 \bottomrule
\end{tabular}
         \end{subtable}

\end{table}

\begin{table}[htbp]
\caption{\\ \centering \textbf{Remote work effects on the extensive margin of generative AI hiring}}  

\vspace{-0.1cm} \small This table shows estimates of the extensive margin remote share effect on generative AI hiring at the firm- and  occupation-by-firm level in specifications of the form
\begin{center}$100 \times \mathbbm{1}[\text{GenAIJobShare(Oct `23--Sep. `24)}_{i}>0] =   {\beta} \text{ RemoteWorkShare(`21-`22)}_{i}  + FEs  + \text{Controls}_{i}  +\varepsilon_{i}$\end{center}
for the generative AI share for the 12 months ending Sep. 2024 period and the remote share during 2021-2022 (excl. Q4 2022). The dependent variable indicates whether there are \textit{any} generative AI mentions in a firm's hiring. The instrument consists of a firm's exposure to MSA teleworkability through its hiring labor markets (measured in 2019), interacted with the teleworkability of the firm's hiring in 2019 in column (2), and with the occupation's teleworkability in column (4).  The definition of baseline characteristics included as control can be found in Section \ref{sec:instruments}, and fixed effects are noted in each column. T-test statistics based on heteroskedasticity-robust standard errors clustered at the firm level (columns 1 and 2) or double-clustered at the company and 6-digit occupation level (columns 3 and 4) in parentheses: * p$<$0.10, ** p$<$0.05, *** p$<$0.01. \label{tab:firmocc_extensive}  \vspace{.2cm}
 
 \centering
 \begin{adjustbox}{max width=\textwidth}
\begin{tabular}{@{}l*{4}{c}@{}}
\toprule
     \emph{Dependent variable:}                & \multicolumn{4}{c}{100 $\times$ $\mathbbm{1}$[Generative AI Job Share $>$ 0]}   \\   
      \addlinespace
      \cmidrule(lr){2-5}
  \emph{Unit:}                & \multicolumn{2}{c}{Firm}    & \multicolumn{2}{c}{Firm $\times$ Occupation}   \\   
      \addlinespace
      \cmidrule(lr){2-3}   \cmidrule(lr){4-5}      
      
  \textit{Estimation:} & OLS & IV & OLS  & IV  \\
                                        &\multicolumn{1}{c}{(1)}   &\multicolumn{1}{c}{(2)}   &\multicolumn{1}{c}{(3)} &\multicolumn{1}{c}{(4)}      \\ 
                   \midrule
Remote Job Share (\%)&       0.083***&       0.945***&       0.002** &       0.911***\\
                    &    (11.008)   &    (12.735)   &     (2.437)   &     (3.203)   \\
\midrule \addlinespace Observations&      87,032   &      87,032   &   1,495,607   &   1,314,930   \\
1st-stage KP F-stat.&               &         417   &               &          30   \\
  \midrule
Firm baseline characteristics \ 	&   X 	& X &  &       \\
2-dig. Industry FEs & X & X &        &  \\
Firm $\times$ Occ. adv. educ. requirements &  &   & X  & X \\
Occupation FEs	&  &  &  X &  X     \\
Firm FEs & & &X & X \\
\midrule
\addlinespace
\textit{Instrument} &&  $\substack{\text{MSA Telework. Exposure} \times \\ \text{Firm Teleworkability} }$ && $\substack{\text{MSA Telework. Exposure} \times \\ \text{Occ. Teleworkability} }$ \\
 \bottomrule
\end{tabular}
\end{adjustbox}
\end{table}

\begin{table}[htbp]
\caption{\\ \centering \textbf{Placebo remote work effects on pre-Covid hiring trends}}  

\vspace{-0.1cm} \footnotesize This figure shows a ``placebo'' test of the IV approach that consists of coefficients estimated using IV for the effect of remote work prevalence in   2021/2022 (excl. Q4 2022) job postings on the trend in the characteristics of the job postings 2017-2019,  in a regression of the form
$$ \Delta\text{Skill(2017-2019)}_{i} = \alpha + {\beta} \text{ RemoteWorkShare(`21-`22)}_{i}  + \text{Controls}_{i} +\varepsilon_{i}, $$
where the controls and instruments in all regressions are the same as in column (2) and column (4) of Table \ref{tab:firmocc_competition}. The dependent variable is the change in the characteristics of the job postings at the firm or occupation-by-firm level from 2017 to 2019. The 95\% confidence intervals shown are based on heteroskedasticity-robust standard errors clustered at the company level (columns 1-3) or double-clustered at the company and 6-digit occupation level (columns 4-6). \label{tab:r2gplacebo}  \vspace{.2cm}
 
 \centering
 \begin{adjustbox}{max width=\textwidth}
\begin{tabular}{@{}l*{6}{c}@{}}
\toprule
  \emph{Unit:}                & \multicolumn{3}{c}{Firm}    & \multicolumn{3}{c}{Firm $\times$ Occupation}   \\   
      \addlinespace
      \cmidrule(lr){2-4}   \cmidrule(lr){5-7}       
      
  \textit{Dep. var.:} & $\Delta_{17-19}$ AI Skill \% & $\Delta_{17-19}$ Remote & \%  $\Delta_{17-19}$ Adv. Deg. \%  & $\Delta_{17-19}$ AI Skill \%  & $\Delta_{17-19}$ Remote \% &  $\Delta_{17-19}$ Adv. Deg. \%  \\
                                        &\multicolumn{1}{c}{(1)}   &\multicolumn{1}{c}{(2)}   &\multicolumn{1}{c}{(3)} &\multicolumn{1}{c}{(4)} &\multicolumn{1}{c}{(5)} &\multicolumn{1}{c}{(6)}      \\ 
                   \midrule
Remote Job Share (\%)&       -0.00   &       -0.03   &        0.03   &        0.63   &       -0.18   &        1.42   \\
                    &     (-0.36)   &     (-1.21)   &      (0.40)   &      (0.57)   &     (-0.29)   &      (0.52)   \\
\midrule \addlinespace Observations&      52,293   &      52,293   &      52,293   &     376,738   &     376,738   &     376,738   \\
1st-stage KP F-stat.&         284   &         284   &         284   &           0   &           0   &           0   \\
  \midrule
Firm baseline characteristics \ 	&   X 	& X & X &  &  &       \\
2-dig. Industry FEs & X & X  & X & &        &  \\
Firm adv. educ. requir. ('21-'22) & X &  X &  X & & & \\
Firm $\times$ Occ. adv. educ. requir. ('21-'22) &  &  & & X  & X & X \\
Occupation FEs	&  &  & & X &  X   & X  \\
Firm FEs & & & &X & X & X \\
\midrule
\addlinespace
\textit{Instrument} &\multicolumn{3}{c}{$\text{MSA Telework. Exposure} \times \text{Firm Teleworkability} $} &\multicolumn{3}{c}{$\text{MSA Telework. Exposure} \times \text{Occ. Teleworkability} $}  \\
 \bottomrule
\end{tabular}
\end{adjustbox}
\end{table}

\FloatBarrier

\section{Data collection} \label{sec:rtodata}

\subsection*{Return-to-Office Data from Flex Index}

I obtain information on firms' return-to-office (RTO) policies by using a webscraper to collect the crowdsourced public information from the website Flex Index (\url{https://www.flexindex.com}).\footnote{The data was collected on March 3rd, 2025 by my research assistant Yi Li, who provided invaluable assistance in this task.}

In total, I am able to collect the RTO data of 13,251 companies from the Flex Index site. This data includes each firm's basic characteristics, including headquarter state and company size, and a classification of its office attendance requirements for employees.

 After removing duplicates, I match the Flex Index companies to Compustat data to identify the standard company code (GVKEY) associated with the Flex Index companies---which are listed only by name. The matching process proceeds in the following order:
\begin{enumerate}
\item Fuzzy match company names, as well as state and/or city between the Compustat HQ location and the Flex Index company location.
\item Include all exact 1:1 company name matches in the final data
\item Drop imprecise matches that do not have city and state information
\item For remaining imprecise matches, drop all with fewer than 1000 employees, as such small firms are unlikely to be included in Compustat.
\item Manually review all remaining imprecise matches to see if both company name and location are plausibly the same, i.e. the company name only has generic differences, e.g. omitting ``Corp.'' or ``Inc.'', and the company location is in the same or a neighboring metro area in both data sets. Include in the final data if both location and name are plausible matches.
\end{enumerate} 
The final dataset includes data on 1,336 companies with office requirements (8 Flex Index classifications, see Table \ref{tab:flex-index}) for which it was possible to determine a Compustat GVKEY, and Compustat company name.

Table \ref{tab:flex-index} shows details on the different RTO policy classifications. I classify a company as having provided a  ``Return-to-Office'' mandate, if its RTO policy requires  either a full-time presence, or requires a non-zero minimum number of days or time in the office (categories 3-7).

\begin{table}[h]
\centering
\begin{tabular}{|c|l|p{9cm}|}
\hline
\textbf{No.} & \textbf{Flex Index Category} & \textbf{Description} \\
\hline
1 & Fully Remote & Organization does not have offices; all employees work remotely. \\
\hline
2 & Employee's Choice & Each employee can choose how often (or never) to come to the office. \\
\hline
3 & Minimum Days a week & There is a minimum number of days employees must come to the office each week. \\
\hline
4 & Specific Days a week & There are specific days of the week employees must come to the office. \\
\hline
5 & Minimum \& Specific Days a week &  \\
\hline
6 & Minimum \% of Time per week & There is a minimum \% of time that employees must be in the office. \\
\hline
7 & Full Time in Office & Employees are expected to be in the office full time. \\
\hline
8 & No Requirements &  \\
\hline
\end{tabular}
\caption{Flex Index Return-to-Office Policy Classifications. Source: \url{https://www.flexindex.com/about}}
\label{tab:flex-index}
\end{table}

\FloatBarrier

\section{Additional results} \label{sec:addl_results}

\subsection{Determinants of remote work adoption} \label{sec:x2r}

What type of firm is more likely to adopt remote work? The conceptual framework suggests that firms with greater  technology skills should be more likely to invest in remote work technology. I test this prediction and the effect of other firm characteristics often used in the finance literature in regressions of the form
$$ \text{RemoteWorkShare(`21-`22)}_{i} = \alpha_{ind} + {\beta} \text{ FirmCharacteristic(2019)}_{i}  + \text{Controls}_{i} +\varepsilon_{i}, $$
where all independent variables are standardized and the controls include the 2019 value of the dependent variable, so the coefficients can be interpreted as the effect of a standardized difference in the characteristic on changes in the remote work share. To proxy for the effect of simply hiring for roles that are more suitable for remote work, the control variables also include firm-level teleworkability in 2019 and 2021/2022, as well as the company's share of jobs requiring a college education in 2019, and NAICS 2-digit sector fixed effects.

The results are shown in Figure \ref{fig:skill2rw}: The first row shows the effect of a higher share of firm hiring in computer occupations on remote work adoption: firms that hire  more for computer-related positions are more likely to go remote. The coefficients in rows 2-5 capture various dimensions of technology skill prevalence in pre-pandemic hiring of the firm and show that hiring for data management, deep learning,  machine learning, and data science skills predicts greater remote work adoption. While these findings are not causal, they are consistent with the predictions regarding a technological capability mechanism for greater remote work adoption in the model.

Other organizational and financial characteristics can provide context for understanding why some firms adopt remote work: the figure also shows that firms that are larger or less R\&D-intensive, or that have a higher labor share or labor intensity in production\footnote{The latter is defined as $\ln(\text{Employment} / \text{PP\&E})$.} are more likey to adopt remote work. With regard to financials, measures of higher earnings, such as gross profitability, ROA, and ROE, all predict higher remote work. However, as these measures require Compustat data, the sample size for these estimations is much smaller than for the main job posting sample and these results should therefore be considered only suggestive.

\section{Derivations for Section ~\ref{sec:theory}}
 \label{sec:deriv}

This appendix derives Proposition~1 and the comparative statics that support Proposition~2.
For compactness, the derivations use the reduced-form output representation \eqref{eq:output_reduced}.
Throughout, assume $S_f^0<S^R<S^G$.

\subsection{Reusable capability and tech-skill heterogeneity}

From \eqref{eq:genai_cost},
\[
c_f(G\mid 0)=\Big(S^G-\omega_f S_f^0\Big)_+,
\qquad
c_f(G\mid 1)=\Big(S^G-\omega_f(S_f^0+S^R)\Big)_+,
\]
so
\[
\Delta c_f^G \equiv c_f(G\mid 0)-c_f(G\mid 1)
=
\Big(S^G-\omega_f S_f^0\Big)_+ - \Big(S^G-\omega_f(S_f^0+S^R)\Big)_+ \;\ge\; 0.
\]
In the interior case where $(\cdot)_+$ does not bind, $\Delta c_f^G=\omega_f S^R$.
Because $\omega_f=\omega(H_f)$ with $\omega'(H_f)>0$, we have
$\partial \Delta c_f^G/\partial H_f = \omega'(H_f)S^R>0$.

\subsection{Derivation of proposition 1: technology ladder}

Let $k_{fj}^G=k_{fj}+g_j$ when generative AI is adopted.
Define the gross generative AI-induced log output gain conditional on remote status $R_f$ as
\[
\Delta \ln y_{fj}^G(R_f)
\equiv
\ln y_{fj}(R_f,k_{fj}^G) - \ln y_{fj}(R_f,k_{fj}).
\]
Using \eqref{eq:output_reduced},
\begin{align}
\Delta \ln y_{fj}^G(R_f)
&=
\big[A_{fj}(k_{fj}^G)-A_{fj}(k_{fj})\big]
+
(k_{fj}-k_{fj}^G)\big(r_jR_f
+
D_{fj}q_f(R_f)\big)
\notag\\
&=
\big[A_{fj}(k_{fj}^G)-A_{fj}(k_{fj})\big]
-\underbrace{g_j r_jR_f}_{\substack{\text{Remote time-savings lost}\\  \text{on automated tasks}}}
-\underbrace{g_jD_{fj} q_f(R_f)}_{\substack{\text{Decision-sensitive human}\\ \text{output lost}}}.
\label{eq:genai_gain}
\end{align}

Taking the difference between remote and non-remote gross gains cancels the $A_{fj}(\cdot)$ term and the $\ln M_f$ component inside $q_f(\cdot)$:
\begin{align}
\Delta \ln y_{fj}^G(1) - \Delta \ln y_{fj}^G(0)
&=
-g_j r_j
-\big(q_f(1)-q_f(0)\big)g_jD_{fj}
\notag\\
&=
-g_j r_j
+\rho_f g_jD_{fj}.
\label{eq:gross_diff}
\end{align}
Next, incorporate the reusability-driven cost difference.
From \eqref{eq:genai_cost},
\[
c_f(G\mid 0)=\Big(S^G-\omega_f S_f^0\Big)_+,
\qquad
c_f(G\mid 1)=\Big(S^G-\omega_f(S_f^0+S^R)\Big)_+,
\]
so the cost reduction is
\[
\Delta c_f^G \equiv c_f(G\mid 0)-c_f(G\mid 1)\;\ge\;0,
\]
which simplifies to $\Delta c_f^G=\omega_f S^R$ in the interior case.

Combining \eqref{eq:genai_gain} with the cost difference yields Proposition~1:
\begin{align}
\mathcal{L}_{fj}
&=
\Big[\Delta \ln y_{fj}^G - c_f(G\mid R_f)\Big]_{R_f=1}
-
\Big[\Delta \ln y_{fj}^G - c_f(G\mid R_f)\Big]_{R_f=0}
\notag\\
&=
\big(\Delta \ln y_{fj}^G(1)-\Delta \ln y_{fj}^G(0)\big)
+
\big(c_f(G\mid 0)-c_f(G\mid 1)\big)
\notag\\
&=
-g_j r_j + g_jD_{fj}\rho_f + \Delta c_f^G.
\label{eq:net_diff}
\end{align}

A sufficient condition for a positive technology ladder effect is therefore
\[
\Delta c_f^G  + g_j \rho_f D_{fj} > g_j r_j,
\]
which is equation \ref{eq:ladder_condition} from the main text.

\subsection{Comparative statics supporting Proposition~2}

Under the interior case, using \eqref{eq:net_diff} we have
\[
\mathcal{L}_{fj}(\rho_f,D_{fj},\omega,r_j,g_j)
=
-g_j r_j + g_jD_{fj}\rho_f + \Delta c_f^G(\omega,S^R,S^G,S_f^0).
\]

\paragraph{Coordination penalty and decision sensitivity.}
\[
\frac{\partial \mathcal{L}_{fj}}{\partial \rho_f} = g_jD_{fj} \;>\;0,
\qquad
\frac{\partial \mathcal{L}_{fj}}{\partial D_{fj}} = g_j\rho_f \;\ge\;0.
\]
Thus, the ladder effect is stronger when remote coordination is less effective (higher $\rho_f$)
and when tasks are more decision-sensitive (higher $D_{fj}$).

\paragraph{Remote time savings.}
\[
\frac{\partial \mathcal{L}_{fj}}{\partial r_j} = -g_j \;<\;0,
\]
so stronger remote time-savings benefits reduce the incentive to automate remote work with GenAI, all else equal.

\paragraph{Reusability and firm tech skills.}
Under the interior case, $\Delta c_f^G=\omega_f S^R$, so
\[
\frac{\partial \mathcal{L}_{fj}}{\partial \omega_f}
=
\frac{\partial \Delta c_f^G}{\partial \omega_f}
=
S^R \;>\;0.
\]
Since $\omega_f=\omega(H_f)$ with $\omega'(H_f)>0$,
\[
\frac{\partial \mathcal{L}_{fj}}{\partial H_f}
=
\omega'(H_f)S^R \;>\;0,
\]
so the technology ladder effect is stronger for firms with higher tech-skill intensity.

\paragraph{Remote-work technology investment.}
Again under the interior case,
\[
\frac{\partial \mathcal{L}_{fj}}{\partial S^R}
=
\omega_f \;\ge\; 0,
\]
so larger remote-work technology investments generate larger downstream cost savings for generative AI.

\section{Earnings call analysis details} \label{sec:methodology_calls}

\textbf{Earnings call data.} I focus on firm earnings calls held in 2019-2025, which are sourced from Capital IQ via WRDS. The raw data includes transcript text segmented by speaker turn (e.g., CEO remarks, analyst questions), along with identifiers allowing me to link each transcript to Compustat via GVKEY. After removing duplicate transcripts and calls with missing text, the sample includes transcripts from \~15K unique firms.

\textbf{Keyword filtering.}  As a first step, I identify text segments for detailed classification by whether they contain any relevant keywords indicating that they may be discussing either generative AI or remote work. Rather than relying on an ad hoc list of keywords, I develop technology-specific keyword sets using word embeddings trained on the corpus of earnings calls. I train a Word2Vec model \citep{mikolov2013distributed} on the 2024 transcript corpus and use its mapping to words with similar use to expand a small set of seed terms for each technology, such as ``remote,'' ``hybrid,'' and ``work-from-home'' for remote work, or ``ChatGPT,'' ``LLM,'' and ``generative AI'' for generative AI. I then manually review and filter the expanded lists to remove false positives. The final keyword selection includes the unigrams and bigrams shown in Appendix \ref{sec:keywords}.

\textbf{Segment selection.} Given the length of typical earnings calls (often exceeding 10,000 words), directly processing full transcripts with LLMs would be prohibitively expensive and would introduce noise from irrelevant passages. I therefore filter transcripts to retain only segments likely to contain relevant content based on the identified keywords.

For each transcript and technology considered (generative AI; remote work), I first chunk long segments to approximately 600-800 tokens to respect LLM context limits while maintaining coherent passages. I identify text chunks containing at least one keyword from the relevant technology's keyword list. To preserve context, I also retain a window of one text chunk before and one chunk after each keyword-containing text chunk.  Near-duplicate segments (Jaccard similarity $>$ 0.95) arising from boilerplate language or multiple copies of the same transcript in the data are removed. This filtering is intended to reduce the text volume while retaining the substantive discussions of each technology.

\subsection{Keywords used for earnings call segment selection}
\label{sec:keywords}

Keywords are developed using Word2Vec embeddings trained on the 2024 earnings call corpus, starting from seed terms and expanding to include semantically similar terms. The expanded lists were then manually reviewed to remove false positives. The final keyword lists used for identifying earnings call segments for further analysis are shown below.

\subsubsection{Generative AI Keywords}

\paragraph{Unigrams.} 
agentic, anthropic, bard, chatbot, chatbots, chatgpt, claude, cohere, copilot, copilots, dalle, deepmind, einstein, gemini, genai, generative, gpt, gpt3, gpt4, gpt5, llama, llm, llms, midjourney, openai, perplexity, transformer, transformers, xai.

\paragraph{Bigrams.} 
agentic ai, ai adoption, ai assistant, ai assistants, ai automation, ai capabilities, ai chatbot, ai companion, ai driven, ai enabled, ai investment, ai powered, ai productivity, ai strategy, ai tool, ai tools, ai transformation, amazon bedrock, artificial intelligence, augmented intelligence, aws bedrock, azure openai, causal ai, cognitive automation, computer vision, conversational ai, einstein copilot, foundation model, gen ai, generative ai, generative ai models, generative artificial intelligence, generative pretrained transformers, google vertex, language model, language models, large language model, large language models, microsoft copilot, retrieval augmented generation (rag), salesforce einstein, speech recognition, vertex ai, virtual agent, virtual assistant, voice assistant, workflow automation.

\subsubsection{Remote Work Keywords}

\paragraph{Unigrams.} 
digitalworkplace, homeoffice, remote, remoteaccess, remotely, returntooffice, rto, telecommute, telecommuting, wfh, workfromhome.

\paragraph{Bigrams.} 
digital workplace, distributed team, distributed workforce, home office, hybrid model, hybrid work, location flexible, office reopening, remote employee, remote employees, remote first, remote work, remote working, return to office, virtual team, virtual workplace, work anywhere, work from home, working from home.

\subsection{LLM Classification Prompts}
\label{sec:prompts}

The LLM classification pipeline uses three steps: (1) a relevance gate to filter for excerpts about the firm's own technology adoption, (2) hypothesis adjudication to evaluate evidence of investment and outcomes, and (3) a skeptic audit to reduce false positives. Below we present the prompts used for each step. Prompts are adapted for each technology (remote work vs.\ generative AI) while maintaining parallel structure.

Each separate prompt is submitted to GPT 5.1-nano via the Azure OpenAI API, using the ``2024-12-01-preview'' version of the API. Note that GPT 5.1 does not allow for adjustments to the temperature parameter.

\tcbset{
    promptbox/.style={
        colback=gray!5,
        colframe=gray!50!black,
        fonttitle=\bfseries,
        boxrule=0.5pt,
        arc=2pt,
        left=6pt,
        right=6pt,
        top=6pt,
        bottom=6pt,
        breakable,
        enhanced,
    }
}

\subsubsection{Remote Work Prompts}

\paragraph{Step 1: Relevance Gate (Remote Work)}

\begin{tcolorbox}[colback=gray!5, colframe=gray!50!black, breakable, title={}]
\begin{lstlisting}[style=prompt]
You are filtering earnings-call excerpts for statements about employee work locations
(remote work / hybrid work / work-from-home / return-to-office), not "remote" in other senses
(remote monitoring, remote operations, remote diagnostics, remote customer support, etc.).

IMPORTANT: We want statements about THIS COMPANY'S OWN WORKFORCE arrangements, NOT:
- How remote work trends affect the company's product, market, or customers
- Commentary on the broader economy or industry trends around remote work
- Impact on tenants, clients, or end-users of the company's products

Examples:
- "remote monitoring of equipment" -> "no" (not about workforce)
- "our remote workforce has grown" -> "yes" (about own employees)
- "customers can access our services remotely" -> "no" (about customers, not employees)
- "demand for office space has declined as companies embrace remote work" -> "no" (market commentary, not own workforce)
- "our tenants are reducing footprint due to hybrid work" -> "no" (about customers/tenants, not own employees)
- "we've seen occupancy decline as remote work persists" -> "no" (product/market impact, not own workforce policy)
- "we implemented a hybrid schedule for our employees" -> "yes" (own workforce policy)

Input excerpt:
{EXCERPT}

Return your response as a JSON object with this exact structure:
```json
{
  "is_about_remote_work_arrangement": "<one of: yes, no, unclear>",
  "remote_work_spans": [
    "<phrase 1>",
    "<phrase 2>",
    "<phrase 3>"
  ]
}
```

Notes:
- "remote_work_spans": Include up to 3 short phrases that directly reference THIS COMPANY'S OWN workforce location or hybrid policy. These should be exact quotes or near-exact phrases from the excerpt. Use empty array [] if no relevant phrases exist.
- If the excerpt discusses remote work only as a market trend, product impact, or customer behavior (not the company's own workforce), set is_about_remote_work_arrangement to "no".
\end{lstlisting}
\end{tcolorbox}

\paragraph{Step 2: Hypothesis Adjudication (Remote Work)}

\begin{tcolorbox}[colback=gray!5, colframe=gray!50!black, breakable, title={}]
\begin{lstlisting}[style=prompt]
You are analyzing an earnings call excerpt to adjudicate three hypotheses about 
the company's OWN WORKFORCE remote/hybrid work arrangements.

IMPORTANT SCOPE:
- Only consider statements about THIS COMPANY'S OWN EMPLOYEES
- Ignore market trends, product impacts, customer behavior, or industry commentary
- Focus on what the company says about its own workforce policies, investments, and experiences

Hypotheses to evaluate:
- H_invest_enablement: The company is investing (spending, building, implementing, expanding) in 
  tools/skills/tech/process/talent specifically to support remote/hybrid work.
- H_experience_positive: The company reports that remote/hybrid work IMPROVED outcomes 
  (e.g., productivity, retention, cost savings, culture, innovation, execution speed).
- H_experience_negative: The company reports that remote/hybrid work HARMED outcomes
  (e.g., productivity, collaboration, culture, coordination, innovation).

CRITICAL DISTINCTIONS - READ CAREFULLY:

1. POLICY != INVESTMENT
   - "We have a hybrid policy" -> Policy statement, NOT investment
   - "We allow employees to work remotely" -> Policy statement, NOT investment
   - "We invested in collaboration tools for our hybrid workforce" -> IS investment
   - "We rolled out new VPN infrastructure" -> IS investment (if linked to remote work)
   Investment requires evidence of SPENDING, BUILDING, IMPLEMENTING, or ACQUIRING something.

2. POLICY != POSITIVE EXPERIENCE
   - "We offer remote work options" -> Policy statement, NOT positive experience
   - "Our employees appreciate the flexibility" -> Could be positive experience (if about outcomes)
   - "Remote work has improved our retention rates" -> IS positive experience
   - "We've seen productivity gains from hybrid work" -> IS positive experience
   Experience requires evidence of OBSERVED OUTCOMES or REPORTED RESULTS.

3. BE CONSERVATIVE
   - Prefer "not_addressed" unless the evidence clearly supports the hypothesis
   - Generic IT modernization without explicit remote/hybrid linkage -> not_addressed
   - Vague or implied benefits without concrete statements -> not_addressed

Excerpt:
{EXCERPT}

Analyze the excerpt and return a JSON object with this exact structure:
```json
{
  "evidence_quotes": [
    {
      "quote": "<exact quote from excerpt, max 30 words>",
      "type": "<one of: investment, experience_positive, experience_negative, policy, other>",
      "relevance": "<1 sentence: what does this evidence support or show?>"
    }
  ],
  "investment_dimensions": {
    "tools": <boolean - general productivity/work tools>,
    "security": <boolean - VPN, endpoint security, zero trust>,
    "cloud": <boolean - cloud migration, SaaS for remote access>,
    "training": <boolean - manager training, remote work skills>,
    "process": <boolean - workflow redesign, async processes>,
    "real_estate": <boolean - office consolidation, hoteling, space redesign>,
    "collaboration_software": <boolean - Zoom, Teams, Slack, etc.>,
    "data_infrastructure": <boolean - data access, analytics for distributed teams>,
    "talent_hiring": <boolean - geographic expansion of hiring, remote-first recruiting>
  },
  "experience_dimensions": {
    "productivity": <integer: -1=negative, 0=not mentioned, 1=positive>,
    "creativity": <integer: -1=negative, 0=not mentioned, 1=positive>,
    "cost": <integer: -1=negative, 0=not mentioned, 1=positive>,
    "culture": <integer: -1=negative, 0=not mentioned, 1=positive>,
    "retention": <integer: -1=negative, 0=not mentioned, 1=positive>,
    "execution_speed": <integer: -1=negative, 0=not mentioned, 1=positive>,
    "innovation_speed": <integer: -1=negative, 0=not mentioned, 1=positive>,
    "collaboration": <integer: -1=negative, 0=not mentioned, 1=positive>
  },
  "H_invest_enablement": {
    "label": "<one of: entailed, contradicted, not_addressed>"
  },
  "H_experience_positive": {
    "label": "<one of: entailed, contradicted, not_addressed>"
  },
  "H_experience_negative": {
    "label": "<one of: entailed, contradicted, not_addressed>"
  },
  "has_remote_workers": <integer: 1 if the excerpt indicates this company has/allows remote or hybrid workers, 0 if not or unclear>,
  "brief_reasoning": "<2-3 sentences explaining your assessment, especially noting any ambiguity or why you chose not_addressed>"
}
```

Additional notes:
- "evidence_quotes": Extract up to 5 relevant quotes that support your analysis. Use empty array [] if no relevant evidence exists.
- If the excerpt contains ONLY policy statements with no investment or experience evidence, all H_* labels should be "not_addressed".
- An excerpt can have BOTH positive and negative experience evidence (e.g., "productivity improved but collaboration suffered").
- Set all investment_dimensions to false and experience_dimensions to 0 if no relevant evidence exists.
- "brief_reasoning" is important for interpretability - explain your logic.
\end{lstlisting}
\end{tcolorbox}

\paragraph{Step 3: Skeptic Audit (Remote Work)}

\begin{tcolorbox}[colback=gray!5, colframe=gray!50!black, breakable, title={}]
\begin{lstlisting}[style=prompt]
Act as a hostile auditor. Your job is to find any place the adjudication overreached the evidence.

You are given:
(1) The original excerpt
(2) The adjudication JSON (hypothesis labels, evidence quotes, reasoning)

Your task: Verify that every claim and label is justified by the excerpt text.

Excerpt:
{EXCERPT}

Adjudication JSON:
{ADJUDICATION_JSON}

CHECK FOR THESE SPECIFIC ISSUES:

1. UNSUPPORTED QUOTES: Do the evidence_quotes actually appear in the excerpt? 
   Flag any quotes that are paraphrased, fabricated, or significantly altered.

2. POLICY CONFLATION: Was a policy statement incorrectly used to support H_invest_enablement or H_experience_*?
   - "We have hybrid work" should NOT entail investment or positive experience
   - Only flag if this error actually affected the hypothesis labels

3. MARKET/PRODUCT LEAKAGE: Was commentary about market trends, customers, or product impact 
   incorrectly treated as evidence about the company's own workforce?

4. OVERCONFIDENT LABELS: Should any "entailed" label be downgraded to "not_addressed"?
   - Evidence too weak or ambiguous?
   - Inference required that isn't explicitly supported?

5. ALTERNATIVE INTERPRETATIONS: Could "remote" plausibly mean something other than remote work?
   - Remote monitoring, remote operations, remote locations, etc.

6. MISSED EVIDENCE: Did the adjudication miss important evidence that would change the labels?
   (This is less common but worth checking)

IMPORTANT: Only flag genuine issues. Reasonable inferences supported by clear evidence are NOT overreach.

Return your response as a JSON object with this exact structure:
```json
{
  "overreach_flags": [
    {
      "issue_type": "<one of: unsupported_quote, policy_conflation, market_leakage, overconfident, alternative_interpretation, missed_evidence>",
      "item": "<specific item that was overreached - quote, label, or dimension>",
      "why": "<explanation of the problem>",
      "suggested_fix": "<how to correct it>"
    }
  ],
  "revised_labels": {
    "H_invest_enablement": "<entailed | contradicted | not_addressed | unchanged>",
    "H_experience_positive": "<entailed | contradicted | not_addressed | unchanged>",
    "H_experience_negative": "<entailed | contradicted | not_addressed | unchanged>"
  },
  "revised_investment_dimensions": {
    "tools": <boolean or null if unchanged>,
    "security": <boolean or null if unchanged>,
    "cloud": <boolean or null if unchanged>,
    "training": <boolean or null if unchanged>,
    "process": <boolean or null if unchanged>,
    "real_estate": <boolean or null if unchanged>,
    "collaboration_software": <boolean or null if unchanged>,
    "data_infrastructure": <boolean or null if unchanged>,
    "talent_hiring": <boolean or null if unchanged>
  },
  "revised_experience_dimensions": {
    "productivity": <integer -1/0/1 or null if unchanged>,
    "creativity": <integer -1/0/1 or null if unchanged>,
    "cost": <integer -1/0/1 or null if unchanged>,
    "culture": <integer -1/0/1 or null if unchanged>,
    "retention": <integer -1/0/1 or null if unchanged>,
    "execution_speed": <integer -1/0/1 or null if unchanged>,
    "innovation_speed": <integer -1/0/1 or null if unchanged>,
    "collaboration": <integer -1/0/1 or null if unchanged>
  },
  "revised_has_remote_workers": <integer 0/1 or null if unchanged>,
  "audit_passed": <boolean - true if no significant issues found>,
  "brief_rationale": "<2-3 sentences summarizing your audit findings>"
}
```

Notes:
- "overreach_flags": Include one object per issue found. Use empty array [] if no issues detected.
- "revised_labels": Use "unchanged" for hypotheses where the original label was correct. Only specify a new label if revision is needed.
- "revised_investment_dimensions" and "revised_experience_dimensions": Use null for dimensions that don't need changes. Only include actual values for dimensions that need correction.
- "revised_has_remote_workers": Use null if the original assessment was correct, or 0/1 if it needs revision.
- "audit_passed": Set to true if the adjudication is fundamentally sound, even if minor issues exist.
- Be fair: Don't flag issues that don't materially affect the analysis.
\end{lstlisting}
\end{tcolorbox}

\subsubsection{Generative AI Prompts}

\paragraph{Step 1: Relevance Gate (Generative AI)}

\begin{tcolorbox}[colback=gray!5, colframe=gray!50!black, breakable, title={}]
\begin{lstlisting}[style=prompt]
You are filtering earnings-call excerpts for statements about THIS COMPANY'S OWN USE 
of generative AI, large language models (LLMs), or related technologies (ChatGPT, Copilot, Claude, 
foundation models, AI assistants, etc.).

IMPORTANT: We want statements about THIS COMPANY'S OWN ADOPTION/USE of generative AI, NOT:
- The company's AI products or services sold to customers
- General commentary on AI trends in the industry or economy
- Customer/client adoption of AI (unless describing how it affects this company's operations)
- Traditional ML/analytics that isn't generative AI

Examples:
- "we're using ChatGPT to help our developers write code" -> "yes" (own workforce using genai)
- "we launched a new AI assistant product for our customers" -> "no" (product for customers, not internal use)
- "generative AI is transforming the industry" -> "no" (market commentary)
- "we deployed Copilot across our engineering teams" -> "yes" (own workforce adoption)
- "our customers are asking about AI features" -> "no" (customer behavior)
- "we're training our employees on AI tools" -> "yes" (own workforce)
- "AI revenue grew 30%" -> "no" (product/revenue, not internal adoption)
- "we use LLMs to automate customer support responses" -> "yes" (own operations)
- "we have concerns about data privacy with AI tools" -> "yes" (own adoption considerations)
- "our existing cloud infrastructure positions us well for AI" -> "yes" (enabling infrastructure)

Input excerpt:
{EXCERPT}

Return your response as a JSON object with this exact structure:
```json
{
  "is_relevant": "<one of: yes, no, unclear>",
  "relevant_spans": [
    "<phrase 1>",
    "<phrase 2>",
    "<phrase 3>"
  ]
}
```

Notes:
- "relevant_spans": Include up to 3 short phrases that directly reference THIS COMPANY'S OWN use/adoption of generative AI, obstacles to adoption, or enabling infrastructure. Use exact quotes or near-exact phrases. Use empty array [] if no relevant phrases exist.
- If the excerpt discusses generative AI only as a product, market trend, or customer behavior (not the company's own internal adoption), set is_relevant to "no".
- Include excerpts that discuss obstacles, concerns, or enabling factors for the company's own AI adoption.
\end{lstlisting}
\end{tcolorbox}

\paragraph{Step 2: Hypothesis Adjudication (Generative AI)}

\begin{tcolorbox}[colback=gray!5, colframe=gray!50!black, breakable, title={}]
\begin{lstlisting}[style=prompt]
You are analyzing an earnings call excerpt to adjudicate hypotheses about 
the company's OWN INTERNAL adoption of generative AI / LLMs.

IMPORTANT SCOPE:
- Only consider statements about THIS COMPANY'S OWN USE of generative AI
- Ignore AI products/services sold to customers
- Ignore general market commentary about AI trends
- Focus on: investments, experiences, obstacles, and enabling infrastructure

Hypotheses to evaluate:
- H_invest_enablement: The company is investing (spending, building, implementing, expanding) in 
  generative AI tools/infrastructure/training specifically for internal use by employees.
- H_experience_positive: The company reports that generative AI adoption IMPROVED internal outcomes 
  (e.g., productivity, efficiency, cost savings, speed, quality, innovation).
- H_experience_negative: The company reports that generative AI adoption HARMED internal outcomes
  or faces significant obstacles (e.g., quality issues, security concerns, failed implementations, costs).

CRITICAL DISTINCTIONS:

1. MENTION != INVESTMENT
   - "We're excited about AI" -> Sentiment, NOT investment
   - "We're exploring AI opportunities" -> Exploration, NOT investment
   - "We deployed GitHub Copilot to 5,000 developers" -> IS investment
   - "We're training our workforce on AI tools" -> IS investment
   Investment requires evidence of SPENDING, DEPLOYING, IMPLEMENTING, or TRAINING.

2. MENTION != POSITIVE EXPERIENCE
   - "AI has great potential" -> Opinion, NOT positive experience
   - "AI tools reduced our code review time by 40%" -> IS positive experience
   Experience requires evidence of OBSERVED OUTCOMES or MEASURED RESULTS.

3. OBSTACLES COUNT AS NEGATIVE EXPERIENCE
   - "We have data privacy concerns" -> IS negative/obstacle
   - "Regulatory uncertainty is slowing our adoption" -> IS negative/obstacle
   - "We lack the talent to implement AI" -> IS negative/obstacle

4. PRODUCT vs INTERNAL USE
   - "Our AI product is growing" -> Product (NOT internal adoption)
   - "We use AI internally to improve our products" -> IS internal adoption

Excerpt:
{EXCERPT}

Analyze the excerpt and return a JSON object with this exact structure:
```json
{
  "evidence_quotes": [
    {
      "quote": "<exact quote from excerpt, max 30 words>",
      "type": "<one of: investment, experience_positive, experience_negative, obstacle, enabling_infrastructure, policy, other>",
      "relevance": "<1 sentence: what does this evidence support or show?>"
    }
  ],
  "investment_dimensions": {
    "llm_apis": <boolean - using OpenAI, Anthropic, Google, Cohere or other LLM APIs>,
    "copilot_tools": <boolean - GitHub Copilot, Amazon CodeWhisperer, coding assistants>,
    "chatbots_internal": <boolean - internal chatbots, AI assistants for employees>,
    "training_programs": <boolean - AI literacy, prompt engineering, upskilling>,
    "compute_infrastructure": <boolean - GPU clusters, AI-optimized hardware>,
    "cloud_ai_services": <boolean - Azure OpenAI, AWS Bedrock, Google Vertex>,
    "data_preparation": <boolean - RAG pipelines, vector databases, fine-tuning, data labeling>,
    "governance_frameworks": <boolean - AI policies, ethics review, safety measures, responsible AI>,
    "vendor_partnerships": <boolean - partnerships with AI vendors, consulting engagements>,
    "custom_models": <boolean - building/fine-tuning proprietary models>
  },
  "enabling_infrastructure": {
    "cloud_maturity": <boolean - existing cloud infrastructure that enables AI adoption>,
    "data_infrastructure": <boolean - data lakes, pipelines, data governance already in place>,
    "developer_tools": <boolean - DevOps, CI/CD, modern development practices>,
    "prior_ml_investments": <boolean - existing ML/analytics capabilities being extended>,
    "security_infrastructure": <boolean - existing security/compliance frameworks applicable to AI>
  },
  "obstacles": {
    "data_privacy": <boolean - concerns about data privacy, confidentiality, IP protection>,
    "security_risks": <boolean - cybersecurity concerns, model vulnerabilities>,
    "regulatory_compliance": <boolean - regulatory uncertainty, compliance requirements>,
    "cost_concerns": <boolean - high costs, unclear ROI, budget constraints>,
    "talent_gap": <boolean - lack of AI skills, difficulty hiring AI talent>,
    "integration_complexity": <boolean - difficulty integrating with existing systems>,
    "quality_accuracy": <boolean - concerns about hallucinations, accuracy, reliability>,
    "employee_resistance": <boolean - workforce concerns, change management issues>,
    "vendor_lock_in": <boolean - concerns about dependency on specific vendors>
  },
  "experience_dimensions": {
    "productivity": <integer: -1=negative, 0=not mentioned, 1=positive>,
    "code_quality": <integer: -1=negative, 0=not mentioned, 1=positive>,
    "cost_efficiency": <integer: -1=negative, 0=not mentioned, 1=positive>,
    "speed_time_savings": <integer: -1=negative, 0=not mentioned, 1=positive>,
    "employee_satisfaction": <integer: -1=negative, 0=not mentioned, 1=positive>,
    "innovation": <integer: -1=negative, 0=not mentioned, 1=positive>,
    "output_quality": <integer: -1=negative, 0=not mentioned, 1=positive>,
    "decision_making": <integer: -1=negative, 0=not mentioned, 1=positive>
  },
  "H_invest_enablement": {
    "label": "<one of: entailed, contradicted, not_addressed>"
  },
  "H_experience_positive": {
    "label": "<one of: entailed, contradicted, not_addressed>"
  },
  "H_experience_negative": {
    "label": "<one of: entailed, contradicted, not_addressed>"
  },
  "has_adopted_technology": <integer: 1 if the excerpt indicates this company is actively using generative AI internally, 0 if not or unclear>,
  "remote_work_mentioned": <boolean: true if remote work, hybrid work, or distributed workforce is mentioned in connection with AI adoption>,
  "brief_reasoning": "<2-3 sentences explaining your assessment, especially noting any ambiguity or why you chose not_addressed>"
}
```

Additional notes:
- "evidence_quotes": Extract up to 5 relevant quotes. Use empty array [] if no relevant evidence exists.
- "enabling_infrastructure": Capture existing capabilities that make AI adoption easier.
- "obstacles": Capture barriers, concerns, or challenges to AI adoption.
- "remote_work_mentioned": Set to true if remote/hybrid work is discussed as related to AI (e.g., "AI helps our distributed teams", "remote workers using Copilot").
- If excerpt contains ONLY mentions/sentiment with no investment or experience evidence, all H_* labels should be "not_addressed".
- Obstacles with no action taken -> H_experience_negative = "not_addressed" (just a concern, not an experienced harm).
- Obstacles that caused actual problems -> H_experience_negative = "entailed".
\end{lstlisting}
\end{tcolorbox}

\paragraph{Step 3: Skeptic Audit (Generative AI)}

\begin{tcolorbox}[colback=gray!5, colframe=gray!50!black, breakable, title={}]
\begin{lstlisting}[style=prompt]
Act as a hostile auditor. Your job is to find any place the adjudication overreached the evidence.

You are given:
(1) The original excerpt
(2) The adjudication JSON (hypothesis labels, evidence quotes, dimensions)

Your task: Verify that every claim and label is justified by the excerpt text.

Excerpt:
{EXCERPT}

Adjudication JSON:
{ADJUDICATION_JSON}

CHECK FOR THESE SPECIFIC ISSUES:

1. UNSUPPORTED QUOTES: Do the evidence_quotes actually appear in the excerpt? 
   Flag any quotes that are paraphrased, fabricated, or significantly altered.

2. MENTION CONFLATION: Was a mere mention of AI incorrectly used to support H_invest_enablement or H_experience_*?
   - "We're excited about AI" should NOT entail investment or positive experience
   - Only flag if this error actually affected the hypothesis labels

3. PRODUCT/MARKET LEAKAGE: Was commentary about AI products, revenue, or market trends 
   incorrectly treated as evidence about the company's own internal AI adoption?

4. OVERCONFIDENT LABELS: Should any "entailed" label be downgraded to "not_addressed"?
   - Evidence too weak or ambiguous?
   - Future plans mistaken for current implementation?
   - Inference required that isn't explicitly supported?

5. SCOPE CREEP: Was traditional ML/analytics incorrectly classified as generative AI?
   - Predictive models, recommendation engines -> NOT generative AI
   - LLMs, ChatGPT, Copilot, image generation -> IS generative AI

6. OBSTACLE vs EXPERIENCED HARM: Were stated concerns incorrectly marked as actual negative experiences?
   - "We're concerned about data privacy" -> obstacle, NOT experienced harm
   - "We had a data breach from our AI system" -> IS experienced harm

7. ENABLING INFRASTRUCTURE OVERREACH: Was general IT infrastructure incorrectly linked to AI enablement?
   - Only flag if the excerpt doesn't actually connect the infrastructure to AI

8. REMOTE WORK FLAG: Was remote_work_mentioned set correctly?
   - Should only be true if remote/hybrid work is explicitly connected to AI discussion

IMPORTANT: Only flag genuine issues. Reasonable inferences supported by clear evidence are NOT overreach.

Return your response as a JSON object with this exact structure:
```json
{
  "overreach_flags": [
    {
      "issue_type": "<one of: unsupported_quote, mention_conflation, product_leakage, overconfident, scope_creep, obstacle_vs_harm, infrastructure_overreach, remote_work_flag, missed_evidence>",
      "item": "<specific item that was overreached - quote, label, or dimension>",
      "why": "<explanation of the problem>",
      "suggested_fix": "<how to correct it>"
    }
  ],
  "revised_labels": {
    "H_invest_enablement": "<entailed | contradicted | not_addressed | unchanged>",
    "H_experience_positive": "<entailed | contradicted | not_addressed | unchanged>",
    "H_experience_negative": "<entailed | contradicted | not_addressed | unchanged>"
  },
  "revised_investment_dimensions": {
    "llm_apis": <boolean or null if unchanged>,
    "copilot_tools": <boolean or null if unchanged>,
    "chatbots_internal": <boolean or null if unchanged>,
    "training_programs": <boolean or null if unchanged>,
    "compute_infrastructure": <boolean or null if unchanged>,
    "cloud_ai_services": <boolean or null if unchanged>,
    "data_preparation": <boolean or null if unchanged>,
    "governance_frameworks": <boolean or null if unchanged>,
    "vendor_partnerships": <boolean or null if unchanged>,
    "custom_models": <boolean or null if unchanged>
  },
  "revised_enabling_infrastructure": {
    "cloud_maturity": <boolean or null if unchanged>,
    "data_infrastructure": <boolean or null if unchanged>,
    "developer_tools": <boolean or null if unchanged>,
    "prior_ml_investments": <boolean or null if unchanged>,
    "security_infrastructure": <boolean or null if unchanged>
  },
  "revised_obstacles": {
    "data_privacy": <boolean or null if unchanged>,
    "security_risks": <boolean or null if unchanged>,
    "regulatory_compliance": <boolean or null if unchanged>,
    "cost_concerns": <boolean or null if unchanged>,
    "talent_gap": <boolean or null if unchanged>,
    "integration_complexity": <boolean or null if unchanged>,
    "quality_accuracy": <boolean or null if unchanged>,
    "employee_resistance": <boolean or null if unchanged>,
    "vendor_lock_in": <boolean or null if unchanged>
  },
  "revised_experience_dimensions": {
    "productivity": <integer -1/0/1 or null if unchanged>,
    "code_quality": <integer -1/0/1 or null if unchanged>,
    "cost_efficiency": <integer -1/0/1 or null if unchanged>,
    "speed_time_savings": <integer -1/0/1 or null if unchanged>,
    "employee_satisfaction": <integer -1/0/1 or null if unchanged>,
    "innovation": <integer -1/0/1 or null if unchanged>,
    "output_quality": <integer -1/0/1 or null if unchanged>,
    "decision_making": <integer -1/0/1 or null if unchanged>
  },
  "revised_has_adopted_technology": <integer 0/1 or null if unchanged>,
  "revised_remote_work_mentioned": <boolean or null if unchanged>,
  "audit_passed": <boolean - true if no significant issues found>,
  "brief_rationale": "<2-3 sentences summarizing your audit findings>"
}
```

Notes:
- "overreach_flags": Include one object per issue found. Use empty array [] if no issues detected.
- "revised_*": Use null for fields that don't need changes. Only include actual values for corrections.
- "audit_passed": Set to true if the adjudication is fundamentally sound, even if minor issues exist.
- Be fair: Don't flag issues that don't materially affect the analysis.
\end{lstlisting}
\end{tcolorbox}

\end{appendices}

\end{document}